\documentclass[10pt,aps,prd,floatfix,nofootinbib,superscriptaddress]{revtex4-2}
\usepackage{url}
\usepackage{xcolor}
\usepackage[utf8]{inputenc}
\usepackage{amssymb}
\usepackage{amsmath}
\usepackage{graphicx}
\colorlet{shadecolor}{gray!15}
\definecolor{greenLinks}{rgb}{0, 0.6, 0}
\definecolor{blueLinks}{rgb}{0, 0, 0.6}
\definecolor{redLinks}{rgb}{0.6, 0, 0}
\definecolor{tempText}{rgb}{0.55, 0.10,0.67}
\definecolor{eprintLinks}{rgb}{0.4, 0.4, 0.4}
\definecolor{journalLinks}{rgb}{0.6, 0, 0}
\usepackage{footnote}
\usepackage{multirow}
\usepackage{enumitem}
\def\dg{\dagger}     

\def\slc#1{\setbox0=\hbox{$#1$}                  
    \dimen0=\wd0                                 
    \setbox1=\hbox{/} \dimen1=\wd1               
    \ifdim\dimen0>\dimen1                        
       \rlap{\hbox to \dimen0{\hfil/\hfil}}      
       #1                                        
    \else                                        
       \rlap{\hbox to \dimen1{\hfil$#1$\hfil}}   
       /                                         
    \fi}

\def\be{\begin{equation}}
\def\ee{\end{equation}}
\def\gs{\mathrel{
   \rlap{\raise 0.511ex \hbox{$>$}}{\lower 0.511ex \hbox{$\sim$}}}}
\def\ls{\mathrel{
   \rlap{\raise 0.511ex \hbox{$<$}}{\lower 0.511ex \hbox{$\sim$}}}}

\newcommand{\ba}{\begin{array}{c}}
\newcommand{\baz}{\begin{array}{cc}}
\newcommand{\barrr}{\begin{array}{rrr}}
\newcommand{\bad}{\begin{array}{ccc}}
\newcommand{\bav}{\begin{array}{cccc}}
\newcommand{\baf}{\begin{array}{ccccc}}
\newcommand{\bea}{\begin{equation} \begin{array}{c}}
\newcommand{\eea}{ \end{array} \end{equation}}
\newcommand{\ea}{\end{array}}

\usepackage[normalem]{ulem}

\def\21{$\mathrm{SU(2)_L \otimes U(1)_Y}$ }

\newcommand {\ignore}[1]{}

\newcommand{\vt}{\vert}
\newcommand{\nn}{\nonumber}

\usepackage{hyperref}

\allowdisplaybreaks[2]
\newcommand{\AddrFCEBUAP}{
 Fac. de Cs. de la Electr\'onica, 
 Benem\'erita Universidad Aut\'onoma de Puebla, 
 Apdo. Postal 542,\\ Puebla, Pue. 72000, M\'exico.}

\newcommand{\AddrCECYT}{Instituto Polit\'ecnico Nacional, CECyT No.16, Carretera 
 Pachuca-Actopan Kil\'ometro 1+500, Distrito de Educaci\'on, Salud, Ciencia, 
 Tecnolog\'ia e Innovaci\'on, San Agust\'in Tlaxiaca, Hidalgo, C.P. 42162, 
 M\'exico.}

\newcommand{\AddrUPIIH}{Instituto Polit\'ecnico Nacional, UPIIH, Carretera 
 Pachuca-Actopan Kil\'ometro 1+500, Distrito de Educaci\'on, Salud, Ciencia, 
 Tecnolog\'ia e Innovaci\'on, San Agust\'in Tlaxiaca, Hidalgo, C.P. 42162, 
 M\'exico.}
\begin{document}
\title{Flavored \texorpdfstring{$\mathbf{S}_{3}$}{S3} Non-Minimal Left-Right Symmetric Model} 
%
\author{F. Gonzalez-Canales}
 \email{felix.gonzalezc@correo.buap.mx}
 \affiliation{\AddrFCEBUAP}
%
%
\author{Juan Carlos G\'omez-Izquierdo}
 \email{cizquierdo@ipn.mx}
\affiliation{\AddrCECYT}
\affiliation{\AddrUPIIH}
%
\date{\bf \today} 
\begin{abstract}\vspace{2cm}
 In this work we investigate fermion masses and flavor mixings 
 in a non-minimal left-right symmetric model endowed with the 
 non-Abelian discrete symmetry $S_3$. 
 The scalar sector contains three Higgs bidoublets, while one 
 left-handed and one right-handed triplet generate Majorana
 neutrino masses. Assigning the first two fermion families to 
 an $S_3$ doublet and the third family to a singlet, together 
 with a specific alignment of the bidoublet vacuum expectation 
 values, leads to a common and analytically tractable structure 
 for the fermion mass matrices. We study both the pseudomanifest 
 and manifest left-right realizations, in which the fermion mass 
 matrices are complex symmetric and Hermitian, respectively, 
 and derive analytical expressions for the left- and 
 right-handed CKM and PMNS mixing matrices. 
 A numerical likelihood test shows that the model successfully reproduces
 the observed quark and lepton mixing patterns for both normal and inverted
 neutrino mass orderings at the level of oscillation data. 
 Remarkably, the manifest scenario emerges as a more restrictive realization 
 of the pseudomanifest one, with the main distinction residing in their
 CP-phase structure. We further explore the phenomenological consequences 
 for the neutrinoless double-beta decay, the muon anomalous magnetic moment, 
 and the lepton-flavor-violating decay $\mu\to e\gamma$. The
right-handed contribution exhibits the characteristic dependence on
the heavy-neutrino and $W_R$ mass scales, while flavor corrections are
found to be subleading over most of the allowed parameter space.
These results show that an $S_3$ flavor symmetry can simultaneously
organize fermion mixing and lead to distinctive low-energy signatures
within left-right symmetric theories.
\end{abstract}

%
\maketitle
\section{Introduction}
%
The discovery of the Higgs boson in 2012~\cite{ATLAS:2012yve,CMS:2012qbp} 
represented a major triumph for the Standard Model (SM) of elementary particles, 
which is currently the most successful theory describing the interactions among 
fundamental particles. Nevertheless, several important questions remain 
unanswered, such as the origin 
of neutrino masses\cite{Romao:1991ex,Hirsch:2004he,Murayama:2006qb}, the flavor
puzzle~\cite{Feruglio:2015jfa,Abbas:2023ivi,Nilles:2023shk}, and the 
nature of dark matter~\cite{Profumo:2019ujg}. 

In this context, the Minimal Left-Right Symmetric Model (MLRSM)~\cite{Pati:1974yy,
Mohapatra:1974gc,Senjanovic:1975rk,Senjanovic:1978ev,Mohapatra:1979ia} 
provides an appealing framework that addresses several shortcomings of the SM. 
For instance, it offers a natural explanation for parity violation at low 
energies, which is closely related to the origin of neutrino 
masses~\cite{Senjanovic:2019moe}. 
Consequently, the seesaw 
mechanism~\cite{GellMann:1980vs,Yanagida:1979as,Mohapatra:1980yp,
Mohapatra:1979ia,Minkowski:1977sc,Schechter:1980gr,Schechter:1981cv} 
emerges naturally, providing an elegant explanation for the smallness of neutrino 
masses. Furthermore, within the MLRSM, the SM hypercharge ($Y$) 
acquires a physical interpretation in terms of baryon and lepton number 
difference ($B-L$). These features make the MLRSM an attractive candidate for 
physics beyond the SM.

Over the years, the MLRSM has attracted considerable attention, since many 
physical processes have been proposed to search for characteristic signatures 
that could be observed at the LHC~\cite{Nemevsek:2011hz,Chen:2013fna,
Dev:2015kca,Patra:2015bga,Lindner:2016lpp,Mitra:2016kov,Dev:2016dja,
Chakrabortty:2016wkl,Cottin:2019drg,Bertolini:2019out,Hernandez:2021uxx}. In 
addition, the right-handed CKM matrix, which is the counterpart of the 
left-handed CKM matrix~\cite{Cabibbo:1963yz,Kobayashi:1973fv}, has been 
extensively studied due to its direct impact on the physics of the $W_{R}$ 
boson, neutrinoless double beta 
decay~\cite{Vergados:2002pv,Rodejohann:2011mu,Vergados:2012xy,
Pas:2015eia}, and the theoretical constraints arising from $K$- and $B$-meson 
physics~\cite{Senjanovic:2014pva,Senjanovic:2015yea}. On the other hand, in the 
lepton sector, the right-handed PMNS matrix contributes to the anomalous magnetic 
moment of the muon as well as to lepton-number-violating 
processes~\cite{Lindner:2016bgg}. 
These studies provide additional motivation to further explore this theoretical 
framework from different phenomenological perspectives.

Nonetheless, due to the gauge structure of the MLRSM, the Yukawa sector contains 
a large number of free couplings, which represents an important drawback of the 
model. Although the imposition of parity symmetry partially alleviates this 
issue, it is still desirable to understand the fermion mixing patterns from the 
perspective of flavor symmetries~\cite{Ishimori:2010au,Grimus:2011fk,King:2013eh,
King:2015aea,Feruglio:2019ybq,Chauhan:2022gkz}.\footnote{Remarkably, left-right 
symmetric models have recently been studied within the framework of modular 
symmetries~\cite{Kakoti:2023isn,Kalita:2024vlt,Kakoti:2025eub,Kumar:2025nut}.}
For this reason, the study of fermion mixings in this class of models has 
attracted comparatively limited 
attention~\cite{Rodejohann:2015hka,GomezIzquierdo:2009id,
Gomez-Izquierdo:2017rxi,Das:2018rdf,Garces:2018nar,Borgohain:2018lro,
CarcamoHernandez:2018hst,Bonilla:2020hct}. 
Closely related flavored non-minimal left-right symmetric models were
previously studied in Refs.~\cite{Gomez-Izquierdo:2017rxi,Garces:2018nar},
where the flavor structure was based on the enlarged discrete symmetry
$S_3\otimes Z_2\otimes Z_2^e$. Those constructions employed three Higgs
bidoublets together with three left-handed and three right-handed
triplets, and the leptonic mixing pattern was analyzed through a soft
breaking of the $\mu\leftrightarrow\tau$ symmetry. The present work
differs from that realization in several essential aspects. Here, the
flavor sector is governed solely by $S_3$, the first two fermion
families transform as a doublet while the third family transforms as
$\mathbf{1}_S$, and only one left-handed and one right-handed Higgs
triplet are retained. This reduced field content leads to a different
Majorana sector and to a common analytical structure for the quark,
charged-lepton, and Dirac-neutrino mass matrices.
Beyond these structural differences, the present analysis develops the
left- and right-handed flavor sectors simultaneously in both the
pseudomanifest and manifest left-right realizations. 
We derive explicit analytical expressions for the corresponding 
CKM~\cite{Cabibbo:1963yz,Kobayashi:1973fv} and 
PMNS~\cite{Maki:1962mu,Pontecorvo:1967fh} mixing
matrices and show that the manifest realization can be understood as a
more restrictive limit of the pseudomanifest one, mainly through its
reduced CP-phase structure. 
The numerical analysis is updated using current quark and neutrino data
and accommodates both normal and inverted neutrino mass orderings at
the level of oscillation data.
Furthermore, in contrast with
Ref.~\cite{Garces:2018nar}, the analysis is extended beyond fermion
masses and mixings to the phenomenology of neutrinoless double-beta
decay, the muon anomalous magnetic moment, and the
lepton-flavor-violating decay $\mu\to e\gamma$, including the
constraints imposed by the current MEG~II limit.
These differences motivate a renewed analysis of the flavor structure,
with particular emphasis on obtaining a unified description of the
markedly different CKM and PMNS mixing patterns, which play a crucial 
role in several relevant physical processes.

To this end, in this work we focus on the well-studied non-Abelian group 
$S_{3}$ (see, for instance, the references therein~\cite{Ishimori:2010au,
Grimus:2011fk,King:2013eh,King:2015aea,Feruglio:2019ybq,Chauhan:2022gkz}), 
which constitutes an attractive candidate for describing fermion mixing patterns, 
within the framework of the left-right symmetric theory. In order to consistently 
implement the flavor symmetry, the scalar sector must also be extended, requiring 
the introduction of three Higgs bidoublets. To preserve a minimal matter content 
in the Majorana sector, one left-handed and one right-handed Higgs triplet are 
introduced to generate the corresponding neutrino masses. 
With these ingredients, all fermion fields are treated on the same footing under 
the flavor symmetry, with the three families assigned to the irreducible 
representations $\mathbf{2}\oplus \mathbf{1}_{S}$. In this way, a unified 
treatment of fermion masses is achieved. 
At high energies, parity symmetry is imposed in the model, although it is
spontaneously broken at lower energies. In addition, we adopt a specific
alignment of the Higgs-bidoublet vacuum expectation values as a
phenomenological working assumption. This alignment, specified explicitly
below, selects a symmetric direction in the $S_3$-doublet scalar sector and
considerably reduces the number of independent entries in the fermion mass
matrices. Its dynamical realization would require a dedicated analysis of
the complete scalar potential, which is not attempted in the present work.
Together, these assumptions determine the flavor structure analyzed below.
 We then study both the pseudomanifest 
 and manifest left-right realizations, in which the fermion mass 
 matrices are complex symmetric and Hermitian, respectively, 
 and derive analytical expressions for the left- and 
 right-handed CKM and PMNS mixing matrices. A numerical likelihood test 
 shows that the model successfully reproduces the observed quark and 
 lepton mixing patterns for both normal and inverted neutrino mass orderings.
 In the literature, these mass orderings are also commonly referred to as
 the normal and inverted hierarchies, NH and IH, 
 respectively~\cite{Garces:2018nar}.
 Remarkably, the manifest scenario emerges as a more restrictive realization 
 of the pseudomanifest one, with the main distinction residing in their
 CP-phase structure. We further explore the phenomenological consequences 
 for the neutrinoless double beta decay, the muon anomalous magnetic moment, 
 and the lepton-flavor-violating decay $\mu\to e\gamma$. The
right-handed contribution exhibits the characteristic dependence on
the heavy-neutrino and $W_R$ mass scales, while flavor corrections are
found to be subleading over most of the allowed parameter space.
These results show that an $S_3$ flavor symmetry can simultaneously
organize fermion mixing and lead to distinctive low-energy signatures that can be 
tested in current and future experiments within left-right symmetric theories.

The paper is organized as follows. In Section II, we provide a brief review 
of the Minimal Left-Right Symmetric Model. In Section III, we introduce the 
flavored model and explicitly diagonalize the quark and lepton mass matrices in 
order to construct the CKM and PMNS mixing matrices, whose parameters are 
determined through a numerical $\chi^{2}$ analysis. After obtaining a successful 
fit for the lepton sector, in Section~IV we study the implications of the model 
for neutrinoless double beta decay, the muon anomalous magnetic moment ($g-2$), 
and the lepton-flavor-violating process $\mu \rightarrow e\gamma$. 
Finally, our conclusions are presented in Section~V.
%
\section{Minimal left-right symmetric model: a brief review}
%
The Minimal Left-Right Symmetric Model 
(MLRSM)~\cite{Pati:1974yy,Mohapatra:1974gc,Senjanovic:1975rk,
Senjanovic:1978ev,Mohapatra:1979ia} is based on the gauge symmetry
$\mathbf{SU(3)}_{c}\otimes \mathbf{SU(2)}_{L}\otimes 
\mathbf{SU(2)}_{R}\otimes \mathbf{U(1)}_{B-L}$. 
The parity symmetry, $\mathcal{P}$, is assumed to be an exact symmetry at high
energies and is spontaneously broken at the scale associated with
$SU(2)_R$ breaking, above the electroweak scale.
This is consistent with the absence of right-handed weak currents at
low energies. 
As is well known, right-handed neutrinos (RHNs) are naturally incorporated into 
the matter content of the model, as shown in Table~\ref{Mat1}. 
\begin{table}[h!]
 \begin{center}
  \begin{tabular}{|c|c|c|c|c|c|c|c|c|c|} \hline \hline	
   {\footnotesize Matter} & 
   {\footnotesize ${\bf SU(3)}_{C}$} & 
   {\footnotesize ${\bf SU(2)}_{L}$} & 
   {\footnotesize ${\bf SU(2)}_{R}$} & 
   {\footnotesize ${\bf U(1)}_{B-L}$} \\\hline
   {\footnotesize  $Q_{(L,R)}=
    \begin{pmatrix}
     u \\
     d \\
    \end{pmatrix}_{(L, R)}$ } &  
   {\footnotesize ${\bf 3}$} & 
   {\footnotesize $({\bf 2},{\bf 1})$} & 
   {\footnotesize $({\bf 1}, {\bf 2})$} & 
   {\footnotesize  $\frac{1}{3}$} \\\hline
   {\footnotesize  $(L,R)=
    \begin{pmatrix}
     \nu \\
     \ell \\
    \end{pmatrix}_{(L,R)}$ } &  
   {\footnotesize ${\bf 1}$} & 
   {\footnotesize $({\bf 2}, {\bf 1})$} & 
   {\footnotesize $({\bf 1},{\bf 2})$} & 
   {\footnotesize  $-1$} \\ \hline
   {\footnotesize  $\Delta_{(L,R)}=\begin{pmatrix}
    \frac{\delta^{+}}{2} & \delta^{++} \\
    \delta^{0} & -\frac{\delta^{+}}{2} \\
   \end{pmatrix}$} &  
   {\footnotesize ${\bf 1}$} & {\footnotesize $({\bf 3},{\bf 1})$}   & 
   {\footnotesize $({\bf 1}, {\bf 3})$} & {\footnotesize  $2$} \\\hline
   {\footnotesize $\Phi=
    \begin{pmatrix}
     \phi^{0} & \phi^{'+} \\
     \phi^{-} & \phi^{'0} \\
    \end{pmatrix}$ } &  
   {\footnotesize ${\bf 1}$} & 
   {\footnotesize ${\bf 2}$} & 
   {\footnotesize ${\bf 2}$} & 
   {\footnotesize  $0$} \\ \hline \hline
  \end{tabular}\caption{Matter content in the MLRSM}\label{Mat1}
 \end{center}
\end{table}
Then, the gauge-invariant Yukawa interaction is given by
\begin{equation}\label{yt}
 -\mathcal{L}_{Y} = 
 \bar{Q}_{L}\left[y^{q}\Phi+ \tilde{y}^{q}\tilde{\Phi} 
 \right]Q_{R}+ \bar{L}\left[y^{\ell}\Phi+ 
 \tilde{y}^{\ell}\tilde{\Phi} \right]R+ 
 y^{L}\bar{L}\Delta_{L}L^{c}+y^{R}\bar{R}^{c}\Delta_{R}R 
 + h.c. 
\end{equation}
with $\tilde{\Phi}=-i\sigma_{2}\Phi^{\ast}i\sigma_{2}$  and the 
charged current is explicitly written as
\begin{equation}
 \mathcal{L}_{CC}=\frac{g_{L}}
 {\sqrt{2}}\left[\bar{Q}_{L}\gamma_{\mu} W^{\mu}_{L} Q_{L}+ 
 \bar{L}\gamma_{\mu} W^{\mu}_{L} L \right]+\frac{g_{R}}
 {\sqrt{2}}\left[\bar{Q}_{R}\gamma_{\mu} W^{\mu}_{R} Q_{R}+ 
 \bar{R}\gamma_{\mu} W^{\mu}_{R} R \right] 
 + h.c.
\end{equation}
For simplicity, the family indices have been suppressed in the 
above expressions.

On the other hand, we emphasize that parity symmetry is assumed 
in the above Lagrangian. Consequently, the following transformations are 
required:
$\Psi_{L}\leftrightarrow \Psi_{R}$ for the fermion fields, and
$\Phi\leftrightarrow \Phi^{\dagger}$ together with
$\Delta_{L}\leftrightarrow \Delta_{R}^{\dagger}$ for the scalar fields. 
As a result, the Yukawa sector is considerably simplified, leading to the 
relations $y=y^{\dagger}$, $\tilde{y}=\tilde{y}^{\dagger}$, and 
$y^{R}=y^{L}$. Likewise, parity symmetry implies the equality of the gauge 
couplings, namely $g_{L}=g_{R}$.
A complete analysis of the most general scalar potential invariant under
the gauge symmetry, parity, and the $S_3$ flavor symmetry is beyond the
scope of the present work, whose main purpose is to investigate the
fermion-mass and flavor-mixing sectors. We therefore do not attempt to
derive the vacuum configuration from the minimization conditions of the
full scalar potential. Instead, we assume that the scalar potential admits
a phenomenologically viable stationary configuration characterized by the
usual left-right hierarchy
\(
v_R \gg |k_i|,\,|k_i'| \gg v_L,
\quad i=1,2,3,
\)
together with the bidoublet alignment specified below. The present
analysis should therefore be understood as a study of the flavor
consequences of this vacuum configuration, rather than as a proof that it
corresponds to the global minimum of the complete scalar potential.
Questions concerning vacuum stability, possible competing minima, and the
physical scalar spectrum are left for a dedicated study.

Within this assumption, parity symmetry is broken at the scale at which
the right-handed Higgs triplet $\Delta_R$ acquires its vacuum expectation
value (VEV). At this stage, the right-handed neutrinos acquire Majorana
masses. Electroweak symmetry breaking subsequently occurs when the Higgs
bidoublets develop their corresponding VEVs, generating the Dirac fermion
masses.
Explicitly, 
\begin{equation}\label{eq6}	
 \langle \Delta_{(L, R)} \rangle = 
 \begin{pmatrix}
  0         & 0 \\
  v_{(L,R)} & 0 \\
 \end{pmatrix}, 
 \quad \langle \Phi \rangle = 
 \begin{pmatrix}
  k & 0 \\
  0 & k^{\prime}  \\
 \end{pmatrix}, 
 \quad \langle \tilde{\Phi} \rangle = 
 \begin{pmatrix}
  k^{\prime \ast} & 0 \\
  0               & k^{\ast}  \\
 \end{pmatrix}.
\end{equation}

Once the scalar fields acquire their VEVs, the corresponding fermion mass terms 
take the form
\begin{align}\label{eq7}
 -\mathcal{L}_{Y} = 
 \bar{q}_{ L} {\bf M}_{q} q_{ R} + 
 \bar{e}_{ L} {\bf M}_{e} e_{R} + 
 \frac{1}{2}\left( \begin{matrix} 
  \bar{\nu}_{L} & \bar{\nu}^{c}_{R}
 \end{matrix} \right) 
 \overbrace{\left(\begin{matrix}
  \mathbf{M}_{L}	& \mathbf{M}_{D} \\
  \mathbf{M}^{T}_{D}	& \mathbf{M}_{R}
 \end{matrix} \right)}^{\mathcal{M}_{\nu}} 
 \left(\begin{matrix}
  \nu^{c}_{ L} \\ 
  \nu_{ R}
 \end{matrix} \right) + 
  h.c. 
\end{align}
where ${\bf M}_{q}$ ($q=u,d$) and ${\bf M}_{e}$ denote the quark and
charged-lepton mass matrices, respectively. In addition,
$\mathbf{M}_{(L,R)}=y^{(L,R)}v_{(L,R)}$ and
$\mathbf{M}_{D}=y^{\ell}k+\tilde y^{\ell}k^{\prime *}$ denote the
Majorana and Dirac neutrino mass matrices, respectively. The full
neutrino mass matrix $\mathcal M_\nu$ is diagonalized by a $6\times6$
unitary matrix $\mathcal V$ such that
\begin{equation}
\mathcal{V}^{T}\mathcal{M}_{\nu}\mathcal{V}
=
\begin{pmatrix}
\hat{\mathbf M}_{\nu} & \mathbf{0}\\
\mathbf{0} & \hat{\mathbf M}_{R}
\end{pmatrix}.
\end{equation}
In the seesaw limit, the unitary matrix $\mathcal V$ can be written as
\begin{equation}
 \mathcal{V} = 
 \begin{pmatrix}
  \mathbf{U}	& \mathbf{S} \\
  \mathbf{T}	& \mathbf{V}
 \end{pmatrix} 
 \simeq 
 \begin{pmatrix}
  \mathbf{1} - \frac{1}{2} \theta \theta^{\dagger} & \theta \\
  -\theta^{\dagger} & \mathbf{1} - \frac{1}{2} \theta^{\dagger} \theta 
 \end{pmatrix}
 \begin{pmatrix}
 \mathbf{U}_{\nu L} & \mathbf{0}\\
 \mathbf{0} & \mathbf{U}_{\nu R}
\end{pmatrix}, 
 \quad \mathcal{V} \mathcal{V}^{\dagger} = 
 \mathbf{1} = \mathcal{V}^{\dagger}\mathcal{V}.
\end{equation}
where $\hat{\mathbf M}_{\nu}$ and $\hat{\mathbf M}_{R}$ contain the
physical light- and heavy-neutrino masses, respectively. In the expression
above, terms up to second order in the light-heavy mixing matrix
$\theta$ have been retained, while higher-order contributions are
neglected. 
Then, $\mathbf{U}\approx \mathbf{U}_{\nu L}$, 
$\mathbf{S}\approx \theta \mathbf{U}_{\nu R}$, 
$\mathbf{T}\approx - \theta^{\dagger} \mathbf{U}_{\nu L}$ and 
$\mathbf{V}\approx \mathbf{U}_{\nu R}$. 
Thus, the mass and interaction neutrino eigenstates are related by
\begin{equation}\label{nbi}
 \begin{pmatrix}
  \nu_{L}	\\
  \left( \nu_{R} \right)^{c}		
 \end{pmatrix} = \mathcal{V}
 \begin{pmatrix}
  \tilde{\nu}_{L}	\\
  \left( \tilde{\nu}_{R} \right)^{c}
 \end{pmatrix}, \qquad 
 \begin{pmatrix}
  \left( \nu_{L} \right)^{c}	\\
  \nu_{R}	
 \end{pmatrix} = \mathcal{V}^{\ast} 
 \begin{pmatrix} 
  \left( \tilde{\nu}_{L} \right)^{c} \\
  \tilde{\nu}_{R}	
 \end{pmatrix}
\end{equation}
As a result, we obtain 
$\mathbf{U}^{\dagger}_{\nu L} \mathbf{M}_{\nu} \mathbf{U}^{\ast}_{\nu L} = 
\hat{\mathbf{M}}_{\nu}$ and 
$\mathbf{U}^{\dagger}_{\nu R} \mathbf{M}_{R} \mathbf{U}^{\ast}_{\nu R}= 
\hat{\mathbf{M}}_{R}$  
where the effective light-neutrino mass matrix arises from the combined
type-I and type-II seesaw contributions and is given by
\begin{equation}
 {\bf M}_{\nu}={\bf M}_{L}-{\bf M}_{D} {\bf M}^{-1}_{R} {\bf M}^{T}_{D}.
\end{equation}
Returning to the charged-current Lagrangian, once parity symmetry is
imposed, it can be written as
\begin{equation}
 \mathcal{L}_{CC} = 
 \frac{g}{\sqrt{2}} 
 \left[ 
  \left \{  
   \bar{d}_{L} \gamma_{\mu} u_{L} + 
   \bar{e}_{L} \gamma_{\mu} \nu_{L} 
   \right \} W^{-\mu}_{L} 
   +
   \left\{
    \bar{d}_{R} \gamma_{\mu} u_{R} + 
    \bar{e}_{R} \gamma_{\mu} \nu_{R} 
   \right\} W^{-\mu}_{R}
 \right] + h.c.
\end{equation}
where the $W^{\mp\mu}_{1,2}$ mass eigenstates are related to the 
$W^{\mp\mu}_{L,R}$ interaction ones as
\begin{equation}\label{wbi}
 \begin{pmatrix}
  W^{\mp\mu}_{L}	\\
  W^{\mp\mu}_{R}	
 \end{pmatrix} = 
 \begin{pmatrix}
  \cos{\zeta}  & \sin{\zeta} \\
  -\sin{\zeta} & \cos{\zeta}
 \end{pmatrix} 
 \begin{pmatrix}
  W^{\mp\mu}_{1}	\\
  W^{\mp\mu}_{2}	
 \end{pmatrix} 
\end{equation}
Finally, using Eqs.~(\ref{nbi}) and~(\ref{wbi}), we have
\begin{eqnarray}\label{CCL}
 \mathcal{L}_{CC} & = & 
 \frac{g}{\sqrt{2}} \left[ \bar{d}_{L} \gamma_{\mu} u_{L} + 
 \bar{e}_{L} \gamma_{\mu} \left( \mathbf{U} \tilde{\nu}_{L} + 
 \mathbf{S} \left( \tilde{\nu}_{R} \right)^{c} \right) \right] 
 \left[ \cos{\zeta} ~ W^{-}_{1 \mu} + \sin{\zeta}~W^{-}_{2 \mu} \right] \nn \\
 &&
 + 
 \frac{g}{\sqrt{2}} \left[ \bar{d}_{R} \gamma_{\mu} u_{R} + 
 \bar{e}_{R} \gamma_{\mu} \left( \mathbf{T}^{\ast} 
 \left( \tilde{\nu}_{L} \right)^{c} + 
 \mathbf{V}^{\ast} \tilde{\nu}_{R} \right) \right]
 \left[-\sin{\zeta}~W^{-}_{1 \mu}+\cos{\zeta}~W^{-}_{2 \mu} \right] + 
 h.c.
\end{eqnarray} 
In the MLRSM, the mixing angle $\zeta$ depends on the charged
gauge-boson masses and on the bidoublet VEVs. Existing phenomenological
constraints require this mixing to be small. Therefore, throughout the
phenomenological analysis we work in the limit
$\zeta\simeq0$, so that $W_1\simeq W_L$ and $W_2\simeq W_R$.
Corrections proportional to $\zeta$ are not included in the present
analysis.
%
\section{Flavored left-right symmetric model}
%
Having briefly described the theoretical framework, we now comment on the 
additional ingredients to study fermion masses and mixing patterns. 
To this purpose, $\Phi_{i}$ ($i=1,2,3$), which are responsible for generating the 
fermion mixing structure. In the Majorana sector, one left-handed and one 
right-handed Higgs triplet are retained, as in the conventional MLRSM. In this 
way, we aim to preserve a minimally extended scalar sector.

Additionally, the MLRSM is supplemented with the flavor symmetry 
$S_{3}$. The main motivation for employing this non-Abelian discrete 
symmetry lies in the structure of its reducible representations, 
$\mathbf{3}_{S}$ and $\mathbf{3}_{A}$, which decompose as 
$\mathbf{3}_{S}=\mathbf{1}_{S}\oplus\mathbf{2}$ and 
$\mathbf{3}_{A}=\mathbf{1}_{A}\oplus\mathbf{2}$, respectively. This feature 
allows the three fermion families to be naturally accommodated within these 
decompositions, providing a framework to understand the observed hierarchies 
of quark and charged lepton masses.

Accordingly, the quark, lepton, and Higgs bidoublet fields are assigned under 
$S_{3}$ in the following way: the first and second families transform 
as a doublet $\mathbf{2}$, whereas the third family is assigned to the singlet 
representation $\mathbf{1}_{S}$. The complete field content and its assignments 
are shown in Table~\ref{fMat1}.
\begin{table}[htbp]
 \begin{center}
  \begin{tabular}{|c|c|c|c|c|c|c|c|c|c|}\hline \hline	
   {\footnotesize Matter} & 
   {\footnotesize $Q_{I (L, R)}$} & 
   {\footnotesize $Q_{3 (L, R)}$} & 
   {\footnotesize $(L_{I}, R_{I})$} & 
   {\footnotesize $(L_{3}, R_{3})$} & 
   {\footnotesize $\Phi_{I}$} & 
   {\footnotesize $\Phi_{3}$} &  
   {\footnotesize $\Delta_{(L, R)}$ } \\ \hline
   {\footnotesize  ${\bf S}_{3}$} & 
   {\footnotesize  ${\bf 2}$} & 
   {\footnotesize  ${\bf 1}_{S}$} & 
   {\footnotesize  ${\bf 2}$} & 
   {\footnotesize  ${\bf 1}_{S}$} & 
   {\footnotesize  ${\bf 2}$} & 
   {\footnotesize  ${\bf 1}_{S}$} & 
   {\footnotesize  ${\bf 1}_{S}$}  \\ \hline \hline
  \end{tabular}\caption{Flavored non-minimal left-right model. 
   Here, $I=1,2$.}\label{fMat1}
 \end{center}
\end{table}
As a result of the field assignments under the flavor symmetry, 
the Yukawa mass terms that are invariant under both the $S_{3}$ flavor 
symmetry and the gauge symmetry are given by
\begin{align}
 -\mathcal{L}_{Y} & =
 y^{q}_{1} \left[ 
  \bar{Q}_{1 L} \left( \Phi_{1} Q_{2 R} + \Phi_{2} Q_{1 R} \right) + 
  \bar{Q}_{2 L} \left( \Phi_{1} Q_{1 R} - \Phi_{2} Q_{2 R} \right) \right] + 
  y^{q}_{2} \left[ \bar{Q}_{1 L} \Phi_{3} Q_{1 R} + 
   \bar{Q}_{2 L} \Phi_{3} Q_{2 R} \right]  \nn \\
   & +
  y^{q}_{3} \left[ \bar{Q}_{1 L} \Phi_{1} + \bar{Q}_{2 L} \Phi_{2} \right] 
  Q_{3 R} +
  y^{q}_{4} \bar{Q}_{3 L} \left[ \Phi_{1} Q_{1 R} + \Phi_{2} Q_{2 R} \right] + 
  y^{q}_{5} \bar{Q}_{3 L} \Phi_{3} Q_{3 R} \nn \\ 
  & +
  \tilde{y}^{q}_{1} \left[ 
   \bar{Q}_{1 L} \left( 
    \tilde{\Phi}_{1} Q_{2 R} + 
    \tilde{\Phi}_{2} Q_{1 R} 
   \right) + 
   \bar{Q}_{2 L} \left( 
    \tilde{\Phi}_{1} Q_{1 R} - \tilde{\Phi}_{2} Q_{2 R} 
   \right) \right] + 
  \tilde{y}^{q}_{2} \left[ 
   \bar{Q}_{1 L} \tilde{\Phi}_{3} Q_{1 R} + 
   \bar{Q}_{2 L} \tilde{\Phi}_{3} Q_{2 R}
  \right] \nn \\ 
  & + 
  \tilde{y}^{q}_{3} \left[ 
   \bar{Q}_{1 L} \tilde{\Phi}_{1} + 
   \bar{Q}_{2 L} \tilde{\Phi}_{2} 
  \right] Q_{3 R} + 
  \tilde{y}^{q}_{4} \bar{Q}_{3 L} \left[ 
   \tilde{\Phi}_{1} Q_{1 R} + 
   \tilde{\Phi}_{2} Q_{2 R}
  \right] + 
  \tilde{y}^{q}_{5} \bar{Q}_{3 L} \tilde{\Phi}_{3} Q_{3 R} \nn \\ 
  & + 
  y^{\ell}_{1} \left[ 
   \bar{L}_{1} \left( \Phi_{1} R_{2} + \Phi_{2} R_{1 } \right) + 
   \bar{L}_{2} \left( \Phi_{1} R_{1} -\Phi_{2} R_{2} \right) 
  \right] + 
  y^{\ell}_{2} \left[
   \bar{L}_{1} \Phi_{3} R_{1} + 
   \bar{L}_{2} \Phi_{3} R_{2} 
  \right] + 
  y^{\ell}_{3} \left[ 
   \bar{L}_{1} \Phi_{1} + 
   \bar{L}_{2} \Phi_{2} 
  \right] R_{3} \nn \\ 
  & +
  y^{\ell}_{4} \bar{L}_{3} \left[ 
   \Phi_{1} R_{1} + \Phi_{2} R_{2} 
  \right] + 
  y^{\ell}_{5} \bar{L}_{3} \Phi_{3} R_{3} + 
  \tilde{y}^{\ell}_{1} \left[ 
   \bar{L}_{1} \left( 
    \tilde{\Phi}_{1} R_{2} + 
    \tilde{\Phi}_{2} R_{1} 
   \right) + 
   \bar{L}_{2} \left( 
    \tilde{\Phi}_{1} R_{1} - 
    \tilde{\Phi}_{2} R_{2} 
   \right)
  \right] \nn \\ 
  & + 
  \tilde{y}^{\ell}_{2} \left[ 
   \bar{L}_{1} \tilde{\Phi}_{3} R_{1} + 
   \bar{L}_{2} \tilde{\Phi}_{3} R_{2} 
  \right] + 
  \tilde{y}^{\ell}_{3} \left[  
   \bar{L}_{1} \tilde{\Phi}_{1} + 
   \bar{L}_{2} \tilde{\Phi}_{2} 
  \right] R_{3} + 
  \tilde{y}^{\ell}_{4} \bar{L}_{3} \left[ 
   \tilde{\Phi}_{1} R_{1} +
   \tilde{\Phi}_{2} R_{2} 
  \right]  + 
  \tilde{y}^{\ell}_{5} \bar{L}_{3} \tilde{\Phi}_{3} R_{3} \nn \\ 
  & +
  y^{L}_{1} \left[ 
   \bar{L}_{1} \Delta_{L} L^{c}_{1} + 
   \bar{L}_{2} \Delta_{L} L^{c}_{2} 
  \right] + 
  y^{L}_{2} \bar{L}_{3} \Delta_{L} L^{c}_{3} + 
  y^{R}_{1} \left[ 
   \bar{R}^{c}_{1} \Delta_{R} R_{1} +  
   \bar{R}^{c}_{2} \Delta_{R} R_{2}	
  \right] + 
  y^{R}_{2} \bar{R}^{c}_{3} \Delta_{R} R_{3} + h.c.\label{eq2}
\end{align}
As mentioned previously, after the usual spontaneous symmetry breaking in 
the left-right symmetric theory, the fermion mass matrices acquire the following 
structure\footnote{The mass matrix $\mathbf{M}_{f}$ commonly appears in minimal 
extensions of the Standard Model based on the $S_{3}$ flavor 
symmetry~\cite{Kubo:2003iw,Kubo:2003pd,Kubo:2005sr,Felix:2006pn,
Mondragon:2007af,Mondragon:2007nk,Mondragon:2007jx,Canales:2011ug,
Canales:2012ix,Canales:2012dr,GonzalezCanales:2012kj,
GonzalezCanales:2012za,Canales:2013ura,Canales:2013cga}. As will be discussed 
below, within left-right symmetric theories this mass matrix may be either 
complex symmetric or Hermitian.} 
\begin{equation}\label{eq8} 
 {\bf M}_{f} = 
 \begin{pmatrix}
  a_{f}+b^{\prime}_{f} & b_{f} & c_{f} \\ 
  b_{f} & a_{f}-b^{\prime}_{f} & c^{\prime}_{f} \\ 
  h_{f} & h^{\prime}_{f} & g_{f}
 \end{pmatrix}, 
 \qquad 
 {\bf M}_{(L, R)} = 
 \begin{pmatrix}
  a_{(L, R)} & 0 & 0 \\ 
  0 & a_{(L, R)} & 0 \\ 
  0 & 0 & b_{(L, R)}
 \end{pmatrix},
\end{equation}
where $f=u,\,d,\,D,\,e$ denotes the up-type quarks, down-type quarks, Dirac 
neutrinos, and charged leptons, respectively. 
In addition, the matrix elements are given by
\begin{align}
 &
 a_{u} = y^{q}_{2} k_{3} + \tilde{y}^{q}_{2} k^{\prime \ast}_{3}, \quad 
 b^{\prime}_{u} = y^{q}_{1} k_{2} + \tilde{y}^{q}_{1} k^{\prime \ast}_{2}, \quad 
 b_{u} = y^{q}_{1} k_{1} + \tilde{y}^{q}_{1} k^{\prime \ast}_{1}, \quad 
 c_{u} = y^{q}_{3} k_{1} + \tilde{y}^{q}_{3} k^{\prime \ast}_{1}, \quad
 c^{\prime}_{u} = y^{q}_{3} k_{2} + \tilde{y}^{q}_{3} k^{\prime \ast}_{2} ; 
 \nn \\ 
 & 
 h_{u} = y^{\dg q}_{3} k_{1} + \tilde{y}^{\dg q}_{3} k^{\prime \ast}_{1}, \quad 
 h^{\prime}_{u} = y^{\dg q}_{3} k_{2} + \tilde{y}^{\dg q}_{3} 
 k^{\prime \ast}_{2}, \quad 
 g_{u} = y^{q}_{5} k_{3} + \tilde{y}^{q}_{5} k^{\prime \ast}_{3}, \quad
 a_{d} = y^{q}_{2} k^{\prime}_{3} + \tilde{y}^{q}_{2} k^{\ast}_{3}, \quad 
 b^{\prime}_{d} = y^{q}_{1} k^{\prime}_{2} + \tilde{y}^{q}_{1} k^{\ast}_{2}; 
 \nn \\ 
 & 
 b_{d} = y^{q}_{1} k^{\prime}_{1} + \tilde{y}^{q}_{1} k^{\ast}_{1}, \quad 
 c_{d} = y^{q}_{3} k^{\prime}_{1} + \tilde{y}^{q}_{3} k^{\ast}_{1}, \quad 
 c^{\prime}_{d} = y^{q}_{3} k^{\prime}_{2} + \tilde{y}^{q}_{3} k^{\ast}_{2}, 
 \quad 
 h_{d} = y^{\dg q}_{3} k^{\prime}_{1} + \tilde{y}^{\dg q}_{3} k^{\ast}_{1}, \quad 
 h^{\prime}_{d} = y^{\dg q}_{3} k^{\prime}_{2} + \tilde{y}^{\dg q}_{3} 
 k^{\ast}_{2}; \nn \\ & 
 g_{d} = y^{q}_{5} k^{\prime}_{3} + \tilde{y}^{q}_{5} k^{\ast}_{3}, \quad
 a_{D} = y^{\ell}_{2} k_{3} + \tilde{y}^{\ell}_{2} k^{\prime \ast}_{3}, \quad 
 b^{\prime}_{D} = y^{\ell}_{1} k_{2} + \tilde{y}^{\ell}_{1} k^{\prime \ast}_{2}, 
 \quad 
 b_{D} = y^{\ell}_{1} k_{1} + \tilde{y}^{\ell}_{1} k^{\prime \ast}_{1}, \quad 
 c_{D} = y^{\ell}_{3} k_{1} + \tilde{y}^{\ell}_{3} k^{\prime \ast}_{1}; \nn \\ 
 &
 c^{\prime}_{D} = y^{\ell}_{3} k_{2} + \tilde{y}^{\ell}_{3} k^{\prime \ast}_{2}, 
 \quad 
 h_{D} = y^{\dg \ell}_{3} k_{1} + \tilde{y}^{\dg \ell}_{3} k^{\prime \ast}_{1}, 
 \quad 
 h^{\prime}_{D} = y^{\dg \ell}_{3} k_{2} + \tilde{y}^{\dg \ell}_{3} 
  k^{\prime \ast}_{2},\quad 
 g_{D} = y^{\ell}_{5} k_{3} + \tilde{y}^{\ell}_{5} k^{\prime \ast}_{3}, \quad
 a_{e} = y^{\ell}_{2} k^{\prime}_{3} + \tilde{y}^{\ell}_{2} k^{\ast}_{3}; \nn \\ 
 & 
 b^{\prime}_{e} = y^{\ell}_{1} k^{\prime}_{2} + \tilde{y}^{\ell}_{1} 
 k^{\ast}_{2}, \quad 
 b_{e} = y^{\ell}_{1} k^{\prime}_{1} + \tilde{y}^{\ell}_{1} k^{\ast}_{1}, \quad 
 c_{e} = y^{\ell}_{3} k^{\prime}_{1} + \tilde{y}^{\ell}_{3} k^{\ast}_{1}, \quad 
 c^{\prime}_{e} = y^{\ell}_{3} k^{\prime}_{2} + \tilde{y}^{\ell}_{3} 
 k^{\ast}_{2}, \quad 
 h_{e} = y^{\dg \ell}_{3} k^{\prime}_{1} + \tilde{y}^{\dg \ell}_{3} k^{\ast}_{1}; 
 \nn \\ 
 & 
 h^{\prime}_{e} = y^{\dg \ell}_{3} k^{\prime}_{2} + \tilde{y}^{\dg \ell}_{3} 
  k^{\ast}_{2}, \quad 
 g_{e} = y^{\ell}_{5} k^{\prime}_{3} + \tilde{y}^{\ell}_{5} k^{\ast}_{3}, \quad
 a_{(L, R)} = y^{R}_{1} v_{(L, R)}, \quad b_{(L, R)} = y^{R}_{2} v_{(L, R)}.
 \label{eq9}
\end{align}
Here, parity symmetry has been imposed. At this stage we impose the vacuum 
alignment in the $S_3$-doublet bidoublet sector. Writing
\begin{equation}
 \langle\Phi_I\rangle=
 \begin{pmatrix}
  k_I & 0\\
  0 & k'_I
 \end{pmatrix},
 \qquad I=1,2,
\end{equation}
we choose the symmetric direction
\begin{equation}
 k_1=k_2\equiv k_D,
 \qquad
 k'_1=k'_2\equiv k'_D .
 \label{eq:vev-alignment}
\end{equation}
while the VEVs of the $S_3$-singlet bidoublet $\Phi_3$ remain independent. 
This alignment identifies the matrix elements generated by
$\Phi_1$ and $\Phi_2$, reducing the number of independent entries of
Eq.~(\ref{eq8}). Together with the parity conditions and the defining
properties of the pseudomanifest and manifest realizations, it leads to
the reduced mass-matrix structures given below. We emphasize that
the symmetric direction is adopted as a vacuum ansatz and is not
derived here from the minimization of the complete scalar potential.
Consequently, if the VEVs are complex, 
the fermion mass matrices become complex symmetric. 
In the literature, this scenario is commonly referred to as the 
{\bf pseudomanifest left-right symmetric framework}~\cite{Langacker:1989xa,
Harari:1983gq}. On the other hand, if the VEVs are real, the fermion mass 
matrices are Hermitian and the number of CP-violating phases is significantly 
reduced. This case is known as the {\bf manifest left-right symmetric 
framework}~\cite{Beg:1977ti,Langacker:1989xa}. 
Explicitly, the mass matrices have the form
\begin{itemize}
 \item {\bf Pseudomanifest left-right theory (PLRT)}.
  \begin{align}
   {\bf M}_{f}=\begin{pmatrix}
    a_{f} + b_{f} & b_{f}       & c_{f} \\ 
    b_{f}         & a_{f}-b_{f} & c_{f} \\ 
    c_{f}         & c_{f}       & g_{f}
   \end{pmatrix}, 
   \quad 
   {\bf M}_{(L, R)} = 
   \begin{pmatrix}
    a_{(L, R)} & 0          & 0 \\ 
    0          & a_{(L, R)} & 0 \\ 
    0          & 0          & b_{(L, R)}
   \end{pmatrix}. \label{eq10}
  \end{align}
 Here, the ${\bf M}_{f}$ is a complex symmetric matrix. 
 \item {\bf Manifest left-right theory (MLRT)}. 
  \begin{align}
   {\bf M}_{f} = 
   \begin{pmatrix}
    a_{f} + b_{f} & b_{f}         & c_{f} \\ 
    b^{\ast}_{f}  & a_{f} - b_{f} & c_{f} \\ 
    c^{\ast}_{f}  & c^{\ast}_{f}  & g_{f}
   \end{pmatrix}, 
   \quad 
   {\bf M}_{(L, R)} = 
   \begin{pmatrix}
    a_{(L, R)} & 0          & 0 \\ 
    0          & a_{(L, R)} & 0 \\ 
    0          & 0          & b_{(L, R)}
   \end{pmatrix}. \label{eq10.m}
  \end{align}
  Here, ${\bf M}_{f}$ is a Hermitian matrix. Therefore, if both combinations 
  $(a_{f}+b_{f})$ and $(a_{f}-b_{f})$ are simultaneously real, then the 
  parameters $a_{f}$ and $b_{f}$ must necessarily be real as well. As a 
  consequence, $c_{f}$ remains as the only complex parameter in the above mass 
  matrix.
\end{itemize}
As can be observed, the Majorana neutrino mass matrices, $\mathbf{M}_{(L,R)}$, 
are diagonal. This implies that neutrino mixing arises from the interplay between 
the type-I seesaw mechanism and the charged-lepton sector. 
Furthermore, in the right-handed sector, the corresponding PMNS mixing matrix is 
completely determined by the charged-lepton sector, as will be shown below. 

The effective neutrino mass matrix is generated through the combined
type-I and type-II seesaw mechanism and is given by
\begin{equation}\label{eq:seesaw-II}
 \mathbf{M}_{\nu} = 
 \mathbf{M}_{L}
 - 
 \mathbf{M}_{D}\mathbf{M}_{R}^{-1}\mathbf{M}_{D}^{T},
\end{equation}
where $\mathbf{M}_{L}$, $\mathbf{M}_{D}$, and $\mathbf{M}_{R}$ denote the 
left-handed Majorana, Dirac, and right-handed Majorana neutrino mass matrices, 
respectively. In both the {\bf PLRT} and {\bf MLRT} frameworks, the mass matrices 
entering Eq.~(\ref{eq:seesaw-II}), together with the charged-lepton mass matrix 
${\bf M}_{e}$, are given in Eqs.~(\ref{eq10}) and (\ref{eq10.m}). The main 
distinction between the fermion mass matrices ${\bf M}_{f}$ in these two 
scenarios lies in their mathematical structure: in the {\bf PLRT} framework they 
are complex symmetric matrices, whereas in the {\bf MLRT} framework they are 
Hermitian. 
Nevertheless, in both cases the effective neutrino mass matrix $\mathbf{M}_{\nu}$ 
resulting from the seesaw mechanism is a complex symmetric matrix, as required 
for Majorana neutrinos. The explicit form of $\mathbf{M}_{\nu}$ is given by
\begin{equation}\label{eq:seesaw-II-1}
 \mathbf{M}_{\nu}^{(\texttt{j})} = 
  \left( \begin{array}{ccc}
   f_{\nu}^{(\texttt{j})} & a_{\nu}^{(\texttt{j})} & d_{\nu}^{(\texttt{j})} \\
   a_{\nu}^{(\texttt{j})} & b_{\nu}^{(\texttt{j})} & c_{\nu}^{(\texttt{j})} \\
   d_{\nu}^{(\texttt{j})} & c_{\nu}^{(\texttt{j})} & g_{\nu}^{(\texttt{j})}
  \end{array} \right)
\end{equation}
where $\texttt{j}=1$ denotes the {\bf PLRT} framework, while $\texttt{j}=2$ 
denotes the {\bf MLRT} scenario. For the PLRT framework
\begin{equation}\label{eq:seesaw-II-2}
 \begin{array}{l}\vspace{2mm}
  a_{\nu}^{(1)} = 
   -\frac{c_{D}^{2}}{b_{R}} - \frac{b_{D} \left(a_{D} - b_{D}\right)}{a_{R}} 
   - \frac{b_{D} \left(a_{D} + b_{D}\right)}{a_{R}}, \quad 
  b_{\nu}^{(1)} = 
   a_{L} - \frac{c_{D}^{2}}{b_{R}} - \frac{b_{D}^{2}}{a_{R}} 
   - \frac{ \left(a_{D} - b_{D} \right)^{2}}{a_{R}}, \\ \vspace{2mm}
  c_{\nu}^{(1)} = -\frac{c_{D} g_{D}}{b_{R}} - \frac{b_{D} c_{D}}{a_{R}} 
   - \frac{c_{D} \left(a_{D} - b_{D} \right)}{a_{R}}, \quad 
  d_{\nu}^{(1)} = 
   - \frac{c_{D} g_{D}}{b_{R}} - \frac{b_{D} c_{D}}{a_{R}} 
   - \frac{c_{D} \left(a_{D} + b_{D} \right)}{a_{R}}, \\ \vspace{2mm}
  f_{\nu}^{(1)} = 
   a_{L} - \frac{c_{D}^{2}}{b_{R}} - \frac{b_{D}^{2}}{a_{R}} 
   - \frac{\left(a_{D} + b_{D}\right)^{2}}{a_{R}}, \quad
  g_{\nu}^{(1)} = 
   b_{L} - \frac{g_{D}^{2}}{b_{R}} - \frac{2 c_{D}^{2}}{a_{R}} .
 \end{array}
\end{equation}
While for the MLRT framework 
\begin{equation}\label{eq:seesaw-II-3}
 \begin{array}{l}\vspace{2mm}
  a_{\nu}^{(2)} = 
   - \frac{c_{D}^{2}}{b_{R}} - \frac{b_{D}^{*} \left(a_{D} + b_{D}\right)}{a_{R}} 
   - \frac{b_{D} \left(a_{D} - b_{D}\right)}{a_{R}} , \quad 
  b_{\nu}^{(2)} = 
   a_{L} - \frac{c_{D}^{2}}{b_{R}} - \frac{b_{D}^{*2}}{a_{R}}  
   -\frac{\left(a_{D} - b_{D}\right)^{2}}{a_{R}} , \\ \vspace{2mm}
  c_{\nu}^{(2)} = 
   - \frac{c_{D} g_{D}}{b_{R}} -\frac{b_{D}^{*} c_{D}^{*}}{a_{R}} 
   - \frac{ c_{D}^{*} \left(a_{D} - b_{D}\right)}{a_{R}} , \quad 
  d_{\nu}^{(2)} = 
   - \frac{c_{D} g_{D}}{b_{R}} -\frac{c_{D}^{*} b_{D}}{a_{R}} 
   - \frac{c_{D}^{*} \left(a_{D} + b_{D}\right)}{a_{R}} , \\ \vspace{2mm}
  f_{\nu}^{(2)} = 
   a_{L} - \frac{c_{D}^{2}}{b_{R}} - \frac{b_{D}^{2}}{a_{R}} 
   - \frac{ \left(a_{D} + b_{D} \right)^{2}}{a_{R}}, \quad
  g_{\nu}^{(2)} = 
   b_{L} - \frac{g_{D}^{2}}{b_{R}} - \frac{ 2 c_{D}^{*2}}{a_{R}}.
 \end{array}
\end{equation}
Within the PLRT framework, the fermion mass matrices ${\bf M}_{f}$ can be brought 
into diagonal form through a unitary transformation which 
is~\cite{Garces:2018nar,Barradas-Guevara:2026zol}
\begin{equation}
 \begin{array}{l}
  {\bf M}_{q} = {\bf U}_{q}^{\top} \hat{\mathbf{M}}_{q} {\bf U}_{q}   
  \textrm{\; for quarks,} 
  \quad \textrm{and} \quad 
  {\bf M}_{e} = {\bf U}_{e} \hat{\mathbf{M}}_{e} {\bf U}_{e}^{\top}    
  \textrm{\; for charged leptons,} 
 \end{array}   
\end{equation}
with $q=u,d$. On the other hand, within the MLRT framework, we have
\begin{equation}
 \begin{array}{l}
  {\bf M}_{q} = {\bf U}_{q}^{\dagger} \hat{\mathbf{M}}_{q} {\bf U}_{q}   
  \textrm{\; for quarks,} 
  \quad \textrm{and} \quad 
  {\bf M}_{e} = {\bf U}_{e} \hat{\mathbf{M}}_{e} {\bf U}_{e}^{\dagger}    
  \textrm{\; for charged leptons.} 
 \end{array}   
\end{equation}
In the above expressions $\hat{\mathbf{M}}_{f}$, with $f=u,d,e,D$,  is a diagonal 
matrix whose non-negative real entries correspond to the physical fermion masses. 
Also,
\begin{equation}\label{eq:Uq-18}
 \begin{array}{l}\vspace{2mm}
  {\bf U}_{q} = {\bf O}_{q}^{\top} {\bf P}_{q} {\bf R}_{\frac{\pi}{4}} 
  \quad \textrm{and} \quad 
  {\bf U}_{e} = {\bf R}_{\frac{\pi}{4}}^{\top} {\bf P}_{e}^{\top} {\bf O}_{e}
  \qquad \textrm{for PLRT.} \\
  {\bf U}_{q} = {\bf O}_{q}^{\top} {\bf P}_{q} {\bf R}_{\frac{\pi}{4}} 
  \quad \textrm{and} \quad 
  {\bf U}_{e} = {\bf R}_{\frac{\pi}{4}}^{\top} {\bf P}_{e}^{\dagger} {\bf O}_{e}
  \qquad \textrm{for MLRT.}
 \end{array}
\end{equation}
In both theoretical frameworks, the effective neutrino mass matrix is 
diagonalized by a unitary transformation of the 
form~\cite{Barradas-Guevara:2026zol}
\begin{equation}\label{eq:Uq-19}
 {\bf M}_{\nu} =   {\bf U}_{\nu}^{ (\texttt{j})}  \hat{\mathbf{M}}_{\nu}  
 {\bf U}_{\nu}^{(\texttt{j}) \top} 
 \quad \textrm{with} \quad 
 {\bf U}_{\nu}^{ (\texttt{j})} = 
  {\bf R}_{\beta_{\nu}}^{(\texttt{j})\top} {\bf P}_{\nu}^{(\texttt{j})\top} 
  {\bf O}_{\nu}
\end{equation}
In the above expressions, ${\bf R}_{\frac{\pi}{4}}$ and ${\bf R}_{\beta_{\nu}}$ 
represent the rotation matrices that bring the fermion mass matrices to a basis 
where the $(1,3)$ and $(3,1)$ matrix elements are identically zero. Their 
explicit forms are given by
\begin{equation}\label{ec:rotaciones}
 \begin{array}{l}\vspace{2mm}
  {\bf R}_{\frac{\pi}{4}} =
  \left( \begin{array}{ccc}
   \cos \left( \frac{\pi}{4} \right) & -\sin \left( \frac{\pi}{4} \right)  & 0 \\
   \sin \left( \frac{\pi}{4} \right) &  \cos \left( \frac{\pi}{4} \right)  & 0 \\
   0 & 0 & 1
  \end{array} \right)
  \quad \textrm{and} \quad
  {\bf R}_{\beta_{\nu}}^{(\texttt{j})} =
  \left( \begin{array}{ccc}
   \cos \beta_{\nu}^{(\texttt{j})} & -\sin \beta_{\nu}^{(\texttt{j})}  & 0 \\
   \sin \beta_{\nu}^{(\texttt{j})} &  \cos \beta_{\nu}^{(\texttt{j})}  & 0 \\
   0 & 0 & 1
  \end{array} \right).
 \end{array}
\end{equation}
where 
$\tan \beta_{\nu}^{(\texttt{j})} = 
\frac{ |d_{\nu}^{(\texttt{j})}| }{ |c_{\nu}^{(\texttt{j})}| } $ with
$\arg \{ c_{\nu}^{(\texttt{j})} \} = \arg \{ d_{\nu}^{(\texttt{j})} \}$.
The matrices ${\bf O}_{\ell}$ correspond to the real  orthogonal matrices 
entering the decomposition of the unitary matrices ${\bf U}_{\ell}$. 
Here, the index $\ell=u,d,e,\nu$ refers to the up-type quark, down-type quark, 
charged-lepton, and neutrino sectors, respectively. Their explicit form is given 
by
\begin{equation}
 \begin{array}{l}\vspace{2mm}
  {\bf O}_{\ell} =
  \left( \begin{array}{ccc} \vspace{2mm} 
   \sqrt{ \frac{ 
    \sigma_{\ell 2} \sigma_{\ell 3} \xi_{\ell 1} 
   }{ \texttt{D}_{\ell 1} } } &
    - \sqrt{ \frac{ 
    \sigma_{\ell 1} \sigma_{\ell 3} \xi_{\ell 2} 
   }{ \texttt{D}_{\ell 2} } } & 
    \sqrt{ \frac{ \sigma_{\ell 1} \sigma_{\ell 2} \xi_{\ell 3} 
   }{ \texttt{D}_{\ell 3} } } \\ \vspace{2mm} 
   \sqrt{ \frac{ 
    \sigma_{\ell 1} {\cal D}_{\ell} \xi_{\ell 1} 
   }{ \texttt{D}_{\ell 1} } } &
   \sqrt{ \frac{ 
    \sigma_{\ell 2} {\cal D}_{\ell} \xi_{\ell 2} 
   }{ \texttt{D}_{\ell 2} } } &  
   \sqrt{ \frac{ 
    \sigma_{\ell 3} {\cal D}_{\ell} \xi_{\ell 3} 
   }{ \texttt{D}_{\ell 3} } } \\ \vspace{2mm} 
   -\sqrt{ \frac{ 
    \sigma_{\ell 1} \xi_{\ell 3} \xi_{\ell 2} 
   }{ \texttt{D}_{\ell 1} } } &
   -\sqrt{ \frac{ 
    \sigma_{\ell 2} \xi_{\ell 3} \xi_{\ell 1} 
   }{ \texttt{D}_{\ell 2} } } &  
   \sqrt{ \frac{ 
    \sigma_{\ell 3} \xi_{\ell 1} \xi_{\ell 2} 
   }{ \texttt{D}_{\ell 3} } }  
 \end{array} \right),
 \end{array}
\end{equation}
where 
$\sigma_{\ell 1} = m_{\ell 1} - m_{11}^{\ell}$, 
$\sigma_{\ell 2} = m_{\ell 2} + m_{11}^{\ell}$, 
$\sigma_{\ell 3} = m_{\ell 3} - m_{11}^{\ell}$,
$\xi_{\ell 1} = \left( {\cal D}_{\ell} - \sigma_{\ell 1} \right)$,
$\xi_{\ell 2} = \left( {\cal D}_{\ell} + \sigma_{\ell 2} \right)$,
$\xi_{\ell 3} = \left( \sigma_{\ell 3} - {\cal D}_{\ell} \right)$,
$\texttt{D}_{\ell 1} =  {\cal D}_{\ell} 
\left(\sigma_{\ell 2} + \sigma_{\ell 1}\right) 
\left(\sigma_{\ell 3} - \sigma_{\ell 1}\right)$,
$\texttt{D}_{\ell 2} = {\cal D}_{\ell} 
\left(\sigma_{\ell 3} + \sigma_{\ell 2}\right) 
\left(\sigma_{\ell 2} + \sigma_{\ell 1}\right)$,
$\texttt{D}_{\ell 3} = {\cal D}_{\ell} 
\left(\sigma_{\ell 3} + \sigma_{\ell 2} \right) 
\left(\sigma_{\ell 3} - \sigma_{\ell 1}\right)$.
The quantities $\sigma_{\ell i}$, with $i=1,2,3$, correspond to the shifted 
masses of fermions and are related to the physical fermion masses through the 
expressions 
$\sigma_{\ell 1} = m_{\ell 1} - m_{11}^{\ell}$,
$\sigma_{\ell 2} = m_{\ell 2} + m_{11}^{\ell}$, 
$\sigma_{\ell 3} = m_{\ell 3} - m_{11}^{\ell}$.
The parameter $m_{11}^{\ell}$ denotes the $(1,1)$ entry of the rotated fermion 
mass matrix. After extracting this contribution, the resulting mass matrix 
acquires an effective two texture zeros structure, which considerably simplifies 
its analytical diagonalization.~\cite{Barradas-Guevara:2026zol}. 
The explicit form of $m_{11}^{\ell}$ is: 
$m_{11}^{f} = |a_{f} - b_{f}|$ for PLRT, 
$m_{11}^{f} = a_{f} - b_{f}$ for MLRT and 
$m_{11}^{\nu} = 
|b_{\nu}^{(\texttt{j})} \sin^{2}\beta_{\nu}^{(\texttt{j})} + 
f_{\nu}^{(\texttt{j})} \cos^{2}\beta_{\nu}^{(\texttt{j})} - 
a_{\nu}^{(\texttt{j})} \sin 2\beta_{\nu}^{(\texttt{j})}|$ 
for neutrinos in both theoretical frameworks.
The parameter ${\cal D}_{f}$ is explicitly given by 
${\cal D}_{f} = |g_{f}| + |a_{f} - b_{f}|$ for PLRT, 
${\cal D}_{f} = g_{f} + b_{f} - a_{f}$ for MLRT, and 
${\cal D}_{\nu} =|g_{\nu}^{(\texttt{j})}| - 
|b_{\nu}^{(\texttt{j})} \sin^{2}\beta_{\nu}^{(\texttt{j})} + 
f_{\nu}^{(\texttt{j})} \cos^{2}\beta_{\nu}^{(\texttt{j})} - 
a_{\nu}^{(\texttt{j})} \sin 2\beta_{\nu}^{(\texttt{j})}|$ 
for neutrinos in both theoretical frameworks. 
For a normal fermion mass hierarchy, these parameters 
$m_{11}^{\ell}$ and ${\cal D}_{\ell}$ satisfy
$m_{\ell 1}>m_{11}^{\ell}$, 
$\sigma_{\ell 3} > {\cal D}_{\ell} > \sigma_{\ell 1}$, and
${\cal D}_{\ell} \neq \sigma_{\ell 1} - \sigma_{\ell 2} + \sigma_{\ell 3}$.
For an inverted hierarchy, which occurs only in the neutrino sector,
the corresponding conditions are
$m_{\ell 3}>m_{11}^{\ell}$, 
$\sigma_{\ell 1} > {\cal D}_{\ell} > \sigma_{\ell 3}$, and 
${\cal D}_{\ell} \neq \sigma_{\ell 1} - \sigma_{\ell 2} + \sigma_{\ell 3}$.
Finally,
\begin{equation}\label{eq:Uq-20}
 \begin{array}{l}\vspace{2mm}
  {\bf P}_{f} = 
  \textrm{diag} \left( 
   e^{i \frac{ \alpha_{f} }{2} }, 
   e^{i \frac{ \beta_{f} }{2} }, 
   e^{i \frac{ \gamma_{f}}{2} } 
  \right) \quad \textrm{for PLRT},\qquad
  {\bf P}_{f} = 
  \textrm{diag} \left( 1, 1, e^{i \phi_{fc}} \right) \quad \textrm{for MLRT,}
 \end{array}   
\end{equation}
where 
$\alpha_{f} = \arg \left \{ a_{f} - b_{f} \right \}$,
$\beta_{f} = \arg \left \{ a_{f} + b_{f} \right \}$, and
$\gamma_{f} = \arg \left \{ g_{f} \right \}$, while 
$\phi_{fc} = \arg \left \{ c_{f} \right \}$. 
Also, for the PLRT we have
$\arg \left \{ a_{f} - b_{f} \right \} +\arg \left \{ a_{f} + b_{f} \right \} =
2 \arg \left \{ b_{f} \right \}$, 
$\arg \left \{ a_{f} + b_{f} \right \} + \arg \left \{ g_{f} \right \} = 
2 \arg \left \{ c_{f} \right \}$.
For neutrinos in both theoretical frameworks
\begin{equation}\label{Eq:Matrix-P-2}
 \begin{array}{l}\vspace{2mm}
 {\bf P}_{\nu}^{(\texttt{j})} = \textrm{diag} 
 \left( 
  e^{ i \frac{ \alpha_{\nu}^{\texttt{(j)}} }{2} }, 
  e^{ i \frac{ \beta_{\nu}^{\texttt{(j)}}  }{2} }, 
  e^{ i \frac{ \gamma_{\nu}^{\texttt{(j)}}  }{2} } 
 \right) ,
 \end{array}   
\end{equation} 
where 
$\alpha_{\nu}^{\texttt{(j)}} = \arg \left \{ m_{11}^{\nu} \right \}$,
$\beta_{\nu}^{\texttt{(j)}}  = \arg \left \{ m_{22}^{\nu} \right \}$,
$\gamma_{\nu}^{\texttt{(j)}} = \arg \left \{ m_{33}^{\nu} \right \}$, 
$\arg \left \{ m_{11}^{\nu} \right \} + 
 \arg \left \{ m_{22}^{\nu} \right \} = 2 
 \arg \left \{ m_{12}^{\nu} \right \} $,
$\arg \left \{ m_{11}^{\nu} \right \} + 
 \arg \left \{ m_{33}^{\nu} \right \} = 
 \arg \left \{ m_{23}^{\nu} \right \}$.
The $m_{ij}^{\nu}$ are the elements of the rotated neutrino mass matrix.
An important result of our analysis is that the unitary matrices arising in the 
PLRT and MLRT frameworks differ only in the number of independent CP-violating 
phase factors they contain. In particular, from a model-independent perspective, 
the unitary matrix obtained in the MLRT framework emerges naturally as a more 
restrictive realization of the general structure derived in the PLRT case. This 
highlights the fact that the MLRT scenario can be understood as a specific limit 
of the broader PLRT 
framework~\cite{Garces:2018nar,Barradas-Guevara:2026zol}. Also, in general terms 
we can say that in this left-right model, we have a unified treatment of the 
fermion mass matrices since these can be reduced to a matrix with two 
texture zeros. 
%
\subsection{Flavor mixing matrix in quark sector}
%
In this section, our primary objective is to derive the theoretical 
expressions for the left- and right-handed quark flavor mixing matrices, commonly 
known as  the CKM matrices. These mixing matrices originate from the mismatch 
between the  diagonalization of the up- and down-type quark mass matrices. 
Consequently, by  virtue of the singular value decomposition theorem, the 
following bilinear forms  are obtained:
\begin{equation}\label{Eq:bi-linear-form}
 \hat{\mathbf{M}}_{q} \hat{\mathbf{M}}^{\dagger}_{q} = 
 \mathbf{U}^{\dagger}_{q L} 
 \mathbf{M}_{q} \mathbf{M}^{\dagger}_{q} 
 \mathbf{U}_{q L}
 \qquad \textrm{and} \qquad 
 \hat{\mathbf{M}}^{\dagger}_{q} \hat{\mathbf{M}}_{q} = 
 \mathbf{U}^{\dagger}_{q R} 
 \mathbf{M}^{\dagger}_{q} \mathbf{M}_{q} 
 \mathbf{U}_{q R},
\end{equation}
where $\hat{\mathbf{M}}_{q}=\textrm{diag}(m_{q_{1}}, m_{q_{2}}, m_{q_{3}})$ and 
$\mathbf{U}_{q(L, R)}$ are unitary matrices. 
We obtain that the left- and right-handed CKM matrices have the 
form~\cite{Zyla:2020zbs} 
\begin{equation}\label{Eq:CKM-Def-Matrices}
 \mathbf{V}^{L}_{CKM}=\mathbf{U}_{u L}\mathbf{U}_{d L}^{\dagger} 
 \qquad \textrm{and} \qquad  
 \mathbf{V}^{R}_{CKM}=\mathbf{U}_{u R}\mathbf{U}_{d R}^{\dagger}
\end{equation}
As discussed above, the left- and right-handed CKM matrices are related
differently in the PLRT and MLRT frameworks. In the MLRT case, where
the quark mass matrices are Hermitian,
$\mathbf{M}_{q}=\mathbf{M}_{q}^{\dagger}$, the diagonalization matrices
satisfy $\mathbf{U}_{qL}=\mathbf{U}_{qR}$. Consequently,
\(\mathbf{V}^{L}_{\mathrm{CKM}} = \mathbf{V}^{R}_{\mathrm{CKM}}
 = \mathbf{U}_{uL}\mathbf{U}_{dL}^{\dagger}
\).
In this theoretical framework, the left- and right-handed quark mixing matrices 
are equal to each other. 
On the other hand, in the {\bf PLRT} framework, where the quark mass matrices 
are complex symmetric, $\mathbf{M}_{q}=\mathbf{M}_{q}^{\top}$, the unitary 
matrices satisfy the relation $\mathbf{U}_{qL}=\mathbf{U}_{qR}^{*}$. 
As a consequence, the CKM matrices take the form
\(\mathbf{V}^{R}_{\mathrm{CKM}} =
\left(\mathbf{V}^{L}_{\mathrm{CKM}}\right)^{*} =
\mathbf{U}_{uL}^{*}\mathbf{U}_{dL}^{\top}\). 
In this scenario, after performing the polar decomposition of the unitary 
matrices that diagonalize the quark mass matrices, one obtains
\( {\bf V}^{L}_{\mathrm{CKM}} =
{\bf O}_{u}^{\top}\,
{\bf P}^{(u-d)}\,
{\bf O}_{d}\) 
and
\({\bf V}^{R}_{\mathrm{CKM}}
 =
 {\bf O}_{u}^{\top}\,
 {\bf P}^{(u-d)\dagger}\,
 {\bf O}_{d},
\)
where ${\bf O}_{q}$ denotes a real orthogonal matrix and
${\bf P}^{(u-d)}={\bf P}_{u}{\bf P}_{d}^{\dagger}$ is a diagonal matrix of 
phase factors. Hence, the only difference between
${\bf V}_{\mathrm{CKM}}^{L}$ and ${\bf V}_{\mathrm{CKM}}^{R}$ lies in the sign 
of the phase factors.
The explicit form of the left-handed CKM mixing matrix is given 
by~\cite{Garces:2018nar}
\begin{equation}\label{Eq:CKM}
 \begin{array}{l}
  {\bf V}_{\mathrm{CKM}}^{L (\texttt{j})} = e^{ i \zeta_{ \texttt{j} } } \,
  {\bf O}_{u}^{\top} 
  {\bf P}_{\texttt{j}}^{( u-d )} 
  {\bf O}_{d}, 
 \end{array}
\end{equation}
where 
\begin{equation}
 {\bf P}_{ \texttt{j} }^{ (u-d) } = 
 \textrm{diag} 
 \left( 1 , e^{i \Theta_{ \texttt{j} } }, e^{ i \Gamma_{ \texttt{j} } } \right), \
\end{equation} 
here $\texttt{j}=1$ denotes the {\bf PLRT} framework, while 
$\texttt{j}=2$ denotes the {\bf MLRT} scenario, and
\begin{equation}
 \begin{array}{lll}
  \Theta_{1} = 
   \frac{ \alpha_{d1} - \alpha_{u1} + \beta_{u1} - \beta_{d1} }{2}, &
  \Gamma_{1} = 
   \frac{ \alpha_{d1} - \alpha_{u1} + \gamma_{u1} - \gamma_{d1} }{2}, &
   \zeta_{1} = \alpha_{u1} -  \alpha_{d1}, \\
   \Theta_{2} = 0, &
   \Gamma_{2} = \phi_{uc} - \phi_{dc}, &
   \zeta_{2} = 0.
 \end{array}
\end{equation}
It is worth noting that the rotation matrix $R_{\frac{\pi}{4}}$ does not 
contribute to the CKM matrix, since it appears identically in the diagonalization 
of both the up- and down-quark sectors. As a consequence, this common rotation 
cancels out in the construction of the CKM matrix. This feature constitutes a 
non-trivial prediction of the underlying $S_{3}$ flavor symmetry and 
indicates that the experimentally observed quark mixing is generated solely by 
the symmetry-breaking effects encoded in the fermion mass eigenvalues and phase 
factors. Therefore, the flavor symmetry naturally removes an otherwise large 
mixing contribution, providing a possible explanation for the hierarchical 
structure of the CKM matrix~\cite{Kubo:2003iw,Canales:2013cga,Barranco:2010we}.
Consequently, from expressions in Eqs.~(\ref{eq:Uq-18})-(\ref{eq:Uq-20}),
the explicit form of the CKM mixing  in both frameworks 
is~\cite{Garces:2018nar,Barradas-Guevara:2026zol}:
\begin{equation}
 {\bf V}_{\mathrm{CKM}}^{L (\texttt{j})} = 
 e^{ i \zeta_{ \texttt{j} } } \,
 \left( \begin{array}{ccc}\vspace{2mm}
  V_{ud}^{\texttt{(j)}} & V_{us}^{\texttt{(j)}} & V_{ub}^{\texttt{(j)}} 
   \\ \vspace{2mm}
  V_{cd}^{\texttt{(j)}} & V_{cs}^{\texttt{(j)}} & V_{cb}^{\texttt{(j)}} 
   \\ \vspace{2mm}
  V_{td}^{\texttt{(j)}} & V_{ts}^{\texttt{(j)}} & V_{tb}^{\texttt{(j)}} 
 \end{array} \right)
\end{equation} 
where
\begin{equation}\label{Eq:CKM-th-1}
 \begin{array}{ll} 
  V_{ud}^{^{(\texttt{j})}} = 
   \sqrt{ \frac{ 
    \widetilde{\sigma}_{c} 
    \widetilde{\sigma}_{s} 
    \widetilde{\sigma}_{t} 
    \widetilde{\sigma}_{b} 
    \widetilde{\xi}_{d1} 
    \widetilde{\xi}_{u1}
   }{ \widetilde{\texttt{D}}_{u1} \widetilde{\texttt{D}}_{d1} } }
   + 
   \sqrt{ \frac{ 
    \widetilde{\sigma}_{u} 
    \widetilde{\sigma}_{d} 
   }{ \widetilde{\texttt{D}}_{u1} \widetilde{\texttt{D}}_{d1} } } 
   \varepsilon_{ud}^{^{(\texttt{j})}} , &   
  V_{us}^{^{(\texttt{j})}} =   
   -\sqrt{ \frac{ 
    \widetilde{\sigma}_{c} 
    \widetilde{\sigma}_{d} 
    \widetilde{\sigma}_{t} 
    \widetilde{\sigma}_{b} 
    \widetilde{\xi}_{u1} 
    \widetilde{\xi}_{d2} 
   }{ \widetilde{\texttt{D}}_{u1} \widetilde{\texttt{D}}_{d2} } }  
   + 
   \sqrt{ \frac{ 
    \widetilde{\sigma}_{u} \widetilde{\sigma}_{s} 
   }{ 
    \widetilde{\texttt{D}}_{u1} \widetilde{\texttt{D}}_{d2} } 
   } \varepsilon_{us}^{^{(\texttt{j})}} ,
  \end{array}  
 \end{equation}
 \begin{equation}\label{Eq:CKM-th-2}
  \begin{array}{ll} 
   V_{ub}^{^{(\texttt{j})}} =  
    \sqrt{ \frac{ 
	   \widetilde{\sigma}_{c} \widetilde{\sigma}_{d} \widetilde{\sigma}_{s} \widetilde{\sigma}_{t} 
      \widetilde{\xi}_{u1} \widetilde{\xi}_{d3}
	}{ \widetilde{\texttt{D}}_{u1} \widetilde{\texttt{D}}_{d3} } } 
    - 
    \sqrt{ \frac{ 
     \widetilde{\sigma}_{u} \widetilde{\sigma}_{b} 
     }{ 
      \widetilde{\texttt{D}}_{u1} \widetilde{\texttt{D}}_{d3} } } 
     \varepsilon_{ub}^{^{(\texttt{j})}} ,  &  
   V_{cd}^{^{(\texttt{j})}} =   
    -\sqrt{ \frac{ 
	   \widetilde{\sigma}_{u} \widetilde{\sigma}_{s} \widetilde{\sigma}_{t} \widetilde{\sigma}_{b} 
      \widetilde{\xi}_{u2} \widetilde{\xi}_{d1}
	  }{ \widetilde{\texttt{D}}_{u2} \widetilde{\texttt{D}}_{d1} } } 
    + 
    \sqrt{ \frac{ 
     \widetilde{\sigma}_{c} \widetilde{\sigma}_{d} 
    }{ \widetilde{\texttt{D}}_{u2} \widetilde{\texttt{D}}_{d1} } } 
    \varepsilon_{cd}^{^{(\texttt{j})}} , 
  \end{array}  
 \end{equation}
 \begin{equation}\label{Eq:CKM-th-3}
  \begin{array}{ll} 
   V_{cs}^{^{(\texttt{j})}} =  
    \sqrt{ \frac{ 
	   \widetilde{\sigma}_{u} \widetilde{\sigma}_{d} \widetilde{\sigma}_{t} \widetilde{\sigma}_{b} 
      \widetilde{\xi}_{u2} \widetilde{\xi}_{d2}
	}{ \widetilde{\texttt{D}}_{u2} \widetilde{\texttt{D}}_{d2} }  }
    + 
    \sqrt{ \frac{ 
     \widetilde{\sigma}_{c} \widetilde{\sigma}_{s} 
     }{ \widetilde{\texttt{D}}_{u2} \widetilde{\texttt{D}}_{d2} } } 
    \varepsilon_{cs}^{^{(\texttt{j})}} , &  
   V_{cb}^{^{(\texttt{j})}} =   
    -\sqrt{ \frac{ 
      \widetilde{\sigma}_{u} \widetilde{\sigma}_{d} \widetilde{\sigma}_{s} \widetilde{\sigma}_{t}  
      \widetilde{\xi}_{u2} \widetilde{\xi}_{d3}
	}{ \widetilde{\texttt{D}}_{u2} \widetilde{\texttt{D}}_{d3} } } 
    - 
    \sqrt{ \frac{ 
     \widetilde{\sigma}_{c} \widetilde{\sigma}_{b} 
    }{ \widetilde{\texttt{D}}_{u2} \widetilde{\texttt{D}}_{d3} } } 
    \varepsilon_{cb}^{^{(\texttt{j})}} ,
  \end{array}  
 \end{equation}
 \begin{equation}\label{Eq:CKM-th-4}
  \begin{array}{ll} 
   V_{td}^{^{(\texttt{j})}} =    
    \sqrt{ \frac{ 
     \widetilde{\sigma}_{u} \widetilde{\sigma}_{c} \widetilde{\sigma}_{s} \widetilde{\sigma}_{b} 
     \widetilde{\xi}_{u3} \widetilde{\xi}_{d1}
	}{ \widetilde{\texttt{D}}_{u3} \widetilde{\texttt{D}}_{d1} } } 
    - 
    \sqrt{ \frac{ \widetilde{\sigma}_{d} \widetilde{\sigma}_{t} 
    }{ \widetilde{\texttt{D}}_{u3} \widetilde{\texttt{D}}_{d1} } } \varepsilon_{td}^{^{(\texttt{j})}} , & 
   V_{ts}^{^{(\texttt{j})}} =    
    -\sqrt{ \frac{ 
	   \widetilde{\sigma}_{u} \widetilde{\sigma}_{c} \widetilde{\sigma}_{d} \widetilde{\sigma}_{b} 
      \widetilde{\xi}_{u3} \widetilde{\xi}_{d2}
	}{ \widetilde{\texttt{D}}_{u3} \widetilde{\texttt{D}}_{d2} } } 
    - 
    \sqrt{ \frac{ \widetilde{\sigma}_{s} \widetilde{\sigma}_{t} 
    }{ \widetilde{\texttt{D}}_{u3} \widetilde{\texttt{D}}_{d2} } } 
    \varepsilon_{ts}^{^{(\texttt{j})}} , 
  \end{array}  
 \end{equation}
 \begin{equation}\label{Eq:CKM-th-5}
  \begin{array}{ll} 
   V_{tb}^{^{(\texttt{j})}} =     
    \sqrt{ \frac{ 
	   \widetilde{\sigma}_{u} \widetilde{\sigma}_{c} \widetilde{\sigma}_{d} \widetilde{\sigma}_{s} 
      \widetilde{\xi}_{u3} \widetilde{\xi}_{d3}
	}{ \widetilde{\texttt{D}}_{u3} \widetilde{\texttt{D}}_{d3} } } 
    + 
    \sqrt{ \frac{ \widetilde{\sigma}_{t} \widetilde{\sigma}_{b} 
    }{ \widetilde{\texttt{D}}_{u3} \widetilde{\texttt{D}}_{d3} } }
    \varepsilon_{tb}^{^{(\texttt{j})}} , \\  \vspace{2mm}  
  \end{array}  
 \end{equation}
 where
 \begin{equation}\label{Eq:CKM-th-6}
  \begin{array}{ll}  
   \varepsilon_{ud}^{^{(\texttt{j})}} = 
    \sqrt{ 
     \widetilde{\xi}_{u2} \widetilde{\xi}_{u3} 
     \widetilde{\xi}_{d2} \widetilde{\xi}_{d3} } e^{i \Gamma_{\texttt{j}} }
    + 
    \sqrt{ 
     \widetilde{\cal D}_{u} \widetilde{\cal D}_{d} 
     \widetilde{\xi}_{u1} \widetilde{\xi}_{d1} } e^{i \Theta_{\texttt{j}} } , & 
   \varepsilon_{us}^{^{(\texttt{j})}} = 
    \sqrt{ 
     \widetilde{\xi}_{u2} \widetilde{\xi}_{u3} 
     \widetilde{\xi}_{d1} \widetilde{\xi}_{d3}  
     } e^{i \Gamma_{\texttt{j}} } 
    + 
    \sqrt{ 
     \widetilde{\cal D}_{u} \widetilde{\cal D}_{d} 
     \widetilde{\xi}_{u1} \widetilde{\xi}_{d2} 
    } e^{i \Theta_{\texttt{j}} } , 
  \end{array}  
 \end{equation}
 \begin{equation}\label{Eq:CKM-th-7}
  \begin{array}{ll}  
   \varepsilon_{ub}^{^{(\texttt{j})}} =  
    \sqrt{ 
     \widetilde{\xi}_{u2} \widetilde{\xi}_{u3} 
     \widetilde{\xi}_{d1} \widetilde{\xi}_{d2} 
    } e^{ i \Gamma_{\texttt{j}} } 
    - 
    \sqrt{ 
     \widetilde{\cal D}_{u} \widetilde{\cal D}_{d} 
     \widetilde{\xi}_{u1} \widetilde{\xi}_{d3} 
    } e^{i \Theta_{\texttt{j}} } ,  &  
  \varepsilon_{cd}^{^{(\texttt{j})}} = 
    \sqrt{ 
     \widetilde{\xi}_{u1} \widetilde{\xi}_{u3} 
     \widetilde{\xi}_{d2} \widetilde{\xi}_{d3}  
    } e^{ i \Gamma_{\texttt{j}} } 
    + 
    \sqrt{ 
     \widetilde{\cal D}_{u} \widetilde{\cal D}_{d} 
     \widetilde{\xi}_{u2} \widetilde{\xi}_{d1} 
    } e^{i \Theta_{\texttt{j}} } ,  
  \end{array}  
 \end{equation}
 \begin{equation}\label{Eq:CKM-th-8}
  \begin{array}{ll}  
   \varepsilon_{cs}^{^{(\texttt{j})}} = 
    \sqrt{ 
     \widetilde{\xi}_{u1} \widetilde{\xi}_{u3} 
     \widetilde{\xi}_{d1} \widetilde{\xi}_{d3}  
    }  e^{ i \Gamma_{\texttt{j}} } 
    + 
    \sqrt{ 
     \widetilde{\cal D}_{u} \widetilde{\cal D}_{d} 
     \widetilde{\xi}_{u2} \widetilde{\xi}_{d2} 
    } e^{i \Theta_{\texttt{j}} } ,&  
    \varepsilon_{cb}^{^{(\texttt{j})}} = 
    \sqrt{ 
     \widetilde{\xi}_{u1} \widetilde{\xi}_{u3} 
     \widetilde{\xi}_{d1} \widetilde{\xi}_{d2} 
    } e^{ i \Gamma_{\texttt{j}} } 
    -
    \sqrt{ \widetilde{\cal D}_{u} \widetilde{\cal D}_{d} 
    \widetilde{\xi}_{u2} \widetilde{\xi}_{d3} }
    e^{i \Theta_{\texttt{j}}  } , 
  \end{array}  
 \end{equation}
 \begin{equation}\label{Eq:CKM-th-9}
  \begin{array}{ll}  
   \varepsilon_{td}^{^{(\texttt{j})}} = 
    \sqrt{ 
     \widetilde{\xi}_{u1} \widetilde{\xi}_{u2} 
     \widetilde{\xi}_{d2} \widetilde{\xi}_{d3}  
    } e^{ i \Gamma_{\texttt{j}} } 
    - 
    \sqrt{ 
     \widetilde{\cal D}_{u} \widetilde{\cal D}_{d} 
     \widetilde{\xi}_{u3} \widetilde{\xi}_{d1} 
    } e^{i \Theta_{\texttt{j}} }   , & 
   \varepsilon_{ts}^{^{(\texttt{j})}} = 
    \sqrt{ 
     \widetilde{\xi}_{u1} \widetilde{\xi}_{u2} 
     \widetilde{\xi}_{d1} \widetilde{\xi}_{d3} 
    } e^{ i \Gamma_{\texttt{j}} } 
    -
    \sqrt{ 
     \widetilde{\cal D}_{u} \widetilde{\cal D}_{d} 
     \widetilde{\xi}_{u3} \widetilde{\xi}_{d2} 
    } e^{i \Theta_{\texttt{j}} } ,
  \end{array}  
 \end{equation}
 \begin{equation}\label{Eq:CKM-th-10}
  \begin{array}{ll}  
   \varepsilon_{tb}^{^{(\texttt{j})}} = 
    \sqrt{ 
     \widetilde{\xi}_{u1} \widetilde{\xi}_{u2} 
     \widetilde{\xi}_{d1} \widetilde{\xi}_{d2} 
    } e^{ i \Gamma_{\texttt{j}} } 
    + 
    \sqrt{ 
     \widetilde{\cal D}_{u} \widetilde{\cal D}_{d} 
     \widetilde{\xi}_{u3} \widetilde{\xi}_{d3} 
    } e^{i \Theta_{\texttt{j}} } .     
  \end{array}  
 \end{equation} 
 Here, 
 $\widetilde{\sigma}_{u} = \widetilde{m}_{u} - \widetilde{m}_{11}^{u}$,
 $\widetilde{\sigma}_{c} = \widetilde{m}_{c} + \widetilde{m}_{11}^{u}$,
 $\widetilde{\sigma}_{t} = 1                 - \widetilde{m}_{11}^{u}$,
 $\widetilde{\sigma}_{d} = \widetilde{m}_{d} - \widetilde{m}_{11}^{d}$,
 $\widetilde{\sigma}_{s} = \widetilde{m}_{s} + \widetilde{m}_{11}^{d}$,
 $\widetilde{\sigma}_{b} = 1                 - \widetilde{m}_{11}^{d}$,
\begin{equation}\label{ec:Vckm-th-11}
 \begin{array}{l}
  \widetilde{\xi}_{u1} = \widetilde{\cal D}_{u} - \widetilde{\sigma}_{u},
   \quad
  \widetilde{\xi}_{u2} = \widetilde{\cal D}_{u} + \widetilde{\sigma}_{c},
   \quad
  \widetilde{\xi}_{u3} = \widetilde{\sigma}_{t} - \widetilde{\cal D}_{u}, \quad
  \widetilde{\xi}_{d1} = \widetilde{\cal D}_{d} - \widetilde{\sigma}_{d},
   \quad
  \widetilde{\xi}_{d2} = \widetilde{\cal D}_{d} + \widetilde{\sigma}_{s},
   \quad
  \widetilde{\xi}_{d3} = \widetilde{\sigma}_{b} - \widetilde{\cal D}_{d},
 \end{array} 
\end{equation} 
\begin{equation}\label{ec:Vckm-th-12}
 \begin{array}{l}\vspace{2mm}  
  \widetilde{\texttt{D}}_{u1} = \widetilde{\cal D}_{u}
   \left( \widetilde{\sigma}_{c} + \widetilde{\sigma}_{u} \right) 
   \left( \widetilde{\sigma}_{t} - \widetilde{\sigma}_{u} \right) , \quad  
  \widetilde{\texttt{D}}_{u2} = 
   \widetilde{\cal D}_{u} 
   \left( \widetilde{\sigma}_{t} + \widetilde{\sigma}_{c} \right) 
   \left( \widetilde{\sigma}_{c} + \widetilde{\sigma}_{u} \right) , \quad  
  \widetilde{\texttt{D}}_{u3} =      
   \widetilde{\cal D}_{u} 
   \left( \widetilde{\sigma}_{t} + \widetilde{\sigma}_{c}  \right) 
   \left( \widetilde{\sigma}_{t} - \widetilde{\sigma}_{u} \right) , \\   
  \widetilde{\texttt{D}}_{d1} =      
   \widetilde{\cal D}_{d}
   \left( \widetilde{\sigma}_{s} + \widetilde{\sigma}_{d} \right) 
   \left( \widetilde{\sigma}_{b} - \widetilde{\sigma}_{d} \right) , 
  \quad  
  \widetilde{\texttt{D}}_{d2} = 
   \widetilde{\cal D}_{d} 
   \left( \widetilde{\sigma}_{b} + \widetilde{\sigma}_{s} \right) 
   \left( \widetilde{\sigma}_{s} + \widetilde{\sigma}_{d} \right) ,
  \quad  
  \widetilde{\texttt{D}}_{d3} =      
   \widetilde{\cal D}_{d} 
   \left( \widetilde{\sigma}_{b}  + \widetilde{\sigma}_{s}  \right) 
   \left( \widetilde{\sigma}_{b}  - \widetilde{\sigma}_{d} \right) . 
 \end{array}
\end{equation} 
Also, $\widetilde{m}_{u} = \frac{ m_{u} }{ m_{t} }$,  
$\widetilde{m}_{c} = \frac{ m_{c} }{ m_{t} }$, 
$\widetilde{m}_{d} = \frac{ m_{d} }{ m_{b} }$,
$\widetilde{m}_{s} = \frac{ m_{s} }{ m_{b} }$,
$\widetilde{\cal D}_{u} = \frac{ {\cal D}_{u} }{ m_{t} }$,
$\widetilde{\cal D}_{d} = \frac{ {\cal D}_{d} }{ m_{b} }$,
$\widetilde{m}_{11}^{u} = \frac{ m_{11}^{u} }{ m_{t} }$,
$\widetilde{m}_{11}^{d} = \frac{ m_{11}^{d} }{ m_{b} }$,
$\widetilde{\sigma}_{u} = \frac{ \sigma_{u} }{ m_{t} } $,
$\widetilde{\sigma}_{c} = \frac{ \sigma_{c} }{ m_{t} } $,
$\widetilde{\sigma}_{t} = \frac{ \sigma_{t} }{ m_{t} } $,
$\widetilde{\sigma}_{d} = \frac{ \sigma_{d} }{ m_{b} } $,
$\widetilde{\sigma}_{s} = \frac{ \sigma_{s} }{ m_{b} } $,
$\widetilde{\sigma}_{b} = \frac{ \sigma_{b} }{ m_{b} } $,
$\widetilde{\xi}_{u1} = \frac{ \xi_{u1} }{ m_{t} } $,
$\widetilde{\xi}_{u2} = \frac{ \xi_{u2} }{ m_{t} } $,
$\widetilde{\xi}_{u3} = \frac{ \xi_{u3} }{ m_{t} } $,
$\widetilde{\xi}_{d1} = \frac{ \xi_{d1} }{ m_{b} } $,
$\widetilde{\xi}_{d2} = \frac{ \xi_{d2} }{ m_{b} } $,
$\widetilde{\xi}_{d3} = \frac{ \xi_{d3} }{ m_{b} } $.
Finally, the parameter $\widetilde{\cal D}_{q}$, where $q=u,d$, is required to 
satisfy the following conditions:
\begin{equation}
 \begin{array}{l}\vspace{2mm}
  \widetilde{\sigma}_{q3} > \widetilde{\cal D}_{q} > \widetilde{\sigma}_{q1}, 
  \qquad
  \widetilde{\cal D}_{q} \neq 
   + \widetilde{\sigma}_{q1} - \widetilde{\sigma}_{q2} 
   + \widetilde{\sigma}_{q3}.
 \end{array}
\end{equation}
The above expressions are formulated in terms of mass ratios since both the 
PDG and recent analyses indicate that the hierarchical structure of quark masses 
is largely preserved under the renormalization-group evolution. Consequently, the 
quark mass ratios remain approximately invariant over a wide range of energy 
scales.~\cite{ParticleDataGroup:2024cfk,Huang:2021,Fusaoka:1998}.
%
\subsubsection{Likelihood Test \texorpdfstring{$\mathbf{\chi}^{2}$}{chi}}
%
A crucial test of the proposed model is its ability to reproduce the observed
phenomenology of the quark sector. Our first objective is therefore to determine
whether the fermion mass matrices and the corresponding CKM mixing matrices
derived within the {\bf PLRT} and {\bf MLRT} frameworks are consistent with the
current experimental information on quark masses and flavor mixing. To this end,
we perform a likelihood analysis based on the following $\chi^{2}$ function:
\begin{equation}
 \chi^{2} =
 \frac{\left(|V_{us}^{\rm th}|-|V_{us}^{\rm exp}|\right)^2}{\sigma_{us}^{2}}
 +
 \frac{\left(|V_{ub}^{\rm th}|-|V_{ub}^{\rm exp}|\right)^2}{\sigma_{ub}^{2}}
 +
 \frac{\left(|V_{cb}^{\rm th}|-|V_{cb}^{\rm exp}|\right)^2}{\sigma_{cb}^{2}}.
\end{equation}
In the above expression, the superscripts ``th'' and ``exp'' denote the
theoretical predictions and the corresponding experimental measurements,
respectively.

The observables entering the likelihood function are the magnitudes of the CKM
matrix elements $|V_{us}|$, $|V_{ub}|$, and $|V_{cb}|$. These quantities are
among the most precisely measured flavor observables and are extracted from
independent experimental processes, including kaon decays and semileptonic
$B$-meson decays. Owing to their different experimental origins and distinct
sources of systematic uncertainty, they can be regarded as effectively
uncorrelated in the statistical analysis. Their current experimental values
are~\cite{Workman:2022}
\begin{equation}\label{Eq:Exp-Values-mag-ckm}
 \begin{array}{lll}
  |V_{us}| = 0.22431 \pm 0.00085, &
  |V_{ub}| = (3.82 \pm 0.20)\times10^{-3}, &
  |V_{cb}| = (41.1 \pm 1.2)\times10^{-3}.
 \end{array}
\end{equation}

The likelihood function depends on a set of ten free parameters,
\begin{equation}
 \chi^{2}
 =
 \chi^{2}
 \left(
 \widetilde{m}_{u},
 \widetilde{m}_{c},
 \widetilde{m}_{d},
 \widetilde{m}_{s},
 \widetilde{m}_{11}^{u},
 \widetilde{m}_{11}^{d},
 \widetilde{\cal D}_{u},
 \widetilde{\cal D}_{d},
 \Gamma_{\texttt{j}},
 \Theta_{\texttt{j}}
 \right).
\end{equation}
However, it is constrained by only three independent experimental observables.
Consequently, the fit is intrinsically underconstrained and the corresponding
$\chi^{2}$ function has no statistical degrees of freedom. As a result,
confidence intervals or confidence regions cannot be assigned to the fitted
parameters in a statistically meaningful way. Therefore, our analysis is
restricted to determining the Best Fit Point (BFP), defined as the parameter
configuration that minimizes the $\chi^{2}$ function and hence provides the best
overall agreement between the theoretical predictions and the available
experimental data.
For the purpose of characterizing the local structure of the numerical
fit, we define
\(\Delta\chi^2(\boldsymbol{\lambda})
 =
 \chi^2(\boldsymbol{\lambda})-\chi^2_{\rm min}
\),
where $\boldsymbol{\lambda}$ denotes the set of model parameters. In
the two-dimensional slices discussed below, the two displayed
parameters are varied simultaneously while the remaining eight
parameters are kept fixed at their BFP values. The conventional
two-parameter thresholds are used only as reference numerical cuts and
should not be interpreted as statistically rigorous confidence levels,
since the global fit is underconstrained.

The likelihood analysis also requires the values of the quark masses. We adopt
the values reported in Ref.~\cite{Workman:2022} and evolve them to the
top-quark mass scale using the {\it RunDec}
package~\cite{Chetyrkin:2000yt}. The resulting quark mass ratios are
\begin{equation}\label{Eq:Quark-ratios}
 \begin{array}{ll}
  \widetilde{m}_{u} = \left(1.26 \pm 0.22\right)\times10^{-5}, &
  \widetilde{m}_{c} = \left(6.35 \pm 0.13\right)\times10^{-3}, \\
  \widetilde{m}_{d} = \left(1.12 \pm 0.08\right)\times10^{-3}, &
  \widetilde{m}_{s} = \left(2.25 \pm 0.16\right)\times10^{-2}.
 \end{array}
\end{equation}
Since these mass ratios are approximately invariant under
renormalization-group evolution over a broad range of energy
scales~\cite{ParticleDataGroup:2024cfk,Huang:2021,Fusaoka:1998}, they are
treated as free parameters within the numerical intervals
\begin{equation}
 \begin{array}{ll}
  \widetilde{m}_{u}\in[0.9\times10^{-5},\,10.0\times10^{-5}], &
  \widetilde{m}_{c}\in[0.9\times10^{-3},\,10.0\times10^{-3}], \\
  \widetilde{m}_{d}\in[0.9\times10^{-3},\,10.0\times10^{-3}], &
  \widetilde{m}_{s}\in[0.9\times10^{-2},\,10.0\times10^{-2}].
 \end{array}
\end{equation}
The remaining free parameters are varied within their corresponding physically
allowed domains, subject to the algebraic constraints imposed by the
singular-value parametrization. These constraints guarantee the consistency of
the fermion mass matrices and ensure that the diagonalization procedure remains
well defined throughout the parameter space.
\begin{figure}[!htbp]
 \begin{center}
  \begin{tabular}{cc}
   \includegraphics[width=8.6cm, height=6.8cm]{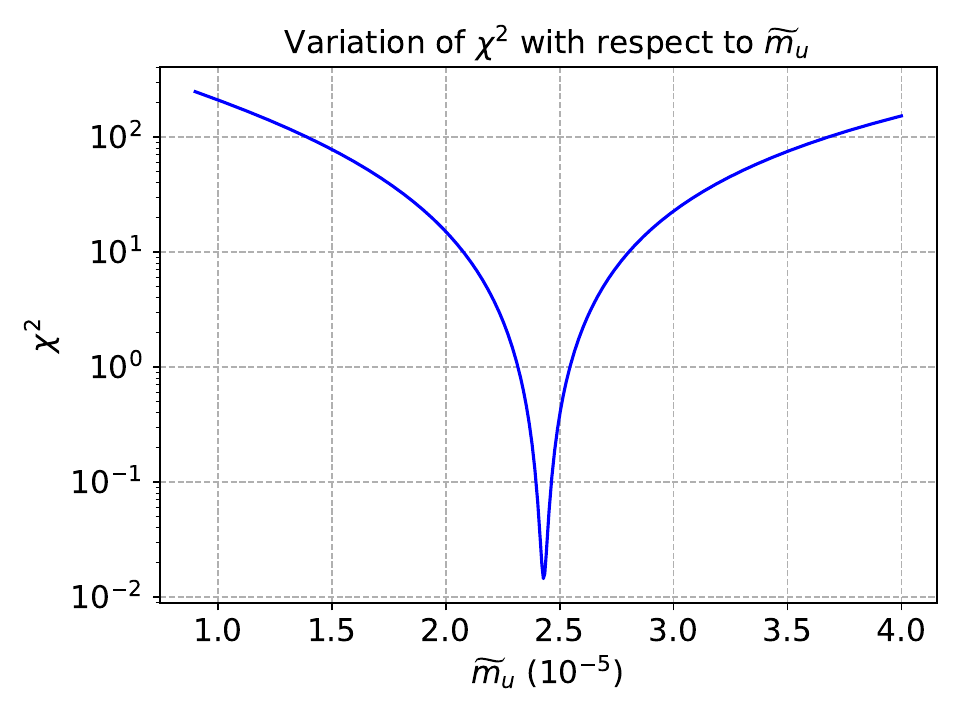} &
   \includegraphics[width=8.6cm, height=6.8cm]{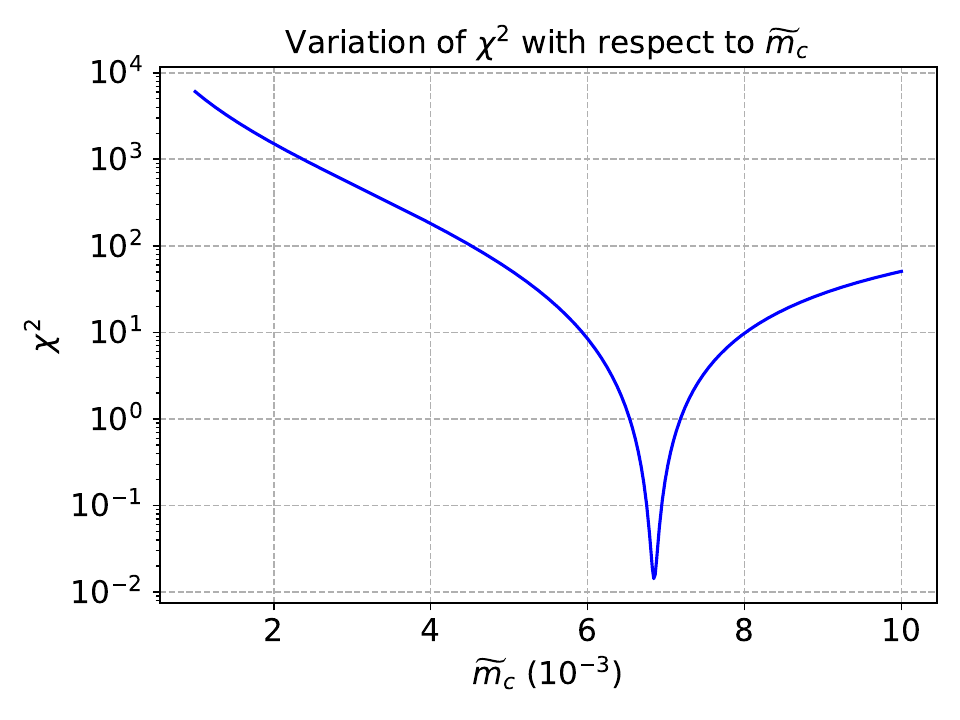} \\
   \includegraphics[width=8.6cm, height=6.8cm]{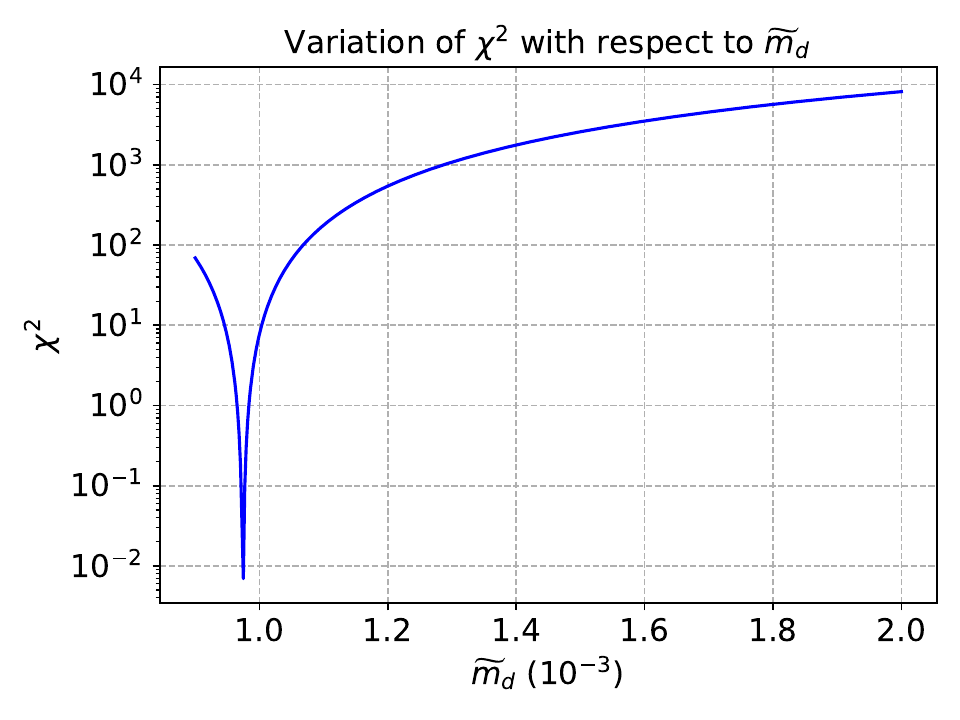} &
   \includegraphics[width=8.6cm, height=6.8cm]{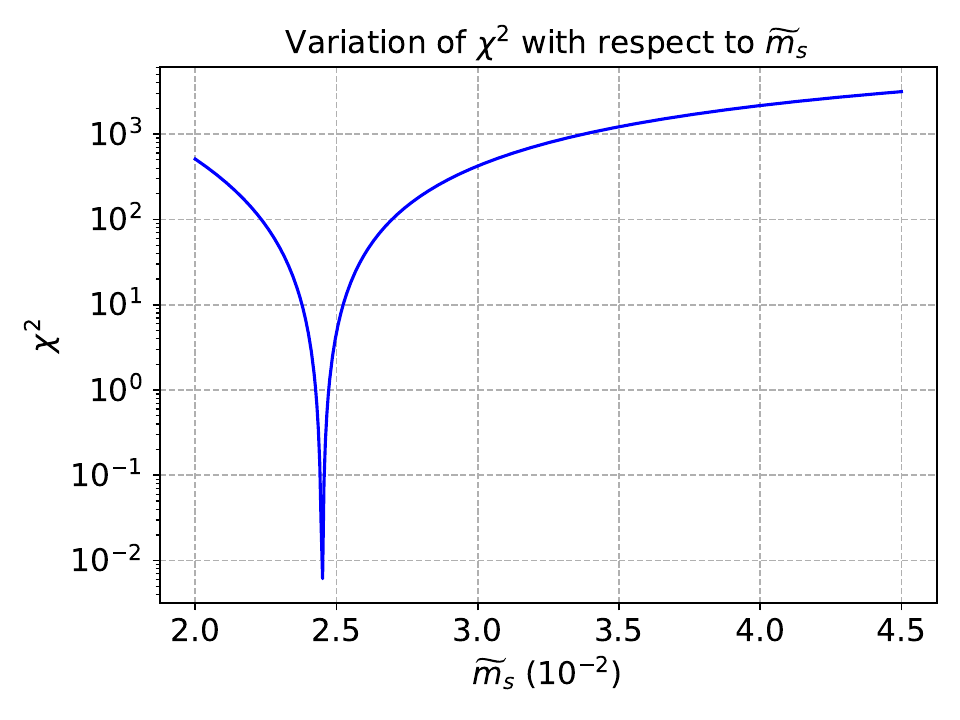} 
  \end{tabular}
  \caption{One-dimensional slices of the $\chi^{2}$ statistic as a function of 
   the quark mass ratios
   $\widetilde{m}_{u}$, $\widetilde{m}_{d}$, $\widetilde{m}_{c}$, and
   $\widetilde{m}_{s}$. Each panel is obtained by varying a single mass ratio
   while keeping the remaining free parameters fixed at their Best Fit Point
   (BFP) values. The location of the minima reveals the preferred regions of the
   parameter space and illustrates the sensitivity of the likelihood function to
   each quark mass ratio.}\label{Fig:mt-q-chi2}
 \end{center}
\end{figure}

Finally, we carry out a comprehensive scan of the allowed parameter space by 
evaluating the $\chi^{2}$ function over a large number of randomly generated 
parameter configurations. The parameter set that yields a minimum of the 
$\chi^{2}$ function is identified as the Best Fit Point (BFP). The resulting 
values of the ten free parameters are
\begin{equation}
 \begin{array}{l}
  \widetilde{m}_{u} = 2.4510 \times 10^{-5}, \quad 
  \widetilde{m}_{c} = 6.7886 \times 10^{-3}, \quad 
  \widetilde{m}_{d} = 0.9763 \times 10^{-3}, \quad 
  \widetilde{m}_{s} = 2.4444 \times 10^{-2}, \\
  \widetilde{m}_{11}^{u} = 1.3507 \times 10^{-9}, \quad  
  \widetilde{m}_{11}^{d} = 1.0850 \times 10^{-8}, \quad 
  \widetilde{\cal D}_{u} = 9.1694 \times 10^{-1}, \quad 
  \widetilde{\cal D}_{d} = 9.1620 \times 10^{-1}, \\ 
  \Theta_{\texttt{j}} = 111.196^{\circ}, \quad 
  \Gamma_{\texttt{j}} = 119.803^{\circ}, \quad 
  \chi^{2} = 4.67 \times 10^{-2}
 \end{array}
\end{equation}
Figure~\ref{Fig:mt-q-chi2} shows the dependence of the $\chi^{2}$ statistic on 
each of the quark mass ratios included in the likelihood analysis. Each panel is 
obtained by performing a one-dimensional scan of the corresponding mass ratio 
while fixing the remaining nine free parameters at their Best Fit Point (BFP) 
values. 
Consequently, the resulting curves represent one-dimensional
$\chi^{2}$ slices that illustrate the sensitivity of the fit to
variations of each mass ratio in the vicinity of the minimum.
In all cases, a well-defined minimum is observed near the corresponding BFP 
value, indicating that the quark mass ratios are tightly constrained by the 
structure of the model and play a non-trivial role in reproducing the 
experimentally measured CKM matrix elements. The asymmetric shape exhibited by 
some of the slices reflects the nonlinear dependence of the $\chi^{2}$ function 
on the mass ratios, as well as the pronounced hierarchical structure of the quark 
mass spectrum.

\begin{figure}
 \begin{tabular}{cc}
  \includegraphics[width=0.48\linewidth]{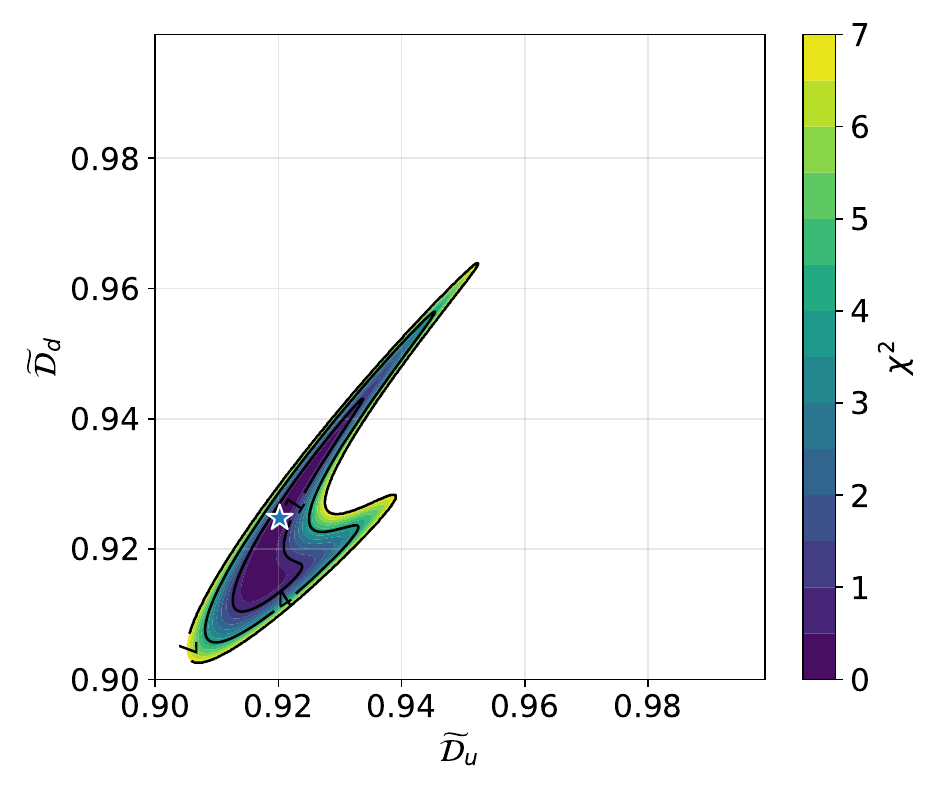} &
  \includegraphics[width=0.48\linewidth]{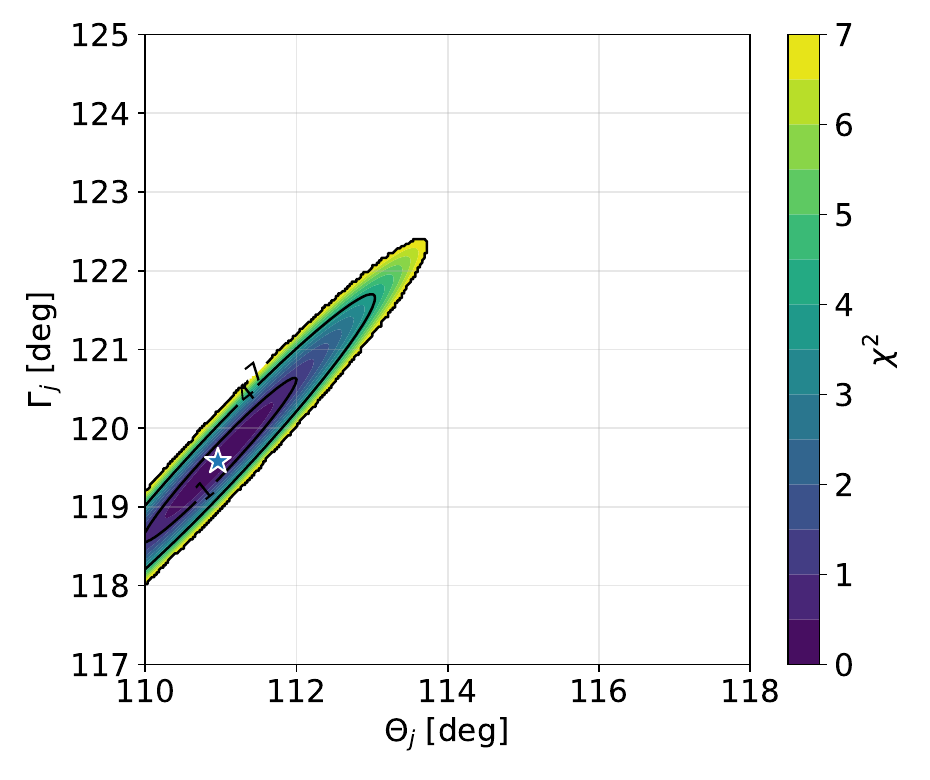}
 \end{tabular}
 \caption{\label{Fig:2DProfiles} Two-dimensional slices of the $\chi^{2}$ 
 function in the
 $(\widetilde{\cal D}_{u},\widetilde{\cal D}_{d})$ (left panel) and 
 $(\Theta_{\texttt{j}},\Gamma_{\texttt{j}})$ (right panel) planes.
 For each slice, the displayed parameters are varied simultaneously over their 
 allowed ranges, while the remaining eight free parameters are fixed at their BFP 
 values. The color map represents the value of the $\chi^{2}$ function, the black 
 contour denotes the $\chi^{2} = 1$ region, and the white star marks the 
 global BFP. The elongated contours reveal the correlations between the 
 corresponding parameters and characterize the local structure of the likelihood 
 function around its global minimum.}
\end{figure} 
Figure~\ref{Fig:2DProfiles} displays the two-dimensional $\chi^{2}$ slices in the
$(\widetilde{D}_{u},\widetilde{D}_{d})$ and
$(\Theta_{\texttt{j}},\Gamma_{\texttt{j}})$ parameter planes.
These slices are obtained by simultaneously varying the two parameters shown
in each panel while fixing the remaining eight free parameters at their BFP
values. Unlike the one-dimensional slices, these contour plots provide direct
information on the local correlations between pairs of parameters and on the
geometry of the $\chi^{2}$ function around its minimum.
For the parameter pairs
$(\widetilde{D}_{u},\widetilde{D}_{d})$, 
$(\widetilde{m}_{11}^{u}, \widetilde{m}_{11}^{d})$, and 
$(\Theta_{\texttt{j}},\Gamma_{\texttt{j}})$, using the reference
two-parameter threshold $\Delta\chi^{2}\leq6.18$, we obtain
\begin{equation}
 \begin{array}{lll}\vspace{2mm}
 \widetilde{\cal D}_{u} = 
  \left( 9.169_{-0.020}^{+0.821} \right) \times 10^{-1}, & 
 \widetilde{m}_{11}^{u} = 
  \left( 1.351_{-0.351}^{+1.865}  \right) \times 10^{-9}, & 
 \Theta_{\texttt{j}}(^{\circ}) = 111.20_{-16.61}^{+23.52} ,\\
 \widetilde{\cal D}_{d} = 
  \left( 9.162_{-0.162}^{+0.821}\right) \times 10^{-1} ,& 
 \widetilde{m}_{11}^{d} = 
  \left( 1.085_{-0.915}^{+0.985}\right) \times 10^{-8},& 
 \Gamma_{\texttt{j}}(^{\circ}) = 119.20_{-13.41}^{+20.20} .
 \end{array}    
\end{equation}
Here, $\chi^{2} = 4.67 \times 10^{-2}$. For each two-dimensional $\Delta\chi^2$ 
contour, we also compute the induced ranges of the nine CKM matrix 
elements. These ranges should be interpreted as the variation of the CKM 
predictions within fixed-$\Delta\chi^2$ regions rather than as statistically 
rigorous confidence intervals, since the global fit is underconstrained.

When the two parameters
$(\widetilde D_u,\widetilde D_d)$ are varied simultaneously,
while the remaining eight free parameters are fixed at their BFP values,
we obtain
\begin{equation}
 \left | {\bf V}_{\mathrm{CKM}}^{L (\texttt{j})} \right | = 
  \left( \begin{array}{ccc}\vspace{2mm}
   0.97447 \pm 0.00014 &  
   0.22449 \pm 0.00063 &  
   \left(3.88 \pm 0.40\right) \times 10^{-3} \\ \vspace{2mm}
   0.22437 \pm 0.00058 &  
   0.97364 \pm 0.00022 & 
   \left( 4.11 \pm 0.30 \right) \times 10^{-2} \\ \vspace{2mm}
   \left( 8.38 \pm 2.90\right) \times 10^{-3} &  
   \left( 4.04 \pm 0.31 \right) \times 10^{-2} & 
   0.999148 \pm 0.00018
  \end{array} \right) .
 \end{equation} 
These results for the magnitudes of all nine CKM elements are in good agreement  
with the results obtained in the global fit~\cite{Workman:2022}.
%
%
\subsection{Leptonic Sector}
%
%
The matrices ${\bf M}_{\nu}$ and ${\bf M}_{e}$ can be brought to
diagonal form through unitary transformations,
\begin{equation}
 \mathbf{U}^{\top}_{\nu L} \, \mathbf{M}_{\nu} \, \mathbf{U}_{\nu L}  = 
 \hat{ \mathbf{M} }_{\nu}, \qquad  
 \mathbf{U}^{\dagger}_{e L} \mathbf{M}_{e} 
 \mathbf{M}^{\dagger}_{e} \mathbf{U}_{e L} = \hat{\mathbf{M}}_{e} 
 \hat{\mathbf{M}}^{\dagger}_{e}
 \qquad \textrm{and} \qquad 
 \mathbf{U}^{\dagger}_{e R} \mathbf{M}^{\dagger}_{e} 
 \mathbf{M}_{e} \mathbf{U}_{e R} = 
 \hat{\mathbf{M}}^{\dagger}_{e} \hat{\mathbf{M}}_{e}, 
\end{equation}
where ${\bf U}_{\nu L}$ and ${\bf U}_{e (L,R)}$ are unitary matrices, while 
$\hat{\bf M}_{e} = \textrm{diag} ( m_{e}, m_{\mu}, m_{\tau} )$ and 
$\hat{\bf M}_{\nu}$ is a diagonal matrix whose elements are the physical 
neutrino masses. 
Similarly, the diagonal Majorana mass matrix $\mathbf M_R$ satisfies
$\mathbf{U}^{\top}_{\nu R} \, \mathbf{M}_{R} \, \mathbf{U}_{\nu R} = 
\hat{ \mathbf{M} }_{R}$ 
where the diagonal entries of $\hat{\mathbf M}_R$ are the physical
heavy-neutrino masses. The left- and right-handed PMNS mixing matrices
are then defined as~\cite{Zyla:2020zbs} 
\begin{equation}
 \begin{array}{l}
  {\bf U}^{L}_{\textrm{PMNS}} = {\bf U}_{e L}^{\dagger} {\bf U}_{\nu L}, 
  \qquad  
  {\bf U}^{R}_{\textrm{PMNS}} = {\bf U}_{e R}^{\dagger} {\bf U}^{\ast}_{\nu R}.
 \end{array}
\end{equation}
Let us emphasize that the structure of
${\bf U}^{R}_{\textrm{PMNS}}$ is determined by the charged-lepton sector
because $\mathbf{M}_{R}$ is diagonal, as discussed above.
The explicit form of the PMNS matrix depends on whether the {\bf PLRT}
or {\bf MLRT} framework is considered.
In analogy with the quark sector, a relation between the left- and 
right-handed leptonic mixing matrices can be established by exploiting 
the properties of the charged lepton mass matrix in the MLRT and PLRT 
scenarios. In the MLRT framework, where the charged-lepton mass matrix 
is Hermitian, ${\bf M}_e = {\bf M}_e^\dagger$, one has
${\bf U}_{eL} = {\bf U}_{eR}$.  
Therefore,
${\bf U}_{\mathrm{PMNS}}^{R} = 
{\bf U}_{\mathrm{PMNS}}^{L}
{\bf U}_{\nu L}^{\dagger}
{\bf U}_{\nu R}^{*}$.
On the other hand, in the PLRT framework, where the charged-lepton 
mass matrix is complex symmetric, ${\bf M}_e={\bf M}^{\top}_e$, the 
diagonalization matrices satisfy ${\bf U}_{eR} = {\bf U}_{eL}^{*}$,
leading to
${\bf U}_{\mathrm{PMNS}}^{R} = 
 \left({\bf U}_{\mathrm{PMNS}}^{L}\right)^{*}
 {\bf U}_{\nu L}^{T}
 {\bf U}_{\nu R}^{*}$.
Since in the present model the right-handed Majorana neutrino mass 
matrix ${\bf M}_R$ is diagonal, one may choose
${\bf U}_{\nu R}=\mathbb{I}_{3 \times 3}$,
up to unphysical phase conventions. Consequently, the above relations 
reduce to
${\bf U}_{\mathrm{PMNS}}^{R}
 = 
 {\bf U}_{\mathrm{PMNS}}^{L}
 {\bf U}_{\nu L}^{\dagger}
 $, 
 ${\bf U}_{\mathrm{PMNS}}^{R}
 = 
 {\bf U}_{e L}^{\dagger}
 $, so
 ${\bf U}_{\mathrm{PMNS}}^{R}
 = 
 {\bf O}_{e L}^{\top} 
 {\bf P}_{e} 
 {\bf R}_{\frac{\pi}{4}}$  
 for MLRT. And
 ${\bf U}_{\mathrm{PMNS}}^{R}
 =
 \left({\bf U}_{\mathrm{PMNS}}^{L}\right)^{*}
 {\bf U}_{\nu L}^{T}$,
 ${\bf U}_{\mathrm{PMNS}}^{R}
 = 
 {\bf U}_{e L}^{\top}
 $, so
 ${\bf U}_{\mathrm{PMNS}}^{R}
 = 
 {\bf O}_{e L}^{\top} 
 {\bf P}_{e} 
 {\bf R}_{\frac{\pi}{4}}$  
 for PLRT.
Thus, unlike in the quark sector, the left- and right-handed PMNS 
matrices are not, in general, directly equal or complex conjugates of 
each other. This difference originates from the distinct diagonalization 
matrices associated with the light and heavy Majorana neutrino sectors. 
In particular, because $M_R$ is diagonal, the right-handed leptonic 
mixing matrix is entirely determined by the charged-lepton sector.
From the unitary matrices in Eqs.~(\ref{eq:Uq-18}) and~(\ref{eq:Uq-19}) 
the explicit form of the left-handed PMNS flavor mixing matrix is given by 
\begin{equation}
 \begin{array}{l}\vspace{2mm}
  {\bf U}_{\textrm{PMNS}}^{L(1)} =
  {\bf O}_{e}^{\top} 
  {\bf P}_{e}^{\dagger} 
  {\bf R}_{\beta \frac{\pi}{4}}
  {\bf P}_{\nu}^{1} 
  {\bf O}_{\nu} 
  \quad \textrm{for PLRT}, 
  \quad \textrm{and} \quad
  {\bf U}_{\textrm{PMNS}}^{L(2)} =
  {\bf O}_{e}^{\top} 
  {\bf P}_{e}
  {\bf R}_{\beta \frac{\pi}{4}}
  {\bf P}_{\nu}^{2} 
  {\bf O}_{\nu} 
  \quad \textrm{for MLRT}, 
 \end{array}
\end{equation}
where 
${\bf R}_{\beta \frac{\pi}{4}} = {\bf R}_{\frac{\pi}{4}} 
{\bf R}_{\beta_{\nu}}^{(\texttt{j})\top}$
\begin{equation}
 {\bf R}_{\beta \frac{\pi}{4}} = 
  \left( \begin{array}{ccc} \vspace{2mm}
   \cos \left( \frac{\pi}{4} - \beta_{\nu}^{(\texttt{j})} \right) & 
   -\sin \left( \frac{\pi}{4} - \beta_{\nu}^{(\texttt{j})} \right)  & 
   0 \\ \vspace{2mm}
   \sin \left( \frac{\pi}{4} - \beta_{\nu}^{(\texttt{j})} \right) &  
   \cos \left( \frac{\pi}{4} - \beta_{\nu}^{(\texttt{j})} \right) & 
   0 \\ \vspace{2mm}
   0 & 0 & 1
  \end{array} \right).
\end{equation}
In the particular case when the rotation angle is the same in both lepton 
sectors, $\beta_{\nu}^{(\texttt{j})} = \frac{\pi}{4}$, we have
\({\bf R}_{\beta \frac{\pi}{4}} = \textrm{diag} \left(1, 1, 1 \right) \).
Then the left-handed PMNS mixing matrix takes the form
\begin{equation}
 \begin{array}{l}\vspace{2mm}
  {\bf U}_{\textrm{PMNS}}^{L(\texttt{j})} =
  e^{ i \Xi_{\texttt{j}} }
  {\bf O}_{e}^{\top} 
  {\bf P}^{(\nu - e)}_{\texttt{j}} 
  {\bf O}_{\nu} ,
 \end{array}
\end{equation}
where  
\begin{equation}
 \begin{array}{l}
  {\bf P}_{j}^{(\nu -e)} = \textrm{diag} 
  \left( 
   1, e^{i \varsigma_{ \texttt{j} } }, e^{i \varrho_{ \texttt{j} } } 
  \right),
 \end{array}
\end{equation}
with 
\begin{equation}\label{Eq:Def-phase-factors}
 \begin{array}{lll}\vspace{2mm}
  \Xi_{1} = \frac{1}{2} \left( \alpha_{\nu 1} - \alpha_{e 1} \right), & 
  \varsigma_{1} = \frac{1}{2} 
   \left( \beta_{\nu 1} - \beta_{e} - \alpha_{\nu 1} + \alpha_{e 1} \right), &
  \varrho_{1} = \frac{1}{2} 
   \left( 
    \gamma_{\nu 1} - \gamma_{e 1} - \alpha_{\nu 1} + \alpha_{e 1} 
   \right), \\
  \Xi_{2} =  \frac{1}{2} \alpha_{\nu 2}, & 
  \varsigma_{2} = 
   \frac{1}{2} \left( \beta_{\nu 2} - \alpha_{\nu 2} \right) , &
  \varrho_{2} = 
   \frac{1}{2} \left( 2 \phi_{e c} + \gamma_{e 2} - \alpha_{\nu 2} \right).
 \end{array}
\end{equation}
It is worth noting that the choice $\beta_{\nu}^{(\texttt{j})}=\pi/4$ is not 
arbitrary. 
In fact, it emerges naturally when the effective neutrino mass matrix preserves 
the residual~$S_{3}$ structure inherited from the underlying flavor symmetry.
Under this condition, the matrix elements responsible for the rotation satisfy
$d_{\nu}=c_{\nu}$, 
leading to
$\tan\beta_{\nu}^{(\texttt{j})}=1$ and therefore
$\beta_{\nu}^{(\texttt{j})}=\pi/4$. This result reflects a maximal mixing
within the $\mathbf{2}$ representation of $S_{3}$ and constitutes
the neutrino-sector analogue of the universal $R_{\pi/4}$ rotation that
appears in the quark sector. The explicit form of the PMNS mixing 
matrix elements in both theoretical frameworks is:
\begin{equation}
 {\bf U}_{\textrm{PMNS}}^{L(\texttt{j})} =
  e^{ i \Xi_{\texttt{j}} }
 \left( \begin{array}{ccc}
  V_{e1}^{\mathrm{th}} & 
  V_{e2}^{\mathrm{th}} & 
  V_{e3}^{\mathrm{th}} \\
  V_{\mu 1}^{\mathrm{th}} & 
  V_{\mu 2}^{\mathrm{th}} & 
  V_{\mu 3}^{\mathrm{th}} \\
  V_{\tau 1}^{\mathrm{th}} & 
  V_{\tau 2}^{\mathrm{th}} & 
  V_{\tau 3}^{\mathrm{th}}
 \end{array}  \right),
\end{equation} 
The entries of the PMNS mixing matrix are given by
\begin{equation}
 \begin{array}{l}
  V_{e1}^{(\texttt{j})} =
   \sqrt{ \frac{
    \widetilde{\sigma}_{\mu} \widetilde{\sigma}_{\tau} 
    \widetilde{\sigma}_{\nu2} \widetilde{\sigma}_{\nu3} 
    \widetilde{\xi}_{\ell 1} \widetilde{\xi}_{\nu 1}
   }{ \widetilde{\texttt{D}}_{\ell 1} \widetilde{\texttt{D}}_{\nu 1} } }
   +
   \sqrt{ \frac{
    \widetilde{\sigma}_{e} \widetilde{\sigma}_{\nu1} 
   }{ \widetilde{\texttt{D}}_{\ell 1} \widetilde{\texttt{D}}_{\nu 1} } }
   \varepsilon_{e1}^{(\texttt{j})}, \quad
  V_{e2}^{(\texttt{j})}  =
   -\sqrt{ \frac{
     \widetilde{\sigma}_{\mu} \widetilde{\sigma}_{\tau} 
     \widetilde{\sigma}_{\nu1} \widetilde{\sigma}_{\nu3} 
     \widetilde{\xi}_{\ell 1} \widetilde{\xi}_{\nu 2}
   }{ \widetilde{\texttt{D}}_{\ell 1} \widetilde{\texttt{D}}_{\nu 2} } }
   +
   \sqrt{ \frac{
    \widetilde{\sigma}_{e} \widetilde{\sigma}_{\nu2} 
   }{ \widetilde{\texttt{D}}_{\ell 1} \widetilde{\texttt{D}}_{\nu 2} } }
   \varepsilon_{e2}^{(\texttt{j})},
 \end{array}
\end{equation}
\begin{equation}\label{ec:V-e3-th}
 \begin{array}{l}
  V_{e3}^{(\texttt{j})} =
   \sqrt{ \frac{
    \widetilde{\sigma}_{\mu} \widetilde{\sigma}_{\tau} 
    \widetilde{\sigma}_{\nu1} \widetilde{\sigma}_{\nu2} 
    \widetilde{\xi}_{\ell 1} \widetilde{\xi}_{\nu 3}
   }{ \widetilde{\texttt{D}}_{\ell 1} \widetilde{\texttt{D}}_{\nu 3} } }
   +
   \sqrt{ \frac{
    \widetilde{\sigma}_{e} \widetilde{\sigma}_{\nu3} 
   }{ \widetilde{\texttt{D}}_{\ell 1} \widetilde{\texttt{D}}_{\nu 3} } }
   \varepsilon_{e3}^{(\texttt{j})}, \quad
  V_{\mu 1}^{(\texttt{j})} =
   -\sqrt{ \frac{
    \widetilde{\sigma}_{e} \widetilde{\sigma}_{\tau} 
    \widetilde{\sigma}_{\nu2} \widetilde{\sigma}_{\nu3} 
    \widetilde{\xi}_{\ell 2} \widetilde{\xi}_{\nu 1}
   }{ \widetilde{\texttt{D}}_{\ell 2} \widetilde{\texttt{D}}_{\nu 1} } }
   +
   \sqrt{ \frac{
    \widetilde{\sigma}_{\mu} \widetilde{\sigma}_{\nu1} 
   }{ \widetilde{\texttt{D}}_{\ell 2} \widetilde{\texttt{D}}_{\nu 1} } }
   \varepsilon_{\mu 1}^{(\texttt{j})},
 \end{array}
\end{equation}
\begin{equation}
 \begin{array}{l}
  V_{\mu 2}^{(\texttt{j})}=
   \sqrt{ \frac{
    \widetilde{\sigma}_{e} \widetilde{\sigma}_{\tau} 
    \widetilde{\sigma}_{\nu1} \widetilde{\sigma}_{\nu3} 
    \widetilde{\xi}_{\ell 2} \widetilde{\xi}_{\nu 2}
   }{ \widetilde{\texttt{D}}_{\ell 2} \widetilde{\texttt{D}}_{\nu 2} } }
   +
   \sqrt{ \frac{
    \widetilde{\sigma}_{\mu} \widetilde{\sigma}_{\nu2} 
   }{ \widetilde{\texttt{D}}_{\ell 2} \widetilde{\texttt{D}}_{\nu 2} } }
   \varepsilon_{\mu 2}^{(\texttt{j})}, \quad
  V_{\mu 3}^{(\texttt{j})}=
   - \sqrt{ \frac{
    \widetilde{\sigma}_{e} \widetilde{\sigma}_{\tau} 
    \widetilde{\sigma}_{\nu1} \widetilde{\sigma}_{\nu2} 
    \widetilde{\xi}_{\ell 2} \widetilde{\xi}_{\nu 3}
   }{ \widetilde{\texttt{D}}_{\ell 2} \widetilde{\texttt{D}}_{\nu 3} } }
   +
   \sqrt{ \frac{
    \widetilde{\sigma}_{\mu} \widetilde{\sigma}_{\nu3} 
   }{ \widetilde{\texttt{D}}_{\ell 2} \widetilde{\texttt{D}}_{\nu 3} } }
   \varepsilon_{\mu 3}^{(\texttt{j})},
 \end{array}
\end{equation}
\begin{equation}
 \begin{array}{l}
  V_{\tau 1}^{(\texttt{j})}=
   \sqrt{ \frac{
    \widetilde{\sigma}_{e} \widetilde{\sigma}_{\mu} 
    \widetilde{\sigma}_{\nu2} \widetilde{\sigma}_{\nu3} 
    \widetilde{\xi}_{\ell 3} \widetilde{\xi}_{\nu 1}
   }{ \widetilde{\texttt{D}}_{\ell 3} \widetilde{\texttt{D}}_{\nu 1} } }
   +
   \sqrt{ \frac{
    \widetilde{\sigma}_{\tau} \widetilde{\sigma}_{\nu1} 
   }{ \widetilde{\texttt{D}}_{\ell 3} \widetilde{\texttt{D}}_{\nu 1} } }
  \varepsilon_{\tau 1}^{(\texttt{j})}, \quad
  V_{\tau 2}^{(\texttt{j})}=
   - \sqrt{ \frac{
    \widetilde{\sigma}_{e} \widetilde{\sigma}_{\mu} 
    \widetilde{\sigma}_{\nu1} \widetilde{\sigma}_{\nu3} 
    \widetilde{\xi}_{\ell 3} \widetilde{\xi}_{\nu 2}
   }{ \widetilde{\texttt{D}}_{\ell 3} \widetilde{\texttt{D}}_{\nu 2} } }
   +
   \sqrt{ \frac{
    \widetilde{\sigma}_{\tau} \widetilde{\sigma}_{\nu2} 
   }{ \widetilde{\texttt{D}}_{\ell 3} \widetilde{\texttt{D}}_{\nu 2} } }
  \varepsilon_{\tau 2}^{(\texttt{j})},
 \end{array}
\end{equation}
\begin{equation}
 \begin{array}{l}
  V_{\tau 3}^{(\texttt{j})} =
   \sqrt{ \frac{
    \widetilde{\sigma}_{e} \widetilde{\sigma}_{\mu} 
    \widetilde{\sigma}_{\nu1} \widetilde{\sigma}_{\nu2} 
    \widetilde{\xi}_{\ell 3} \widetilde{\xi}_{\nu 3}
   }{ \widetilde{\texttt{D}}_{\ell 3} \widetilde{\texttt{D}}_{\nu 3} } }
   +
   \sqrt{ \frac{
    \widetilde{\sigma}_{\tau} \widetilde{\sigma}_{\nu3} 
   }{ \widetilde{\texttt{D}}_{\ell 3} \widetilde{\texttt{D}}_{\nu 3} } }
   \varepsilon_{\tau 3}^{(\texttt{j})} .
 \end{array}
\end{equation}
where
\begin{equation}
 \begin{array}{l}
 \varepsilon_{e1}^{(\texttt{j})} =
    \sqrt{ \widetilde{\cal D}_{\ell} \widetilde{\cal D}_{\nu} 
    \widetilde{\xi}_{\ell 1} \widetilde{\xi}_{\nu 1} } 
    e^{i \varsigma_{ \texttt{j} } }
    +
    \sqrt{ \widetilde{\xi}_{\ell 2} \widetilde{\xi}_{\ell 3} 
    \widetilde{\xi}_{\nu 2} 
     \widetilde{\xi}_{\nu 3} } 
    e^{i \varrho_{ \texttt{j} } }, \quad
  \varepsilon_{e2}^{(\texttt{j})} = 
    \sqrt{ \widetilde{\cal D}_{\ell} \widetilde{\cal D}_{\nu} 
    \widetilde{\xi}_{\ell 1} \widetilde{\xi}_{\nu 2} }  
    e^{i \varsigma_{ \texttt{j} } }
    +
    \sqrt{ \widetilde{\xi}_{\ell 2} \widetilde{\xi}_{\ell 3} 
     \widetilde{\xi}_{\nu 1} 
     \widetilde{\xi}_{\nu 3} } 
     e^{i \varrho_{ \texttt{j} } },
 \end{array}
\end{equation}
\begin{equation}
 \begin{array}{l}
  \varepsilon_{e3}^{(\texttt{j})} = 
    \sqrt{ \widetilde{\cal D}_{\ell} \widetilde{\cal D}_{\nu} 
    \widetilde{\xi}_{\ell 1} \widetilde{\xi}_{\nu 3} }  
    e^{i \varsigma_{ \texttt{j} } }
    -
    \sqrt{ \widetilde{\xi}_{\ell 2} \widetilde{\xi}_{\ell 3} 
    \widetilde{\xi}_{\nu 1} \widetilde{\xi}_{\nu 2} }  
    e^{i \varrho_{ \texttt{j} } } , \quad
  \varepsilon_{\mu 1}^{(\texttt{j})} =
    \sqrt{ \widetilde{\cal D}_{\ell} \widetilde{\cal D}_{\nu} 
    \widetilde{\xi}_{\ell 2} \widetilde{\xi}_{\nu 1} }  
    e^{i \varsigma_{ \texttt{j} } }
    +
    \sqrt{ \widetilde{\xi}_{\ell 1} \widetilde{\xi}_{\ell 3} 
    \widetilde{\xi}_{\nu 2} \widetilde{\xi}_{\nu 3} }  
    e^{i \varrho_{ \texttt{j} } },
 \end{array}
\end{equation}
\begin{equation}
 \begin{array}{l} 
  \varepsilon_{\mu 2}^{(\texttt{j})} = 
    \sqrt{ \widetilde{\cal D}_{\ell} \widetilde{\cal D}_{\nu} 
    \widetilde{\xi}_{\ell 2} \widetilde{\xi}_{\nu 2} }  
    e^{i \varsigma_{ \texttt{j} } }
    +
    \sqrt{ \widetilde{\xi}_{\ell 1} \widetilde{\xi}_{\ell 3} 
    \widetilde{\xi}_{\nu 1} \widetilde{\xi}_{\nu 3} }  
    e^{i \varrho_{ \texttt{j} } } , \quad
  \varepsilon_{\mu 3}^{(\texttt{j})} = 
    \sqrt{ \widetilde{\cal D}_{\ell} \widetilde{\cal D}_{\nu} 
    \widetilde{\xi}_{\ell 2} \widetilde{\xi}_{\nu 3} } 
    e^{i \varsigma_{ \texttt{j} } }
    -
    \sqrt{ \widetilde{\xi}_{\ell 1} \widetilde{\xi}_{\ell 3} 
    \widetilde{\xi}_{\nu 1} \widetilde{\xi}_{\nu 2} } 
    e^{i \varrho_{ \texttt{j} } },
 \end{array}
\end{equation}
\begin{equation}
 \begin{array}{l} 
  \varepsilon_{\tau 1}^{(\texttt{j})} =
    \sqrt{ \widetilde{\cal D}_{\ell} \widetilde{\cal D}_{\nu} 
    \widetilde{\xi}_{\ell 3} \widetilde{\xi}_{\nu 1} }  
    e^{i \varsigma_{ \texttt{j} } }
    -
    \sqrt{ \widetilde{\xi}_{\ell 1} \widetilde{\xi}_{\ell 2} 
    \widetilde{\xi}_{\nu 2} \widetilde{\xi}_{\nu 3} }  
    e^{i \varrho_{ \texttt{j} } } , \quad
  \varepsilon_{\tau 2}^{(\texttt{j})} =
    \sqrt{ \widetilde{\cal D}_{\ell} \widetilde{\cal D}_{\nu} 
    \widetilde{\xi}_{\ell 3} \widetilde{\xi}_{\nu 2} } 
    e^{i \varsigma_{ \texttt{j} } }
    -
    \sqrt{ \widetilde{\xi}_{\ell 1} \widetilde{\xi}_{\ell 2} 
    \widetilde{\xi}_{\nu 1} \widetilde{\xi}_{\nu 3} } 
    e^{i \varrho_{ \texttt{j} } },
 \end{array}
\end{equation}
\begin{equation}
 \begin{array}{l}
  \varepsilon_{\tau 3}^{(\texttt{j})} =
    \sqrt{ \widetilde{\cal D}_{\ell} \widetilde{\cal D}_{\nu} 
    \widetilde{\xi}_{\ell 3} \widetilde{\xi}_{\nu 3} }  
    e^{i \varsigma_{ \texttt{j} } }
    +
    \sqrt{ \widetilde{\xi}_{\ell 1} \widetilde{\xi}_{\ell 2} 
    \widetilde{\xi}_{\nu 1} \widetilde{\xi}_{\nu 2} }  
   e^{i \varrho_{ \texttt{j} } } .
 \end{array}
\end{equation}
Here, 
$\widetilde{\sigma}_{e} = \widetilde{m}_{e} - \widetilde{m}_{11}^{\ell}$,
$\widetilde{\sigma}_{\mu} = \widetilde{m}_{\mu} + \widetilde{m}_{11}^{\ell}$,
$\widetilde{\sigma}_{\tau} = 1 - \widetilde{m}_{11}^{\ell}$,
$\widetilde{\xi}_{\ell 1} = 
 \widetilde{\cal D}_{\ell} - \widetilde{\sigma}_{e}$, 
$\widetilde{\xi}_{\ell 2} = 
 \widetilde{\cal D}_{\ell} +\widetilde{\sigma}_{\mu}$, 
$\widetilde{\xi}_{\ell 3} = 
 \widetilde{\sigma}_{\tau} - \widetilde{\cal D}_{\ell}$,
\begin{equation}\label{ec:Vpmns-th-5}
 \begin{array}{l}\vspace{2mm}
  \widetilde{\texttt{D}}_{\ell 1} =      
   \widetilde{\cal D}_{\ell}
   \left( \widetilde{\sigma}_{\mu}  + \widetilde{\sigma}_{e} \right) 
   \left( \widetilde{\sigma}_{\tau} - \widetilde{\sigma}_{e} \right) , \quad
 \widetilde{\texttt{D}}_{\ell 2} = 
   \widetilde{\cal D}_{\ell} 
   \left( \widetilde{\sigma}_{\tau} + \widetilde{\sigma}_{\mu} \right) 
   \left( \widetilde{\sigma}_{\mu}  + \widetilde{\sigma}_{e} \right) ,  \quad
  \widetilde{\texttt{D}}_{\ell 3} =      
   \widetilde{\cal D}_{\ell} 
   \left( \widetilde{\sigma}_{\tau} + \widetilde{\sigma}_{\mu}  \right) 
   \left( \widetilde{\sigma}_{\tau} - \widetilde{\sigma}_{e} \right) , \\
   \widetilde{\texttt{D}}_{\nu 1} =      
   \widetilde{\cal D}_{\nu}
   \left( \widetilde{\sigma}_{\nu 2} + \widetilde{\sigma}_{\nu 1} \right) 
   \left( \widetilde{\sigma}_{\nu 3} - \widetilde{\sigma}_{\nu 1} \right) ,  \quad
  \widetilde{\texttt{D}}_{\nu 2} = 
   \widetilde{\cal D}_{\nu} 
   \left( \widetilde{\sigma}_{\nu 3} + \widetilde{\sigma}_{\nu 2} \right) 
   \left( \widetilde{\sigma}_{\nu 2} + \widetilde{\sigma}_{\nu 1} \right) , \quad 
  \widetilde{\texttt{D}}_{\nu 3} =      
   \widetilde{\cal D}_{\nu} 
   \left( \widetilde{\sigma}_{\nu 3}  + \widetilde{\sigma}_{\nu 2}  \right) 
   \left( \widetilde{\sigma}_{\nu 3}  - \widetilde{\sigma}_{\nu 1} \right) , 
 \end{array}
\end{equation}  
where
$\widetilde{m}_{e} = \frac{ m_{e} }{ m_{\tau} }$,
$\widetilde{m}_{\mu} = \frac{ m_{\mu} }{ m_{\tau} }$,
$\widetilde{m}_{11}^{\ell} = \frac{ m_{11}^{\ell} }{ m_{\tau} }$,
$\widetilde{\sigma}_{e} = \frac{ \sigma_{e} }{ m_{\tau} }$,
$\widetilde{\sigma}_{\mu} = \frac{ \sigma_{\mu} }{ m_{\tau} }$,
$\widetilde{\sigma}_{\tau} = \frac{ \sigma_{\tau} }{ m_{\tau} }$,
$\widetilde{\xi}_{\ell 1} = \frac{ \xi_{\ell 1} }{ m_{\tau} }$,
$\widetilde{\xi}_{\ell 2} = \frac{ \xi_{\ell 2} }{ m_{\tau} }$,
$\widetilde{\xi}_{\ell 3} = \frac{ \xi_{\ell 3} }{ m_{\tau} }$,
$\widetilde{\cal D}_{\ell} = \frac{ {\cal D}_{\ell} }{ m_{\tau} }$. 
For the normal neutrino mass ordering,
$m_{\nu 3} > m_{\nu 2} > m_{\nu 1}$, the following relations hold: 
$\widetilde{\sigma}_{\nu 1} = \widetilde{m}_{\nu 1} - \widetilde{m}_{11}^{\nu}$, 
$\widetilde{\sigma}_{\nu 2} = \widetilde{m}_{\nu 2} + \widetilde{m}_{11}^{\nu}$,
$\widetilde{\sigma}_{\nu 3} = 1 - \widetilde{m}_{11}^{\nu}$,
$\widetilde{\xi}_{\nu 1} = \widetilde{\cal D}_{\nu} - \widetilde{\sigma}_{\nu 1}$,
$\widetilde{\xi}_{\nu 2} = \widetilde{\cal D}_{\nu} + \widetilde{\sigma}_{\nu 2}$,
$\widetilde{\xi}_{\nu 3} = \widetilde{\sigma}_{\nu 3} - \widetilde{\cal D}_{\nu}$,
where
$\widetilde{m}_{\nu 1} = \frac{ m_{\nu 1} }{ m_{\nu 3} }$,
$\widetilde{m}_{\nu 2} = \frac{ m_{\nu 2} }{ m_{\nu 3} }$,
$\widetilde{m}_{11}^{\nu} = \frac{ m_{11}^{\nu} }{ m_{\nu 3} }$,
$\widetilde{\sigma}_{\nu 1} = \frac{ \sigma_{\nu 1} }{ m_{\nu 3} }$,
$\widetilde{\sigma}_{\nu 2} = \frac{ \sigma_{\nu 2} }{ m_{\nu 3} }$,
$\widetilde{\sigma}_{\nu 3} = \frac{ \sigma_{\nu 3} }{ m_{\nu 3} }$,
$\widetilde{\xi}_{\nu 1} = \frac{ \xi_{\nu 1} }{ m_{\nu 3} }$,
$\widetilde{\xi}_{\nu 2} = \frac{ \xi_{\nu 2} }{ m_{\nu 3} }$,
$\widetilde{\xi}_{\nu 3} = \frac{ \xi_{\nu 3} }{ m_{\nu 3} }$,
$\widetilde{\cal D}_{\nu} = \frac{ {\cal D}_{\nu} }{ m_{\nu 3} }$. 
For the inverted neutrino mass ordering, $m_{\nu 2} > m_{\nu 1} > m_{\nu 3}$: 
$\widetilde{\sigma}_{\nu 1} = \widetilde{m}_{\nu 1} - \widetilde{m}_{11}^{\nu}$,
$\widetilde{\sigma}_{\nu 2} = 1 + \widetilde{m}_{11}^{\nu}$,
$\widetilde{\sigma}_{\nu 3} = \widetilde{m}_{\nu 3} - \widetilde{m}_{11}^{\nu}$,
$\widetilde{\xi}_{\nu 1} = \widetilde{\cal D}_{\nu} - \widetilde{\sigma}_{\nu 1}$,
$\widetilde{\xi}_{\nu 2} = \widetilde{\cal D}_{\nu} + \widetilde{\sigma}_{\nu 2}$,
$\widetilde{\xi}_{\nu 3} = \widetilde{\sigma}_{\nu 3} - \widetilde{\cal D}_{\nu}$,
where
$\widetilde{m}_{\nu 1} =      \frac{ m_{\nu 1} }{ m_{\nu 2} }$,
$\widetilde{m}_{\nu 3} =      \frac{ m_{\nu 3} }{ m_{\nu 2} }$,
$\widetilde{m}_{11}^{\nu} =   \frac{ m_{11}^{\nu} }{ m_{\nu 2} }$,
$\widetilde{\sigma}_{\nu 1} = \frac{ \sigma_{\nu 1} }{ m_{\nu 2} }$,
$\widetilde{\sigma}_{\nu 2} = \frac{ \sigma_{\nu 2} }{ m_{\nu 2} }$,
$\widetilde{\sigma}_{\nu 3} = \frac{ \sigma_{\nu 3} }{ m_{\nu 2} }$,
$\widetilde{\xi}_{\nu 1} = \frac{ \xi_{\nu 1} }{ m_{\nu 2} }$,
$\widetilde{\xi}_{\nu 2} = \frac{ \xi_{\nu 2} }{ m_{\nu 2} }$,
$\widetilde{\xi}_{\nu 3} = \frac{ \xi_{\nu 3} }{ m_{\nu 2} }$,
$\widetilde{\cal D}_{\nu} = \frac{ {\cal D}_{\nu} }{ m_{\nu 2} }$. 

For the normal neutrino mass ordering, the 
parameter $\widetilde{\mathcal D}_{\ell}$, with $\ell=e,\nu$, is required to 
satisfy the following condition:
$\widetilde{m}_{11}^{l} < \widetilde{m}_{l1}$, 
$\widetilde{\sigma}_{l3} > \widetilde{\cal D}_{l} > \widetilde{\sigma}_{l1}$, 
and
$\widetilde{\cal D}_{l} \neq 
 + \widetilde{\sigma}_{l1} - \widetilde{\sigma}_{l2} + \widetilde{\sigma}_{l3}$.
For the inverted neutrino mass ordering, the following condition is 
required to hold:
$\widetilde{m}_{11}^{\nu} < \widetilde{m}_{\nu3}$,
$\widetilde{\sigma}_{\nu 1} > \widetilde{\cal D}_{\nu} > 
\widetilde{\sigma}_{\nu 3}$, 
and
$\widetilde{\cal D}_{\nu} \neq + \widetilde{\sigma}_{\nu 1} - 
\widetilde{\sigma}_{\nu 2} + \widetilde{\sigma}_{\nu 3}$.

As in the quark sector, the main distinction between the PMNS matrices obtained 
within the {\bf PLRT} and {\bf MLRT} frameworks resides in their phase structure. 
Nevertheless, both scenarios contain the same number of effective physical 
phases, which are related through the definitions given in 
Eq.~(\ref{Eq:Def-phase-factors}). Consequently, a numerical analysis based solely 
on the PMNS matrix is unable to discriminate between the two frameworks. 
This result indicates that neutrino mixing observables alone do not provide 
sufficient information to distinguish whether the underlying theory corresponds 
to the {\bf PLRT} or the {\bf MLRT} scenario.
%
\subsubsection{Likelihood Test }
%
\begin{figure}
 \begin{tabular}{cc}
  \includegraphics[width=0.48\linewidth]{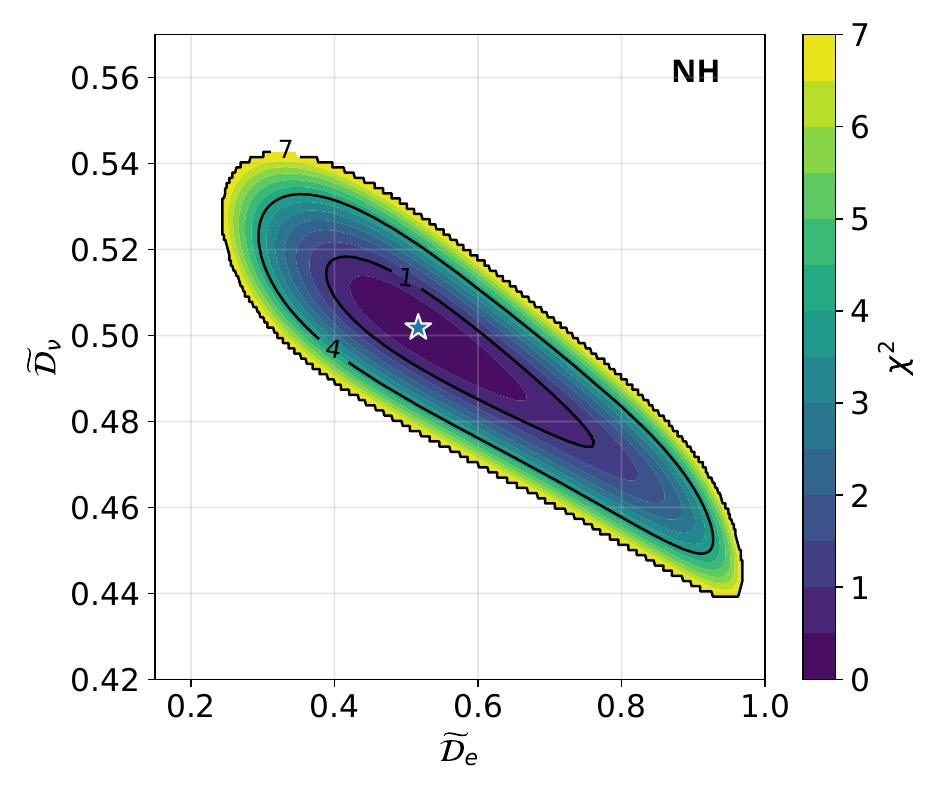} &
  \includegraphics[width=0.48\linewidth]{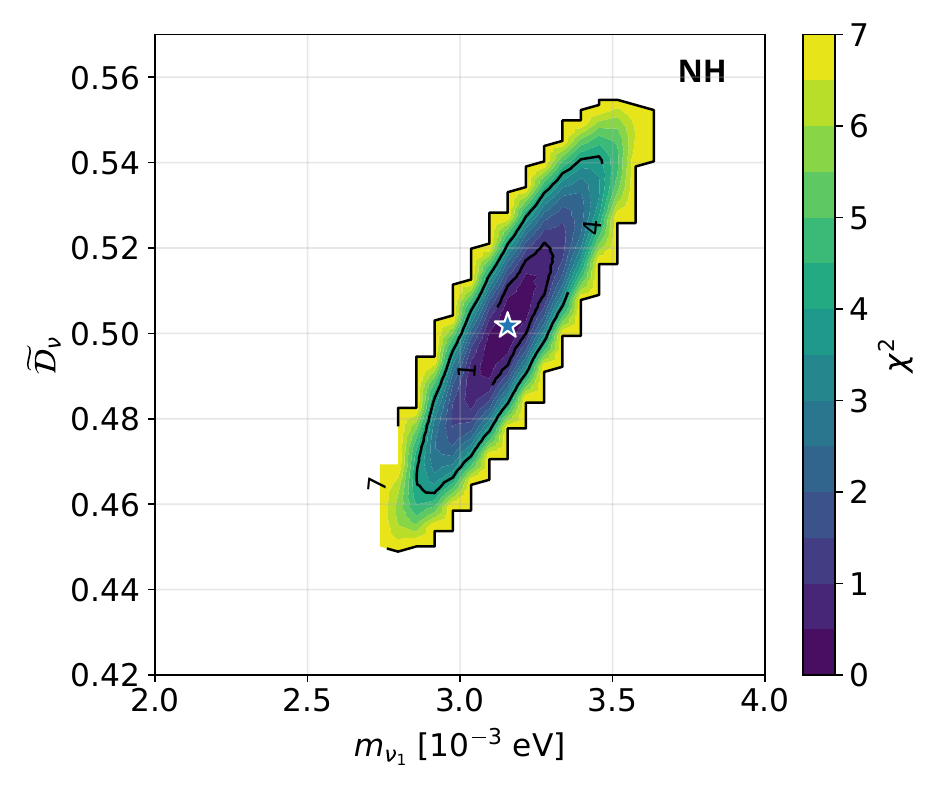} \\
  \includegraphics[width=0.48\linewidth]{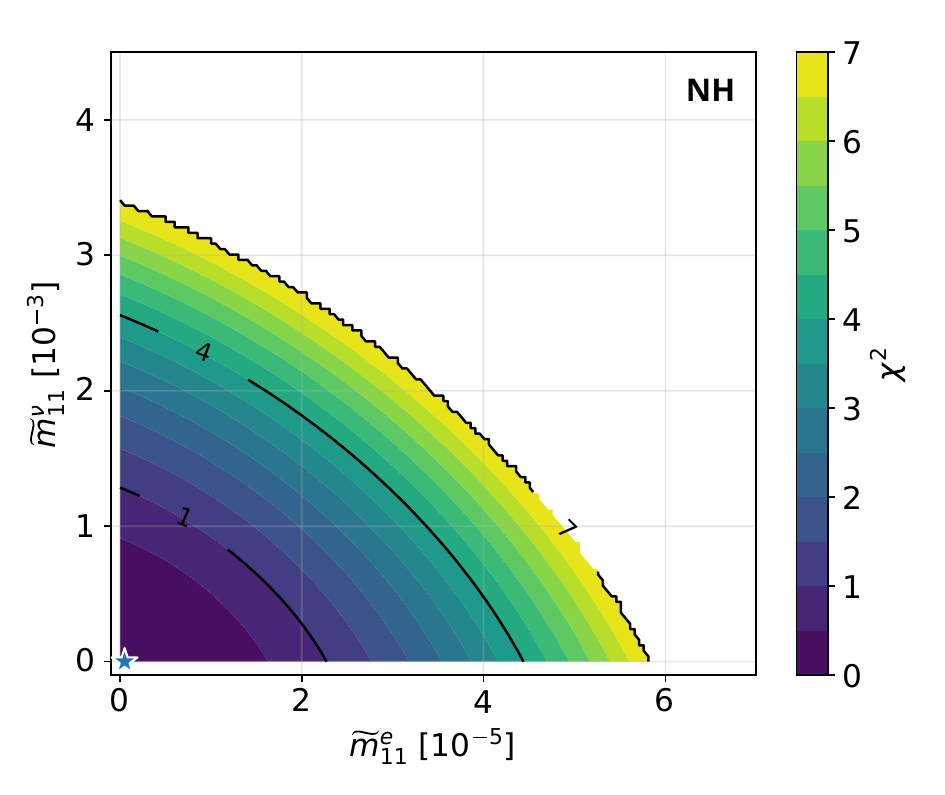} &
  \includegraphics[width=0.48\linewidth]{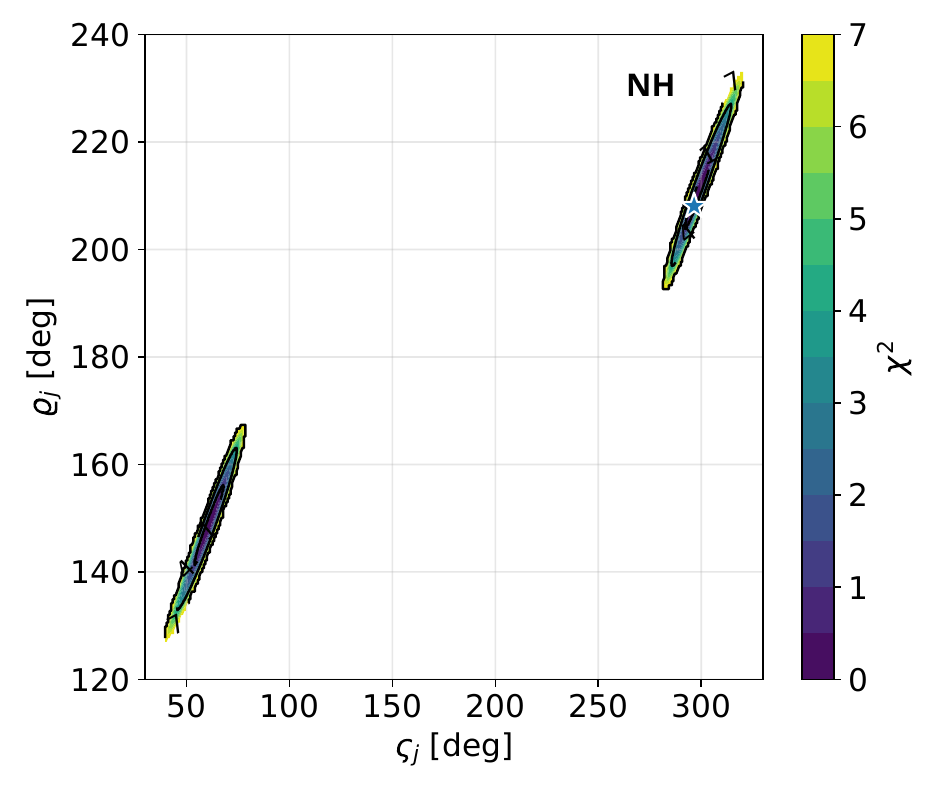}
 \end{tabular}
\caption{\label{Fig:2DProfiles-leptones} The two-dimensional $\chi^2$ slices for 
the normal neutrino mass ordering (NH). The star indicates the BFP, 
while the colored regions represent the variation of $\chi^2$ around the global 
minimum. }
\end{figure} 
To confront the model with the latest neutrino oscillation data, we
construct a $\chi^2$ function using the experimental determinations of
the three leptonic mixing angles reported by 
NuFIT~6.1~\cite{NuFIT6.1,Esteban:2024}. 
The neutrino mass-squared differences are
used as fixed experimental inputs to reconstruct the neutrino mass
spectrum for a given value of the lightest neutrino mass, while they do
not enter directly into the $\chi^2$ function defined below.

By minimizing the $\chi^2$ function over the free parameters of the
model, we determine the Best-Fit Point (BFP). Since the number of free
parameters exceeds the number of observables entering the fit, the
system is statistically underconstrained. Consequently, the numerical
regions discussed below should not be interpreted as conventional
confidence regions of the full parameter space. Instead, we use
one- and two-dimensional $\chi^2$ slices around the BFP to study the
local sensitivity to the model parameters and their correlations. 
In this case the $\chi^{2}$ function has the form:

\begin{equation}\label{ec:def-chi-2-leptones}
 \chi^{2} = 
  \frac{ \left( \sin^{2} \theta_{12}^{\textrm{exp}} 
   -
   \sin^{2} \theta_{12}^{\textrm{th}} \right)^{2} 
   }{ \sigma_{\theta_{12}}^{2} }
  +
  \frac{ \left( \sin^{2} \theta_{13}^{\textrm{exp}} 
   -
   \sin^{2} \theta_{13}^{\textrm{th}} \right)^{2} 
   }{ \sigma_{\theta_{13}}^{2} }
  +
  \frac{ \left( \sin^{2} \theta_{23}^{\textrm{exp}} 
   -
   \sin^{2} \theta_{23}^{\textrm{th}} \right)^{2} 
   }{ \sigma_{\theta_{23}}^{2} }
\end{equation} 
where $\theta_{ij}$ are the lepton flavor mixing angles.
In the above expression, the ``th'' superscript denotes the theoretical 
expressions of leptonic mixing angles, while the ``exp'' superscript denotes 
the experimental data for lepton mixing angles with an uncertainty 
$\sigma_{\theta_{ij}}$. The corresponding experimental best-fit values and their
$1\sigma$ uncertainties are~\cite{NuFIT6.1,Esteban:2024}:
\begin{equation}
 \begin{array}{l}\vspace{2mm}
  \sin^{2} \theta^{\textrm{exp}}_{12} = 0.3088^{+0.0067}_{-0.0066} , \qquad
  \sin^{2} \theta^{\textrm{exp}}_{13} = 
   \left \{ \begin{array}{l}\vspace{2mm}
    0.02248^{+0.00055}_{-0.00059} \\
    0.02262^{+0.00057}_{-0.00056}
   \end{array}    \right.  , \qquad 
  \sin^{2} \theta^{\textrm{exp}}_{23} = 
   \left \{ \begin{array}{l} \vspace{2mm}
    0.470^{+0.017}_{-0.014} \\
    0.550^{+0.013}_{-0.016}
   \end{array}    \right. .
 \end{array}
\end{equation} 
The other neutrino oscillation parameters are~\cite{NuFIT6.1,Esteban:2024}: 
\begin{equation}
 \begin{array}{l}\vspace{2mm}
  \Delta m_{21}^{2}\,[10^{-5}\,\mathrm{eV}^{2}] = 7.537^{+0.094}_{-0.10} , \qquad
  \Delta m_{31}^{2}\,[10^{-3}\,\mathrm{eV}^{2}] = 
   \left \{ \begin{array}{l}\vspace{2mm}
    +2.511^{+0.021}_{-0.020} \\
    -2.483^{+0.020}_{-0.020}
   \end{array}    \right.  , \qquad 
  \delta_{\rm CP}\,[^\circ] = 
   \left \{ \begin{array}{l} \vspace{2mm}
    212^{+26}_{-36}\\
    274^{+22}_{-25}
   \end{array}    \right. .
 \end{array}
\end{equation} 
where $\Delta m_{21}^{2} = m_{\nu_{2}}^{2} - m_{\nu_{1}}^{2}$ and 
$\Delta m_{31}^{2} = m_{\nu_{3}}^{2} - m_{\nu_{1}}^{2}$ are the neutrino 
mass-squared differences, while 
$\delta_{\rm CP}$ is the Dirac CP-violating phase. 
The upper and lower entries correspond to the normal and inverted
neutrino mass orderings (NH and IH), respectively.
The neutrino masses are given by
\begin{equation}\label{ec:78-nu}
 \begin{array}{l}\vspace{2mm}
  m_{\nu_{2}} = \sqrt{ \Delta m_{21}^{2} + m_{\nu_{1}}^{2} }, \quad
  m_{\nu_{3}} = \sqrt{ \Delta m_{31}^{2} + m_{\nu_{1}}^{2} }  \quad
  \textrm{for NH}, \\
  m_{\nu_{1}} = \sqrt{ \Delta m_{13}^{2} + m_{\nu_{3}}^{2} }, \quad 
  m_{\nu_{2}} = \sqrt{ \Delta m_{23}^{2} + m_{\nu_{3}}^{2} }  \quad 
  \textrm{for IH}.
 \end{array}
\end{equation}
In the above expressions, the atmospheric mass-squared difference satisfies the 
relation $\Delta m_{23}^{2}=\Delta m_{21}^{2}+\Delta m_{13}^{2}$.
Moreover, $m_{\nu_1}$ and $m_{\nu_3}$ correspond to the lightest neutrino mass 
eigenvalues for the NH and IH, respectively. 
The expressions in Eq.~(\ref{ec:78-nu}) allow two of the neutrino masses to be 
expressed in terms of the experimentally measured neutrino mass-squared 
differences and the lightest neutrino mass.

\begin{figure}
 \begin{tabular}{cc}
  \includegraphics[width=0.48\linewidth]{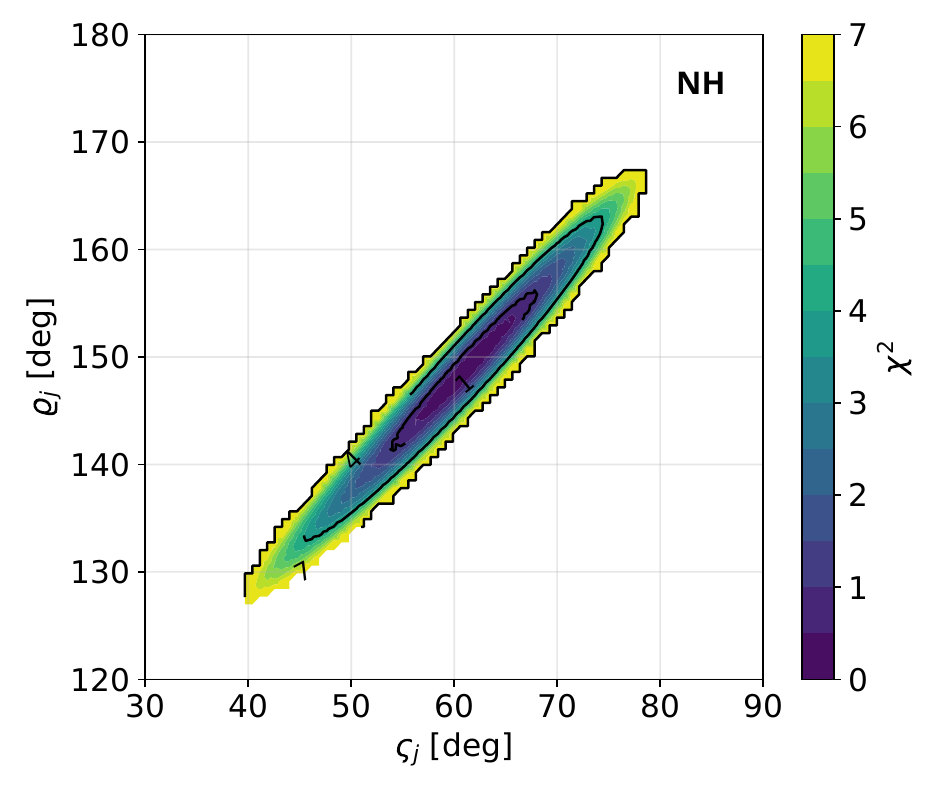} &
  \includegraphics[width=0.48\linewidth]{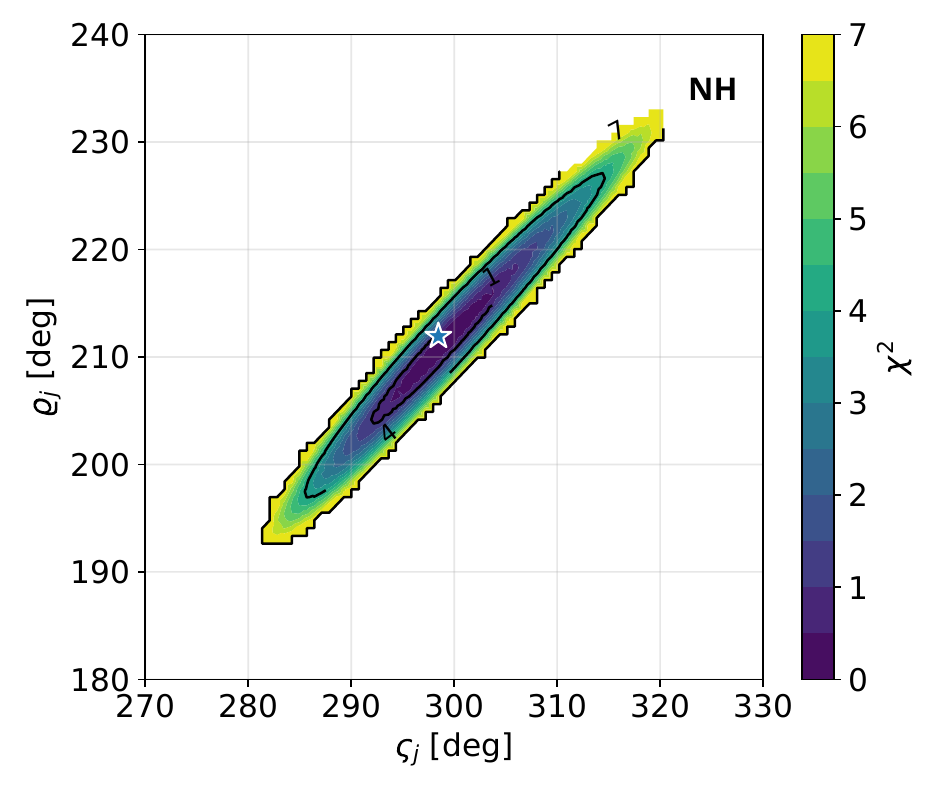}
 \end{tabular}
\caption{\label{Fig:2DProfiles-leptones-1} The two-dimensional $\chi^2$ slices  
 for the normal neutrino mass ordering (NH). The star indicates the 
 BFP, while the colored regions represent the variation of $\chi^2$ around the 
 global minimum.}
\end{figure} 

On the other hand, the charged-lepton masses 
are~\cite{ParticleDataGroup:2024cfk}
\begin{equation}
 \begin{array}{l}\vspace{2mm}
  m_{e}    = 0.5109998928 \pm 0.000000011, \quad
  m_{\mu}  = 105.6583715 \pm 0.0000035, \quad
  m_{\tau} = 1776.82 \pm 0.16.
 \end{array}
\end{equation}
In the numerical analysis that we perform,  the charged lepton masses and the 
neutrino mass-squared differences are not varied, but are instead fixed to their 
experimental central values. The physical motivation for this treatment is that 
charged lepton masses are known with a precision much higher than that required 
for the present analysis, and moreover, their mass ratios are only weakly 
affected by renormalization-group evolution over a broad range of 
energy scales~\cite{ParticleDataGroup:2024cfk,Huang:2021,Fusaoka:1998}.  
Therefore, allowing them to vary within their experimental uncertainties would 
have a negligible impact on the determination of the remaining model parameters.  
Similarly, once the lightest neutrino mass is specified, the 
other two neutrino masses are determined through the measured mass-squared 
differences. Consequently, varying these differences  within their experimental 
uncertainties would mainly introduce additional nuisance directions 
in the numerical analysis, while having only a subleading effect on the 
determination of the flavor parameters considered here.

The lightest neutrino mass determines the absolute neutrino mass scale, whereas 
the measured mass-squared differences determine the remaining two neutrino 
masses. The parameters $\widetilde{m}_{11}^{e}$,
$\widetilde{m}_{11}^{\nu}$, $\widetilde{\mathcal D}_{e}$ and
$\widetilde{\mathcal D}_{\nu}$ control the structure of the corresponding charged 
lepton and neutrino mass matrices, while $\varsigma_{\texttt{j}}$ and 
$\varrho_{\texttt{j}}$ contain the independent phase information relevant to 
lepton mixing. Hence, the $\chi^{2}$ function explored in the numerical analysis 
is effectively a function of these seven 
independent parameters,
$\left\{
m_{\nu_{\rm lightest}},
\widetilde{m}_{11}^{e},
\widetilde{m}_{11}^{\nu},
\widetilde{\mathcal D}_{e},
\widetilde{\mathcal D}_{\nu},
\varsigma_{\texttt{j}},
\varrho_{\texttt{j}} 
\right\}.$
The lightest neutrino mass determines the absolute neutrino-mass scale, 
whereas the measured mass-squared differences determine the remaining two 
neutrino masses.  Since the $\chi^2$ function depends on seven free 
parameters, while only  three experimental observables are included in the fit, 
Eq.~(\ref{ec:def-chi-2-leptones}), the system is statistically 
underconstrained. Therefore, when all seven parameters are simultaneously varied, 
the analysis only allows us to identify the BFP, but not to determine 
statistically meaningful confidence intervals for each individual 
parameter.
For the purpose of exploring the local structure of the numerical fit,
we define
\( \Delta\chi^2(\boldsymbol{\lambda}) = 
\chi^2(\boldsymbol{\lambda})-\chi^2_{\rm min}\),
where $\boldsymbol{\lambda}$ denotes the set of model parameters. In
the two-dimensional scans shown below, the two displayed parameters
are varied simultaneously, while the remaining five parameters are
kept fixed at their BFP values. We use the reference values
\(\Delta\chi^2=2.30,\qquad 6.18,\qquad 11.83\),
which, for a regular likelihood with two fitted parameters, would
conventionally correspond to the $68.27\%$, $95.45\%$, and $99.73\%$
confidence levels, respectively. In the present underconstrained fit,
however, these values are used only as convenient numerical thresholds
to characterize the local $\chi^2$ structure and should not be
interpreted as statistically rigorous confidence levels.

\begin{figure}
 \begin{tabular}{cc}
  \includegraphics[width=0.48\linewidth]{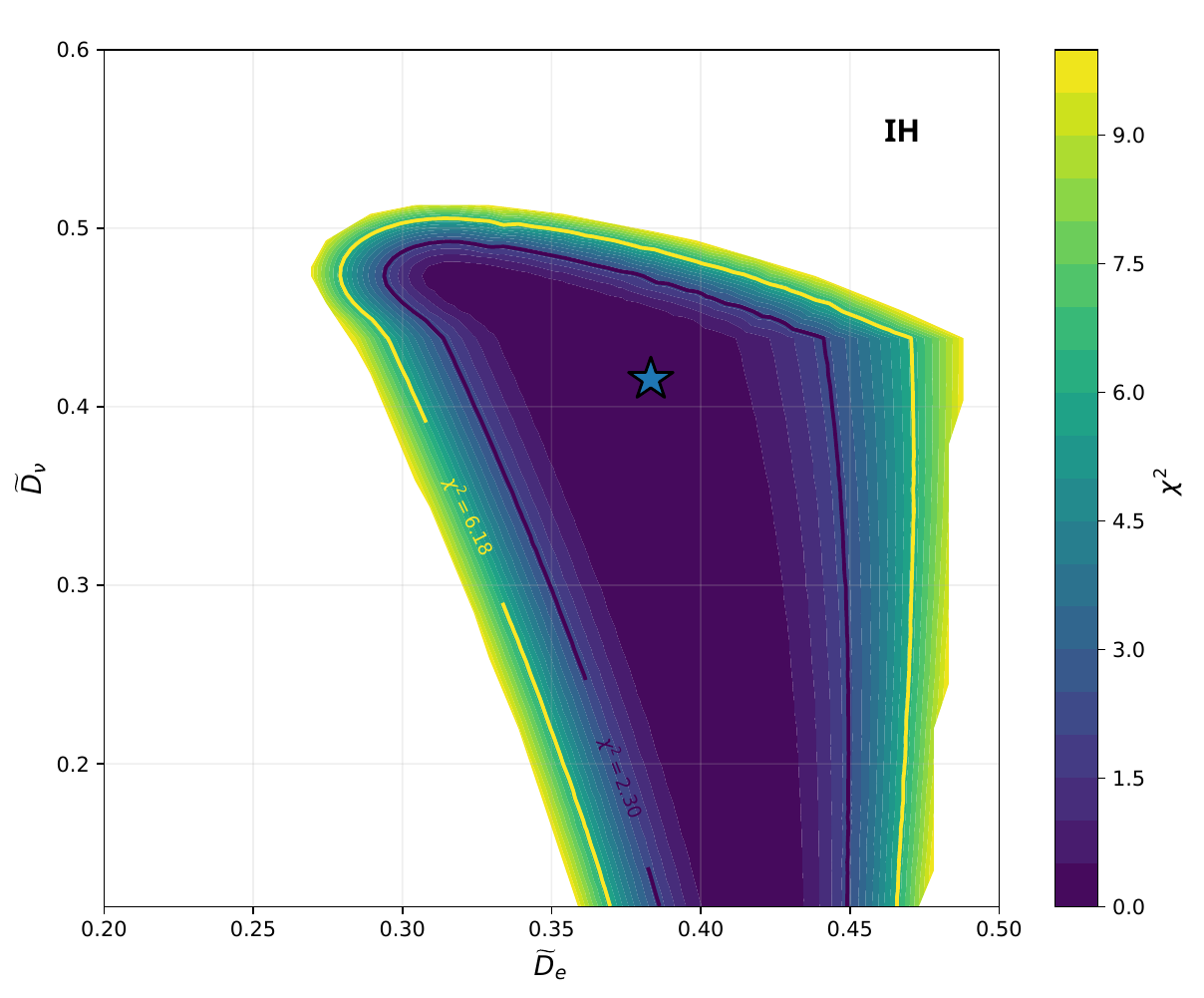} &
  \includegraphics[width=0.48\linewidth]{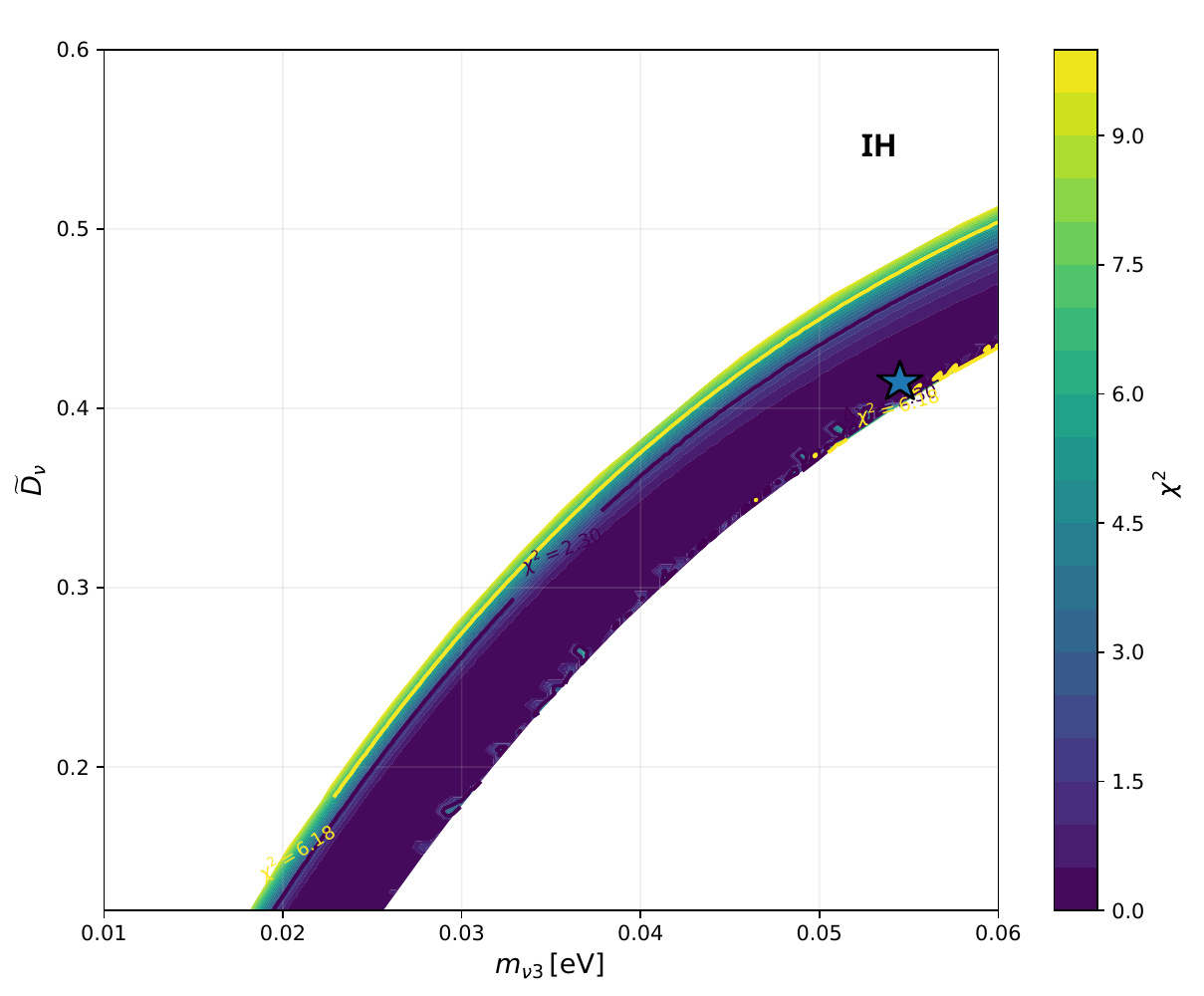} \\
  \includegraphics[width=0.48\linewidth]{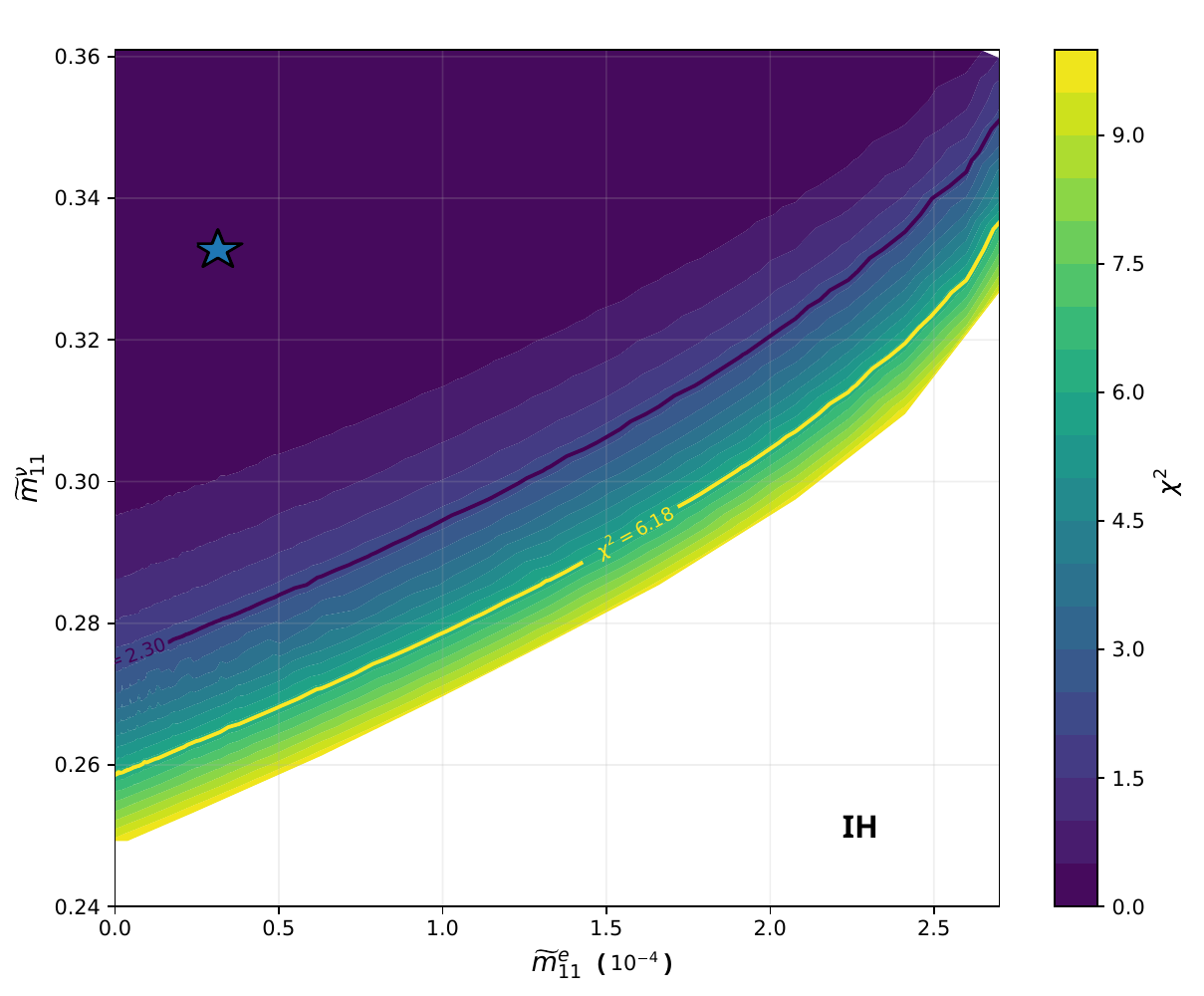} &
  \includegraphics[width=0.48\linewidth]{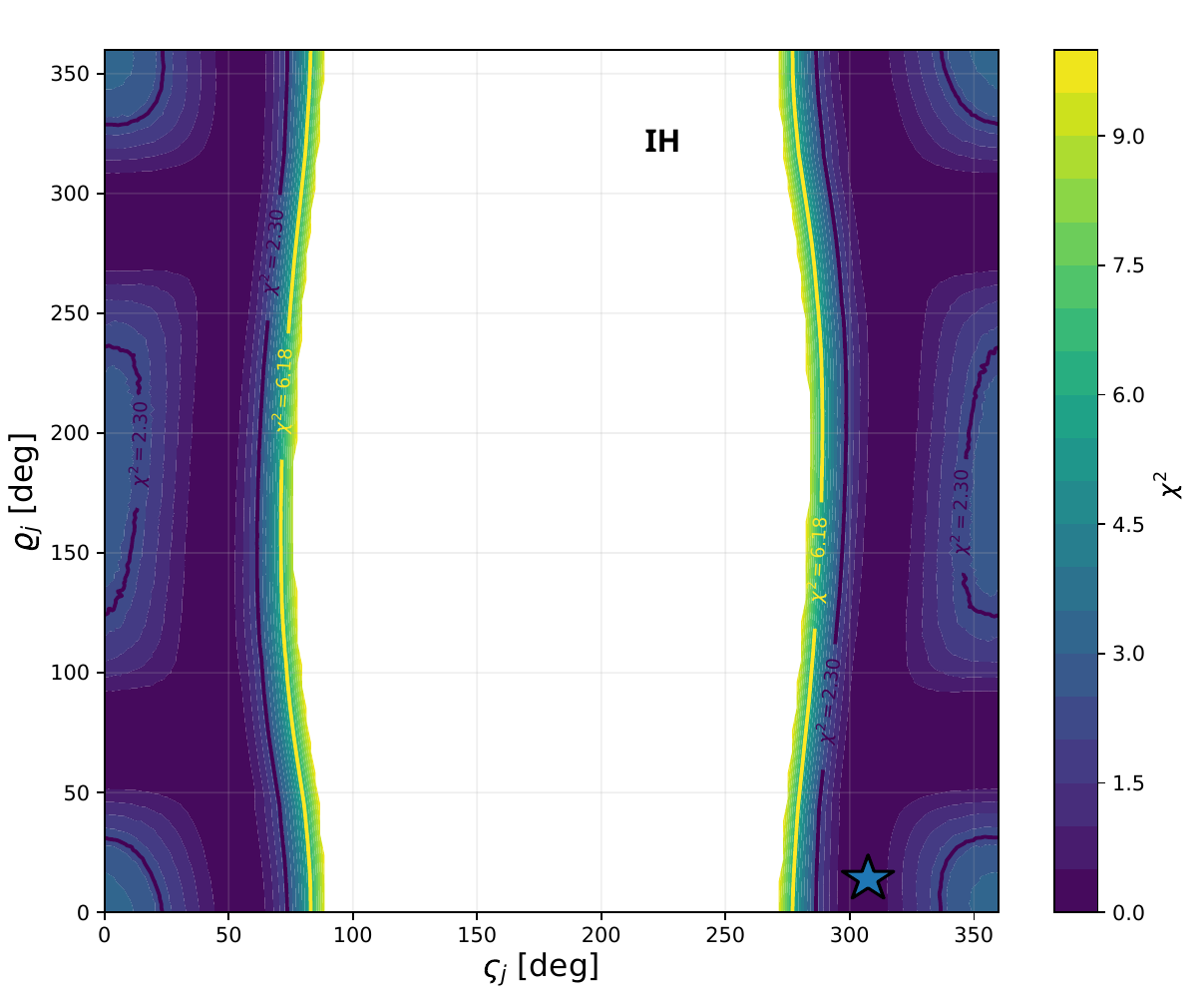}
 \end{tabular}
 \caption{\label{Fig:2DProfiles-leptones-IH} The two-dimensional $\chi^2$ slices 
 for IH. The star indicates the BFP, while the colored regions represent
 the variation of $\chi^2$ around the global minimum. }
\end{figure} 

In the numerical analysis performed assuming the normal neutrino mass
ordering (NH), $m_{\nu_1}<m_{\nu_2}<m_{\nu_3}$.
We obtain that the seven free parameters in the BFP, with a minimum $\chi^{2}$ 
value of $\chi_{\min}^{2}=1.41558\times10^{-4}$, have the values
\begin{equation}
 \begin{array}{l}\vspace{2mm}
  m_{\nu_1} = 3.155\times10^{-3}~\textrm{eV}, \quad
  \widetilde m_{11}^{e} = 2.16\times10^{-7}, \quad
  \widetilde m_{11}^{\nu} = 3.25\times10^{-8}, \\ 
  \widetilde{\mathcal D}_{e} = 0.5192,  \quad
  \widetilde{\mathcal D}_{\nu} = 0.5015,  \quad
  \varsigma_{\texttt{j}} = 298.79^{\circ},  \quad
  \varrho_{\texttt{j}} = 211.05^{\circ}.
 \end{array}
\end{equation}
The corresponding mixing angles are
\begin{equation}
 \sin^2\theta_{12}=0.308771, \qquad
 \sin^2\theta_{13}=0.022486, \qquad
 \sin^2\theta_{23}=0.469933.
\end{equation}

In Figures~\ref{Fig:2DProfiles-leptones} and~\ref{Fig:2DProfiles-leptones-1}, we 
show the two-dimensional $\chi^2$ slices for the free parameters of the model in 
the vicinity of the BFP. These slices illustrate the local correlations among the
free parameters.

The numerical ranges induced by the innermost reference region,
$\Delta\chi^2\leq2.30$, for the seven free parameters of the model
and the three flavor mixing angles are:
\begin{equation}
 \begin{array}{l}\vspace{2mm}
  m_{\nu_1} = \left(3.155^{+0.241}_{-0.179}\right)\times10^{-3}~\textrm{eV}, 
  \quad
  \widetilde m_{11}^{e} \lesssim 3.36\times10^{-5}, \quad 
  \widetilde m_{11}^{\nu} \lesssim 1.92\times10^{-3}, \\
  \widetilde{\mathcal D}_{e} = 0.5192^{+0.3550}_{-0.1812}, \quad
  \widetilde{\mathcal D}_{\nu} = 0.5015^{+0.0292}_{-0.0418}, \quad
  \varsigma_{\texttt{j}} = \left( 298.79^{+11.43}_{-10.22} \right)^{\circ}, 
  \quad 
  \varrho_{\texttt{j}} = \left( 211.05^{+11.87}_{-10.49} \right)^{\circ}.
 \end{array}
\end{equation}

\begin{equation}
 \sin^2\theta_{12} = 0.3088^{+0.0085}_{-0.0102}, \qquad
 \sin^2\theta_{13} = 0.02249^{+0.00085}_{-0.00086}, \qquad
 \sin^2\theta_{23} = 0.4699^{+0.0218}_{-0.0170}.
\end{equation}
These ranges were obtained by projecting the two-dimensional slices
satisfying $\Delta\chi^2\leq2.30$ onto the corresponding parameter
axes. For the phases $\varsigma_{\texttt{j}}$ and
$\varrho_{\texttt{j}}$, the quoted intervals correspond to the
disconnected branch containing the global best-fit point; an additional
disconnected branch is also present within the same reference region,
$\Delta\chi^2\leq2.30$, as shown in
Figure~\ref{Fig:2DProfiles-leptones-1}. 
In the cases of $\widetilde{m}_{11}^{e}$ and 
$\widetilde{m}_{11}^{\nu}$, the allowed regions extend down to the lower 
boundary of the numerical scan, indicating that values compatible with zero 
cannot be excluded. Therefore, these two parameters are more appropriately 
reported in terms of upper bounds rather than as central values with 
two-sided uncertainties.

The numerical ranges of the elements of the real orthogonal matrices
${\bf O}_e$ and ${\bf O}_\nu$ are obtained by propagating the parameter
regions defined by the two-dimensional slices satisfying the reference
condition $\Delta\chi^2\leq2.30$. 
Therefore, the resulting intervals should be
interpreted as induced ranges rather than as independent statistical
uncertainties. In particular, the matrix elements are strongly correlated
through the orthogonality conditions, and hence they do not represent
independent degrees of freedom. The corresponding numerical results are
\begin{equation}
 {\bf O}_e =
 \begin{pmatrix}
  0.997460^{+0.000295}_{-0.000148} &
 -0.071130^{+0.004255}_{-0.001933} &
  0.003865^{+0.001756}_{-0.002341} \\[2mm]
  0.049962^{+0.014875}_{-0.009657} &
  0.737245^{+0.199214}_{-0.126240} &
  0.673776^{+0.116824}_{-0.329042} \\[2mm]
-0.050775^{+0.025344}_{-0.010423} &
-0.671871^{+0.328073}_{-0.116377} &
 0.738926^{+0.199774}_{-0.126618}
 \end{pmatrix}. \label{eq:Oe_1sigma}
\end{equation}
\begin{equation}
 {\bf O}_{\nu} =
 \begin{pmatrix}
  0.834104^{+0.005378}_{-0.005446} &
 -0.542139^{+0.008130}_{-0.006571} &
  0.101765^{+0.008887}_{-0.004843} \\[2mm]
  0.345206^{+0.009454}_{-0.016849} &
  0.656932^{+0.011569}_{-0.020333} &
  0.670279^{+0.027521}_{-0.016579} \\[2mm]
 -0.430237^{+0.013258}_{-0.023094} &
 -0.523953^{+0.011268}_{-0.017951} &
  0.735098^{+0.015423}_{-0.027404}
 \end{pmatrix}. \label{eq:Onu_1sigma}
\end{equation}
Once the numerical ranges associated with the reference condition
$\Delta\chi^2\leq2.30$ have been determined, their impact on the
leptonic mixing matrix can be evaluated. We propagate the corresponding
parameter ranges to the elements of the PMNS matrix. Since
${\bf U}_{\rm PMNS}$ is in general complex, we report separately its
real and imaginary parts. Thus, we have
\begin{equation}\label{eq:UPMNS_complex_1sigma_split-NH}
 {\bf U}_{\rm PMNS} =
 \operatorname{Re}\!\left({\bf U}_{\rm PMNS}\right)
 +
 i\,\operatorname{Im}\!\left({\bf U}_{\rm PMNS}\right),
\end{equation}
where
\begin{equation}\label{eq:Re_UPMNS_1sigma-1}
 \operatorname{Re}\!\left({\bf U}_{\rm PMNS}\right)
 =
 \begin{pmatrix}
  0.82158^{+0.00598}_{-0.00591} &
 -0.54774^{+0.00903}_{-0.00886} &
  0.14961^{+0.00267}_{-0.00302} \\[2mm]
 -0.18438^{+0.14209}_{-0.06452} &
 -0.02973^{+0.19628}_{-0.10701} &
  0.65389^{+0.03711}_{-0.13860} \\[2mm]
  0.38761^{+0.03315}_{-0.02845} &
  0.54278^{+0.02991}_{-0.04137} &
 -0.24741^{+0.14169}_{-0.20721}
\end{pmatrix},
\end{equation}
and
\begin{equation}
\operatorname{Im}\!\left({\bf U}_{\rm PMNS}\right)
=
\begin{pmatrix}
 -0.02638^{+0.00243}_{-0.00174} &
 -0.04249^{+0.00270}_{-0.00155} &
 -0.01009^{+0.01049}_{-0.02027} \\[2mm]
 -0.37214^{+0.03042}_{-0.02087} &
 -0.60602^{+0.04013}_{-0.02285} &
 -0.17829^{+0.13733}_{-0.26887} \\[2mm]
 -0.03984^{+0.16015}_{-0.07208} &
 -0.18818^{+0.25826}_{-0.11076} &
 -0.67597^{+0.12248}_{-0.04119}
\end{pmatrix}. \label{eq:Im_UPMNS_1sigma}
\end{equation}
The moduli of the matrix elements are
\begin{equation}
 \left|{\bf U}_{\rm PMNS}\right| =
 \begin{pmatrix}
  0.82200^{+0.00593}_{-0.00462} &
  0.54939^{+0.00730}_{-0.00892} &
  0.14995^{+0.00280}_{-0.00291} \\[2mm]
  0.41531^{+0.01862}_{-0.06280} &
  0.60675^{+0.03493}_{-0.02961} &
  0.67777^{+0.01523}_{-0.01205} \\[2mm]
  0.38965^{+0.04757}_{-0.01346} &
  0.57448^{+0.02602}_{-0.03029} &
  0.71983^{+0.01119}_{-0.01469}
 \end{pmatrix}.
 \label{eq:PMNS_modulus_1sigma}
\end{equation}

For the inverted neutrino mass ordering (IH),
$m_{\nu_3}<m_{\nu_1}<m_{\nu_2}$.
We obtain that the seven free parameters in the BFP, with a minimum $\chi^{2}$ 
value of  $\chi_{\min}^{2} = 1.5907417\times10^{-4}$, have the values
\begin{equation}
 \begin{array}{l}\vspace{2mm}
  m_{\nu_3} = 0.0549332~\mathrm{eV},\quad
  \widetilde{m}_{11}^{e} = 3.08403\times10^{-5}, \quad
  \widetilde{m}_{11}^{\nu} = 0.332738, \\
  \widetilde{D}_{e} = 0.384517,  \quad
  \widetilde{D}_{\nu} = 0.411413, \quad
  \varsigma_{\texttt{j}}  = 303.859^\circ, \quad
  \varrho_{\texttt{j}} = 9.11931^\circ.
 \end{array}
\end{equation}
The corresponding mixing angles are
\begin{equation}
 \begin{array}{l}\vspace{2mm}
  \sin^2\theta_{12} = 0.3088268,\quad
  \sin^2\theta_{13} = 0.0226219,\quad
  \sin^2\theta_{23} = 0.5501664.
 \end{array}
\end{equation}
In Figure~\ref{Fig:2DProfiles-leptones-IH}, we show the
two-dimensional $\chi^2$ slices for the free parameters of the model
in the vicinity of the BFP. The numerical ranges induced by the
innermost reference region, $\Delta\chi^2\leq2.30$, for the seven free
parameters of the model and the three flavor mixing angles are:
\begin{equation}
 \begin{array}{l}\vspace{2mm}
  m_{\nu_3} = \left(5.493^{+0.507}_{-1.193}\right)\times10^{-2}~\textrm{eV}, 
  \quad
  \widetilde m_{11}^{e} \lesssim 2.5\times10^{-4}, \quad 
  \widetilde m_{11}^{\nu} = 0.332^{+0.027}_{-0.055}, \\
  \widetilde{\mathcal D}_{e} = 0.3845^{+0.0455}_{-0.0744}, \quad
  \widetilde{\mathcal D}_{\nu} = 0.4114^{+0.0708}_{-0.0776}, \quad
  \varsigma_{\texttt{j}} = \left( 303.86^{+8.18}_{-7.42} \right)^{\circ}, \quad 
  \varrho_{\texttt{j}} = \left( 9.12^{+5.60}_{-9.12} \right)^{\circ}.
 \end{array}
\end{equation}
\begin{equation}
 \begin{array}{l}
  \sin^2\theta_{12} = 0.3088^{+0.0066}_{-0.0067}, \quad
  \sin^2\theta_{13} = 0.02262^{+0.00056}_{-0.00057}, \quad
  \sin^2\theta_{23} = 0.5502^{+0.0143}_{-0.0146}.
\end{array}
\end{equation}
These ranges were obtained by projecting the two-dimensional slices
satisfying $\Delta\chi^2\leq2.30$ onto the corresponding parameter
axes. For $\widetilde{m}_{11}^{e}$, the numerical region extends down
to the lower boundary of the scan, indicating that values compatible
with zero cannot be excluded. Therefore, this parameter is more
appropriately reported as an upper bound rather than as a central
value with a two-sided uncertainty. 

The absolute neutrino-mass scale obtained in the flavor analysis can
also be confronted with cosmological constraints. Defining
\(
 \Sigma_{\nu}=m_{\nu_1}+m_{\nu_2}+m_{\nu_3},
\)
the neutrino mass spectra at the best-fit points yield
\( \Sigma_{\nu}^{\rm NH,BFP}\simeq 0.0626~{\rm eV} \) and
\( \Sigma_{\nu}^{\rm IH,BFP}\simeq 0.2038~{\rm eV} \).
For the inverted ordering, the range of the lightest neutrino mass
induced by the reference region $\Delta\chi^2\leq2.30$,
$0.0430~{\rm eV}\lesssim m_{\nu_3}\lesssim0.0600~{\rm eV}$,
corresponds approximately to
\( 0.175~{\rm eV}\lesssim\Sigma_{\nu}^{\rm IH} \lesssim0.216~{\rm eV}.\)
These results should be compared with cosmological constraints with
some caution, since limits on the sum of neutrino masses depend on the
assumed cosmological model. Within the $\Lambda$CDM framework, the
recent DESI DR2 analysis combined with CMB and neutrino-oscillation
data gives the ordering-dependent bounds
$\Sigma_{\nu}<0.101~{\rm eV}$ for normal ordering and
$\Sigma_{\nu}<0.133~{\rm eV}$ for inverted ordering at $95\%$
probability~\cite{DESI:2025Neutrino}. Consequently, while the NH
best-fit region obtained here is compatible with these constraints,
the IH region selected by our flavor analysis is in strong tension
with the corresponding $\Lambda$CDM cosmological bound.

We emphasize, however, that this tension should not be interpreted 
as a model-independent exclusion of the inverted-ordering solution.
Cosmological neutrino-mass bounds are sensitive to assumptions about
the background cosmology. Therefore, we retain the IH solution as a
viable realization at the level of neutrino oscillation and flavor
data, while explicitly noting its tension with standard
$\Lambda$CDM cosmology.

Following the same procedure as
for the normal-ordering case, using the reference condition
$\Delta\chi^2\leq2.30$, we obtain
\begin{equation}
 {\bf O}_e=
 \begin{pmatrix}
  0.997644_{-0.000373}^{+0.001620} &
  -0.0684331_{-0.0051466}^{+0.0302020} &
  0.0048043_{-0.0019149}^{+0.0011889} \\[2mm]
  0.0406399_{-0.0196380}^{+0.0048442} &
  0.6459788_{-0.0568164}^{+0.0326320} &
  0.7622727_{-0.0286910}^{+0.0448186} \\[2mm]
  -0.0552681_{-0.0076462}^{+0.0232249} &
  -0.7602817_{-0.0445256}^{+0.0288000} &
   0.6472381_{-0.0568419}^{+0.0323495}
 \end{pmatrix}.
\end{equation}
\begin{equation}
 {\bf O}_\nu=
 \begin{pmatrix}
  0.7957468_{-0.0172680}^{+0.0117998} &
  -0.5710577_{-0.0214910}^{+0.0135330} &
  0.2016932_{-0.0276733}^{+0.0077675} \\ [2mm]
  0.5660641_{-0.0123978}^{+0.0237243} &
  0.8197016_{-0.0162914}^{+0.0086162} &
  0.0875254_{-0.0138411}^{+0.0106723} \\[2mm]
  -0.2153102_{-0.0090654}^{+0.0238787} &
  -0.0445232_{-0.0157682}^{+0.0110747} &
   0.9755302_{-0.0022041}^{+0.0054007}
 \end{pmatrix}.
\end{equation}
Also we have
\begin{equation}\label{eq:UPMNS_complex_1sigma_split-IH}
 {\bf U}_{\rm PMNS} =
 \operatorname{Re}\!\left({\bf U}_{\rm PMNS}\right)
 +
 i\,\operatorname{Im}\!\left({\bf U}_{\rm PMNS}\right),
\end{equation}
where
\begin{equation}
 \operatorname{Re}({\bf U}_{\rm PMNS})
 =
 \begin{pmatrix}
  0.81844^{+0.00383}_{-0.00530} & 
  -0.54872^{+0.00562}_{-0.00750} &
  0.14997^{+0.00184}_{-0.00137} \\
  0.31090^{+0.06102}_{-0.08390} &
  0.36752^{+0.08943}_{-0.11912} &
  -0.71460^{+0.02608}_{-0.02038} \\
  0.10663^{+0.07465}_{-0.09740} &
  0.31693^{+0.10607}_{-0.14023} &
  0.66156^{+0.01502}_{-0.02504}
 \end{pmatrix}.
\end{equation}
\begin{equation}
 \operatorname{Im}({\bf U}_{\rm PMNS})
 =
 \begin{pmatrix}
  -0.01722^{+0.04163}_{-0.00720} &
  -0.02727^{+0.05922}_{-0.00468} &
  -0.01150^{+0.02963}_{-0.00663} \\
  -0.27771^{+0.65941}_{-0.10399} &
  -0.43435^{+0.94107}_{-0.07237} &
  -0.16450^{+0.42032}_{-0.09132} \\
  -0.38041^{+0.80242}_{-0.04161} &
  -0.52344^{+1.10900}_{-0.06212} &
  0.04467^{+0.17531}_{-0.26464}
 \end{pmatrix}.
\end{equation}
Finally, the moduli of the matrix elements are
\begin{equation}
 \left|{\bf U}_{\mathrm{PMNS}}\right|
 =
 \begin{pmatrix}
  0.8186196_{-0.0054242}^{+0.0053820} &
  0.5494002_{-0.0059597}^{+0.0058543} &
  0.1504057_{-0.0018937}^{+0.0018598} \\[2mm]
  0.4168717_{-0.0840141}^{+0.0587493} &
  0.5689692_{-0.0281937}^{+0.0245705} &
  0.7332944_{-0.0098193}^{+0.0094788} \\[2mm]
  0.3950695_{-0.0725278}^{+0.0732944} &
  0.6119097_{-0.0230531}^{+0.0241320} &
  0.6630667_{-0.0106420}^{+0.0107076}
 \end{pmatrix}.
\end{equation}
%
\section{Some phenomenological aspects}
%
In the previous section, we supplemented an extended left-right symmetric
model with the $S_{3}$ flavor symmetry and analyzed the CKM and PMNS flavor
mixing matrices in the quark and lepton sectors. Having determined these
matrices numerically, we now study their impact on selected observables.
For the phenomenological analysis presented below, we adopt two
simplifying approximations:
\begin{enumerate}
 \item The mixing between the left- and right-handed charged gauge bosons
  is neglected, $\zeta\simeq0$, consistently with the stringent
  phenomenological constraints on this parameter~\cite{Li:2020flq}.
 \item The light-heavy neutrino mixing is neglected, $\theta\simeq0$. 
  This approximation is appropriate for the regime in
  which the heavy-neutrino scale is sufficiently above the electroweak
  scale and is consistent with existing constraints on heavy-light
  neutrino mixing~\cite{Fernandez-Martinez:2016lgt}.
\end{enumerate}
These assumptions allow us to isolate the purely left-handed and
right-handed contributions considered below. Mixed-current
contributions proportional to $\zeta$, $\theta$, or their products are
therefore outside the scope of the present analysis.
With these assumptions in place, we now determine the allowed ranges of
the observables considered below.
%
\subsection{Neutrinoless double beta decay}
%
It should be kept in mind that the IH region displayed here is obtained
from the flavor and oscillation analysis alone. Its comparatively large
absolute neutrino-mass scale is in tension with current cosmological
bounds within $\Lambda$CDM, as discussed above. Therefore, the
$0\nu\beta\beta$ predictions for IH should be understood conditionally
on either a relaxation of the standard cosmological assumptions or
future changes in the cosmological neutrino-mass constraints.
\begin{figure}
 \begin{tabular}{c}
  \includegraphics[width=0.7\linewidth]{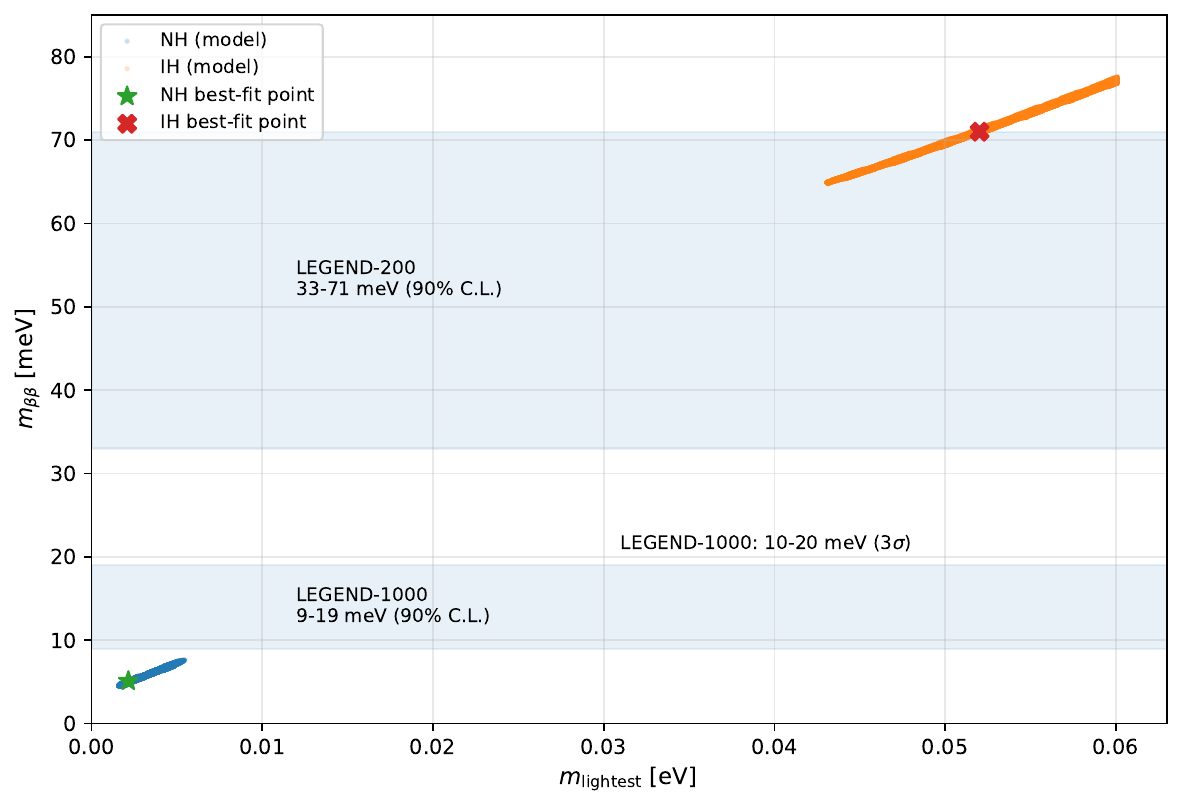} 
 \end{tabular}
 \caption{ Effective Majorana neutrino mass $m_{\beta\beta}$ as a function of the
  lightest neutrino mass for NH and IH. 
  The points correspond to the parameter configurations satisfying
  $\chi^{2}<1$, while the markers indicate the corresponding BFP. 
  The shaded regions illustrate the projected sensitivity to
  $m_{\beta\beta}$ of the LEGEND experimental program. The conversion
  between the neutrinoless double-beta decay half-life and
  $m_{\beta\beta}$ depends on the adopted nuclear matrix elements.
 }\label{fig:mbb_lightest}
\end{figure}

Neutrinoless double-beta decay is a promising process for probing whether
neutrinos are Majorana fermions~\cite{Vergados:2002pv,Rodejohann:2011mu,
Vergados:2012xy,Pas:2015eia}. In general, the decay width for
$0\nu\beta\beta$ can be written as
\begin{equation}
 \Gamma^{0\nu}=\sum_{j} G_{j}\left(Q,Z\right) 
 \vert \mathcal{M}_{j}\eta_{j}\vert^{2}.
\end{equation}
In this expression, $\mathcal{M}_{j}$ denotes the nuclear matrix element
associated with the $j$-th mechanism, $\eta_{j}$ is the corresponding
dimensionless particle-physics parameter, and $G_j(Q,Z)$ is a phase-space
factor that may also depend on the underlying mechanism
~\cite{Vergados:2002pv,Rodejohann:2011mu,Vergados:2012xy,Pas:2015eia}.

Under the approximations $\zeta\simeq0$ and $\theta\simeq0$, one has
$W_1\simeq W_L$, $W_2\simeq W_R$, $\mathbf{S}\simeq\mathbf{T}\simeq0$,
$\mathbf{U}\simeq\mathbf{U}_{\nu L}$, and
$\mathbf{V}\simeq\mathbf{U}_{\nu R}$. Therefore, the leptonic charged
current in Eq.~(\ref{CCL}) reduces to
\begin{equation}
 \mathcal{L}_{CC}
 \simeq
 \frac{g}{\sqrt{2}}
 \left[
  \bar{\tilde e}_{L}\gamma_{\mu}
  \mathbf{U}_{\rm PMNS}^{L}\tilde{\nu}_{L} \, W_{1}^{-\mu}
  +
  \bar{\tilde e}_{R}\gamma_{\mu}
  \mathbf{U}_{\rm PMNS}^{R}\tilde{\nu}_{R}\,
  W_{2}^{-\mu}
 \right]
 + h.c.
 \label{CCL-2}
\end{equation}
In the MLRSM, the additional scalar and fermionic degrees of freedom give
rise to several contributions to the decay amplitude
~\cite{Huang:2013kma,BhupalDev:2014qbx,Borah:2015ufa}.
Under the assumptions adopted in our analysis, some of these contributions
are neglected. The relevant contributions are then those involving the
left- and right-handed PMNS mixing matrices, namely,
\begin{equation}
 \Gamma^{0\nu} = 
 G^{0\nu}_{01} \left[
  \vert \mathcal{M}^{0\nu}_{\nu} \eta_{\nu} \vert^{2} 
  + 
  \vert \mathcal{M}^{0\nu}_{N} \eta^{R}_{\nu R} \vert^{2} 
 \right]
\end{equation}	
with
\begin{equation}\label{ec:104}
 \eta_{\nu} = \frac{1}{m_{e}} 
 \sum_{i=1}^{3} \left( \mathbf{U}^{L}_{PMNS} \right)^{2}_{e i} m_{i};
\end{equation}
\begin{equation}\label{ec:105}
 \eta^{R}_{\nu R} = m_{p}
 \left( \frac{ m_{W_{L}} }{ m_{W_{R}} } \right)^{4} 
 \sum_{i=1}^{3} 
 \frac{ \left( \mathbf{U}^{R}_{PMNS} \right)^{2}_{e i} }{ M_{\nu_{i R}} },
\end{equation}
where $m_{e}$ and $m_{p}$ are the electron and proton masses, respectively.
From Eq.~(\ref{ec:104}), we obtain
\begin{equation}
 |\eta_{\nu}| = \frac{m_{\beta \beta}}{m_{e}} 
\end{equation}
where $m_{\beta\beta}$ is the effective Majorana mass, defined as
\begin{equation}
 m_{\beta\beta}
 =
 \left|
 \sum_{i=1}^{3}
 \left(U_{\rm PMNS}^{L}\right)_{ei}^{2}m_{\nu_i}
 \right|.
\end{equation}
The effective Majorana mass $m_{\beta\beta}$ connects the low-energy
neutrino parameters with the $0\nu\beta\beta$ decay rate when the process
is dominated by the exchange of light Majorana neutrinos
~\cite{Agostini2023}.
For the parameter points satisfying $\chi^2<1$, our numerical analysis
predicts for NH
\( 4.63~{\rm meV} \lesssim m_{\beta\beta}^{\rm NH} \lesssim 7.74~{\rm meV}\), 
with a best-fit value \(m_{\beta\beta}^{\rm NH,BFP} \simeq 5.10~{\rm meV}\).
For IH, a significantly larger and rather narrow interval is obtained,
\( 64.85~{\rm meV} \lesssim m_{\beta\beta}^{\rm IH} \lesssim 77.52~{\rm meV} \),
with
\( m_{\beta\beta}^{\rm IH,BFP} \simeq 71.00~{\rm meV} \).
Therefore, within the parameter region selected by our numerical
analysis, the two neutrino mass orderings lead to clearly separated
predictions for the effective Majorana mass. Notice that the quoted
intervals correspond to the envelope of the accepted points satisfying
$\chi^{2}<1$ and should not be interpreted as conventional statistical
$1\sigma$ confidence intervals.

The predicted regions can be confronted with the current experimental
limits on neutrinoless double-beta decay. The KamLAND-Zen Collaboration,
using the complete KamLAND-Zen dataset, obtained a lower limit on the
$^{136}{\rm Xe}$ neutrinoless double-beta decay half-life of
\(T_{1/2}^{0\nu}>3.8\times10^{26}~{\rm yr} \quad (90\%~{\rm C.L.})\),
which, under the standard light-Majorana-neutrino exchange mechanism,
corresponds to an upper bound on the effective Majorana mass in the
range \(m_{\beta\beta}<28-122~{\rm meV}\).
The range reflects the theoretical uncertainty associated with the
nuclear matrix elements (NMEs) entering the conversion between
$T_{1/2}^{0\nu}$ and $m_{\beta\beta}$~\cite{KamLANDZen2025}.
The first results from LEGEND-200, based on $^{76}{\rm Ge}$, yield
\(T_{1/2}^{0\nu}>1.9\times10^{26}~{\rm yr}\quad (90\%~{\rm C.L.})\),
corresponding to
\(m_{\beta\beta}<75-200~{\rm meV}\), depending on the adopted nuclear 
matrix element~\cite{LEGEND200_2026}.

Consequently, the NH region predicted by the present model lies well
below the sensitivity of current neutrinoless double-beta decay
searches. In contrast, the IH prediction,
\(m_{\beta\beta}^{\rm IH}\simeq 65-78~{\rm meV}\),
lies in the mass range currently being approached by these
experiments. Because the translation from a half-life limit into a
constraint on $m_{\beta\beta}$ depends significantly on the nuclear
matrix elements, the present experimental results do not provide a
nuclear-matrix-element-independent exclusion of the IH region
predicted by the model.
This situation becomes particularly interesting for the next-generation 
neutrinoless double-beta decay experiments.
LEGEND-1000 is designed to probe half-lives at the
$10^{28}$~yr scale, corresponding to a sensitivity to the effective
Majorana mass at the level of
\(m_{\beta\beta}\sim \mathcal{O}(10)~{\rm meV}\),
depending on the nuclear matrix elements employed in the conversion
from the half-life sensitivity~\cite{LEGEND1000}. 
Therefore, the IH region predicted by the present model, 
$m_{\beta\beta}^{\rm IH}\simeq65-78~{\rm meV}$, lies well within the
mass scale expected to be explored by next-generation experiments,
under the assumption that light Majorana-neutrino exchange is the
dominant mechanism responsible for $0\nu\beta\beta$ decay.
On the other hand, the NH prediction,
\(m_{\beta\beta}^{\rm NH}\simeq4.6-7.7~{\rm meV}\),
lies below the nominal projected sensitivity of LEGEND-1000 and would
therefore require experiments capable of probing the few-meV region.

Figure~\ref{fig:mbb_lightest} summarizes these results by showing the
predicted effective Majorana mass as a function of the lightest
neutrino mass for both mass orderings, together with the experimental
sensitivity ranges. The strong separation between the NH and IH
regions constitutes a distinctive phenomenological feature of the
present flavored left-right symmetric scenario and provides a direct
connection between the model predictions and forthcoming searches for
neutrinoless double-beta decay.
\begin{figure}
 \begin{tabular}{cc}
  \includegraphics[width=0.48\linewidth]{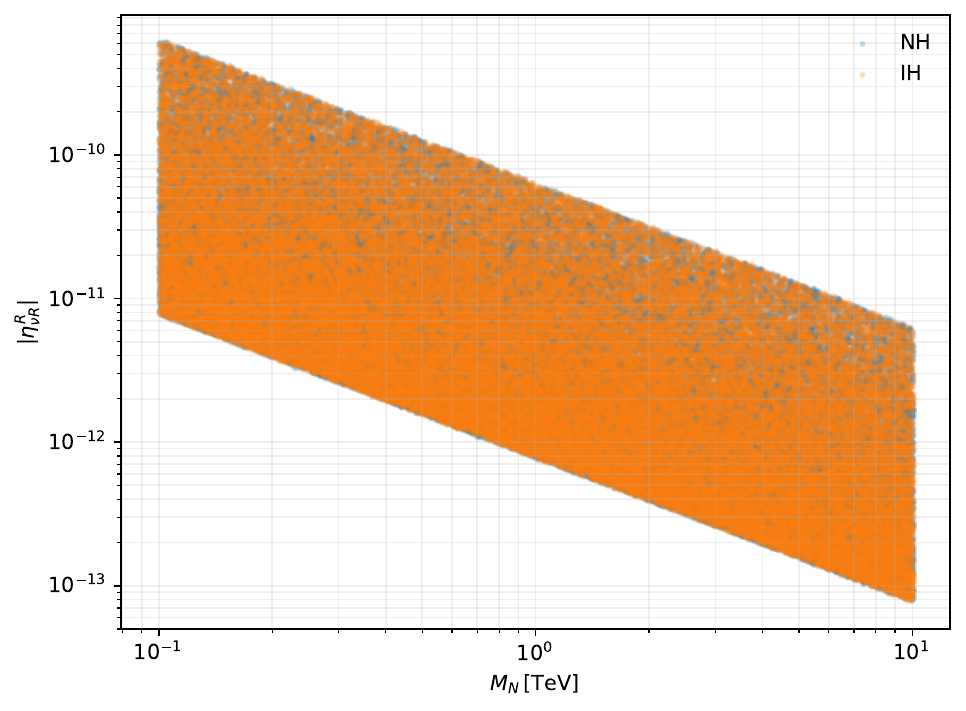} &
  \includegraphics[width=0.48\linewidth]{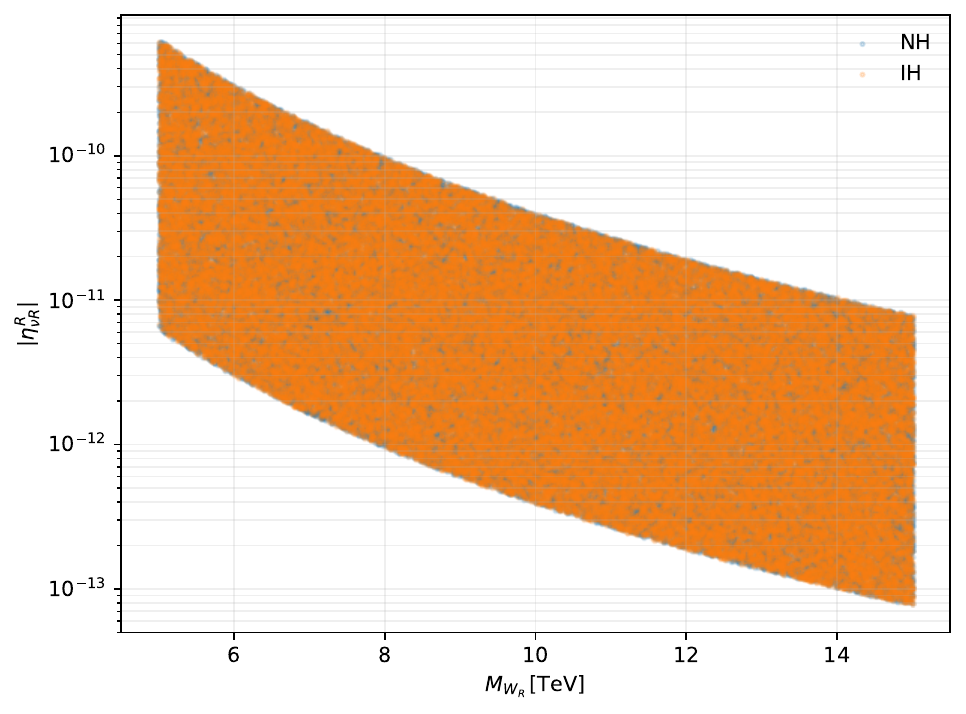} 
 \end{tabular}
\caption{Right-handed contribution to neutrinoless double-beta decay,
 $|\eta_{\nu R}^{R}|$, as a function of the common heavy-neutrino mass scale
 $M_N=M_{\nu_1^R}=M_{\nu_2^R}$ (left panel) and the right-handed charged
 gauge boson mass $M_{W_R}$ (right panel), for the normal (NH) and 
 inverted (IH) neutrino mass orderings. The points correspond to parameter
 configurations satisfying $\chi^2<1$ in the lepton-sector analysis, while
 the remaining heavy-sector parameters are varied within the ranges adopted
 in the numerical scan. \label{fig:eta_RR}
}
\end{figure} 

In this model,  the right-handed PMNS matrix is completely determined
by the charged lepton sector, since the right-handed Majorana neutrino mass
matrix is diagonal. Thus, one has 
${\bf U}_{\rm PMNS}^{R} = {\bf O}_e^{\top} {\bf P}_e {\bf R}_{\pi/4}$, 
where ${\bf O}_e$ is the real orthogonal matrix associated with the charged-lepton
sector, ${\bf P}_e$ contains the corresponding phase factors, and 
${\bf R}_{\pi/4}$ is the maximal rotation defined in Eq.~(\ref{ec:rotaciones}). 
Writing ${\bf P}_e={\rm diag}(p_1,p_2,p_3)$, the elements of the first row of
${\bf U}_{\rm PMNS}^{R}$, which are the ones entering Eq.~(\ref{ec:105}), are
\begin{equation}
 \begin{aligned}
  \left({\bf U}_{\rm PMNS}^{R}\right)_{e1} & 
  =
  \frac{1}{\sqrt{2}} \left(O_{11}p_1+O_{21}p_2\right),\\
  \left({\bf U}_{\rm PMNS}^{R}\right)_{e2} & 
  = \frac{1}{\sqrt{2}} \left(-O_{11}p_1+O_{21}p_2\right),\\
  \left({\bf U}_{\rm PMNS}^{R}\right)_{e3} & =
  O_{31} p_3.
 \end{aligned}
\end{equation}
Therefore, the right-handed contribution to neutrinoless double-beta decay
can be expressed as
\begin{equation}\label{ec:109}
 \eta_{\nu R}^{R} =
 m_p
 \left( \frac{ m_{W_L} }{ m_{W_R} } \right)^4
 \left[
  \frac{(O_{11}p_1+O_{21}p_2)^2}{2M_{\nu_{1R}}}
  +
  \frac{(-O_{11}p_1+O_{21}p_2)^2}{2M_{\nu_{2R}}}
  +
  \frac{O_{31}^2p_3^2}{M_{\nu_{3R}}}
 \right].
\end{equation}
This form makes explicit that the contribution $\eta_{\nu R}^{R}$ depends
simultaneously on the right-handed gauge boson mass scale, the heavy neutrino
mass spectrum, and the flavor structure predicted by the charged lepton
sector. In particular, since eq.~(\ref{ec:105}) involves
$\left({\bf U}_{\rm PMNS}^{R}\right)_{ei}^{2}$ rather than
$\left|\left({\bf U}_{\rm PMNS}^{R}\right)_{ei}\right|^{2}$, the phases 
contained
in ${\bf P}_e$ may lead to constructive or destructive interference among the
different heavy neutrino contributions. The corresponding phase matrices are
those defined in Eq.~(\ref{eq:Uq-20}), namely
${\bf P}_e={\rm diag}(e^{i\alpha_e/2},e^{i\beta_e/2},e^{i\gamma_e/2})$
for PLRT and
${\bf P}_e={\rm diag}(1,1,e^{i\varphi_{ec}})$
for MLRT.
\begin{figure}
 \begin{tabular}{c}
  \includegraphics[width=0.65\linewidth]{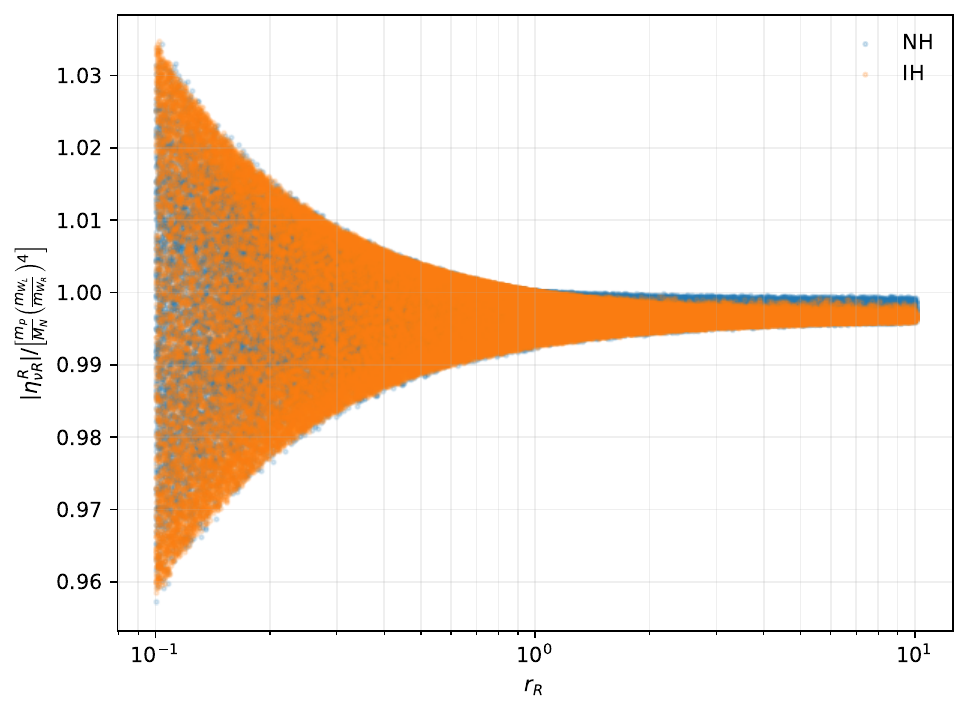} 
 \end{tabular}
 \caption{ Flavor-dependent correction factor
  $\mathcal{F}_{R}=|\eta_{\nu R}^{R}|/
  \left[\frac{m_p}{M_N}\left(\frac{m_{W_L}}{m_{W_R}}\right)^4\right]$
  as a function of the heavy-neutrino mass ratio $r_R=M_{\nu_3^R}/M_N$ 
  for the normal (NH) and inverted (IH) neutrino mass orderings. 
  The proximity of $\mathcal{F}_{R}$ to unity shows that the dependence on 
  $r_R$, $O_{31}$, and $\varphi_{ec}$ provides only a subleading correction to 
  the dominant heavy mass scaling. \label{fig:flavor_factor}
 }
\end{figure} 
Also, the right-handed Majorana neutrino mass matrix has the $S_3$-induced 
structure
${\bf M}_R={\rm diag}(a_R,a_R,b_R)$, with $a_R=y_1^R v_R$ and $b_R=y_2^R v_R$. 
Therefore, taking the physical masses to be positive, 
$M_{\nu_{1R}} = M_{\nu_{2R}} \equiv M_N=|a_R|$ and
$M_{\nu_{3R}} = |b_R|$. Defining $r_R \equiv M_{\nu_{3R}}/M_N=|b_R/a_R|$, 
the twofold degeneracy of the first two right-handed neutrinos considerably 
simplifies Eq.~(\ref{ec:109}), since the interference terms between the 
first two contributions cancel. As a result,
\begin{equation}\label{eq:etaRR-degenerate}
 \eta_{\nu R}^{R}
 =
 \frac{m_p}{M_N}
 \left(\frac{m_{W_L}}{m_{W_R}}\right)^4
 \left[
  O_{11}^{2}p_1^{2}
  +
  O_{21}^{2}p_2^{2}
  +
  \frac{O_{31}^{2}p_3^{2}}{r_R}
 \right].
\end{equation}
Thus, the right-handed contribution is controlled by the overall heavy neutrino
mass scale $M_N$, the mass ratio $r_R$, the right-handed gauge boson mass,
and the flavor structure encoded in ${\bf O}_e$ and ${\bf P}_e$. 
In particular, for the MLRT case, where 
${\bf P}_e={\rm diag}(1,1,e^{i\varphi_{ec}})$, this expression can
be rewritten using the orthogonality of $O_e$ as
\begin{equation}\label{eq:etaRR-MLRT}
 \eta_{\nu R}^{R,\rm MLRT}
 =
 \frac{m_p}{M_N}
 \left(\frac{m_{W_L}}{m_{W_R}}\right)^4
 \left[
  1 - O_{31}^{2}
  +
  \frac{O_{31}^{2}}{r_R}e^{2i\varphi_{ec}}
 \right],
\end{equation}
making explicit the interplay between the heavy neutrino mass hierarchy and
the phase structure of the charged lepton sector.

Figures~\ref{fig:eta_RR} and \ref{fig:flavor_factor} summarize the behavior
of the right-handed contribution to neutrinoless double-beta decay in the
parameter region compatible with the lepton-sector fit. The left panel of
Figure~\ref{fig:eta_RR} shows $|\eta_{\nu R}^{R}|$ as a function of the common
heavy-neutrino mass scale $M_N=M_{\nu_1^R}=M_{\nu_2^R}$, whereas the right panel 
displays its dependence on the right-handed charged gauge-boson mass $M_{W_R}$. 
As follows from Eq.~(\ref{eq:etaRR-MLRT}), the dominant scaling is approximately
\( |\eta_{\nu R}^{R}| \propto M_N^{-1}M_{W_R}^{-4}\).
Accordingly, the left panel exhibits the expected inverse dependence on
$M_N$, while the right panel shows the much stronger suppression associated
with increasing $M_{W_R}$. The vertical spread in each panel originates from
the simultaneous variation of the remaining heavy-sector parameters, while
the normal and inverted light-neutrino mass orderings largely overlap. This
indicates that the magnitude of the right-handed contribution is controlled
mainly by the heavy-neutrino and right-handed gauge-boson mass scales, with
only a weak residual dependence on the light-neutrino ordering.
The latter feature can be understood more clearly from
Figure~\ref{fig:flavor_factor}, where we isolate the flavor-dependent part by
defining
\begin{equation}
 \mathcal{F}_{R} \equiv
 \frac{|\eta_{\nu R}^{R}|}
 {\displaystyle
 \frac{m_p}{M_N}
 \left(\frac{m_{W_L}}{m_{W_R}}\right)^4}.
\end{equation}
For the MLRT framework, Eq.~(\ref{eq:etaRR-MLRT}) gives
\begin{equation}
 \mathcal{F}_{R}
 =
 \left| 1-O_{31}^{2} + \frac{O_{31}^{2}}{r_R}e^{2i\varphi_{ec}} \right|,
\end{equation}
so that this quantity contains the dependence on the heavy neutrino mass
ratio $r_R$, the charged lepton mixing parameter $O_{31}$, and the phase
$\varphi_{ec}$. As shown in Figure~\ref{fig:flavor_factor}, the numerical
points remain close to $\mathcal{F}_{R}=1$ over most of the scanned region.
This behavior is a consequence of the small value of $O_{31}^{2}$ in the
parameter space selected by the lepton sector fit. Therefore, the
flavor-dependent correction is generally subleading, and the right-handed
contribution can be approximated to a good accuracy by
\begin{equation}
|\eta_{\nu R}^{R}|
\simeq
\frac{m_p}{M_N}
\left(\frac{m_{W_L}}{m_{W_R}}\right)^4.
\end{equation}
The nearly coincident NH and IH regions observed in both figures are thus a
natural consequence of the dominant dependence on the heavy right-handed
sector rather than on the ordering of the light neutrino masses.
%
\subsection{Muon anomalous magnetic moment and lepton-flavor violation}
%
\begin{figure}
 \begin{tabular}{c}
  \includegraphics[width=0.65\linewidth]{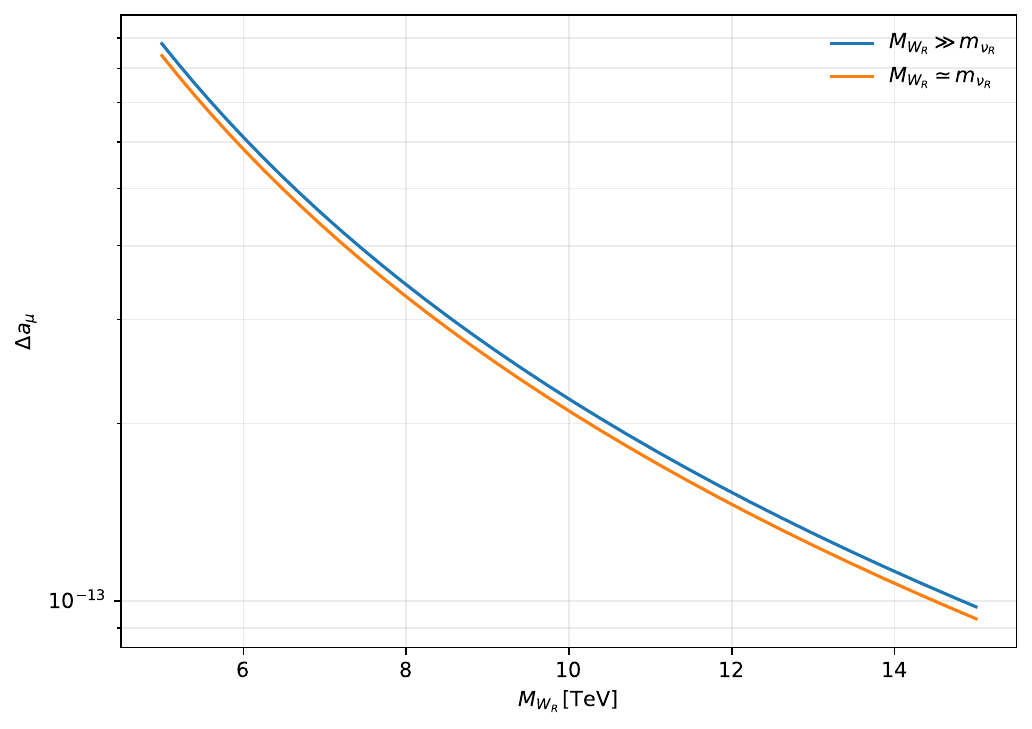} 
 \end{tabular}
 \caption{Contribution of the right-handed neutrinos and the $W_R$ 
  gauge boson to the muon anomalous magnetic moment, 
  $\Delta a_\mu(\nu_R,W_R)$, as a function of the right-handed 
  charged gauge-boson mass $M_{W_R}$. The two curves
  correspond to the limiting regimes $M_{W_R}\gg m_{\nu_R}$ and
  $M_{W_R}\simeq m_{\nu_R}$ appearing in Eq.~(\ref{eq:115}). Parity 
  symmetry, $g_R=g_L$, and the unitarity relation
  $\sum_i |(U_{\rm PMNS}^{R})_{\mu i}|^2=1$ have been used. Both 
  regimes show the characteristic behavior $\Delta a_\mu\propto 
  M_{W_R}^{-2}$ and differ only slightly in their overall 
  normalization. }\label{fig:amu_WR}
\end{figure} 
In this section, we examine the impact of lepton mixing on the muon
anomalous magnetic moment, $(g-2)_{\mu}$ (for a recent review,
see~\cite{Aliberti:2025beg}), and on the lepton-flavor-violating decay
$\mu\rightarrow e\gamma$. Fermilab reported a measurement of the muon
anomalous magnetic moment~\cite{Muong-2:2021ojo} which, when combined
with the BNL data~\cite{Muong-2:2006rrc}, showed a discrepancy with the
SM prediction. More recently, this tension has disappeared
~\cite{Muong-2:2025xyk}, and the Muon $g-2$ Collaboration has reported
the most precise measurement of this quantity to date.
With this motivation, we briefly examine the role of left-right
interactions in these observables. In particular, right-handed charged
currents can contribute to $(g-2)_{\mu}$ and to the
$\mu\rightarrow e\gamma$ decay. The relevant contributions involve the
$W_R$ gauge boson and right-handed neutrinos $\nu_R$. Approximate
expressions were obtained in Ref.~\cite{Lindner:2016bgg} for two
limiting regimes:
\begin{equation}\label{eq:115}
 \Delta a_{\mu} \left(\nu_{R}, W_{R} \right) = 
 \left\{ \begin{array}{lcc}
  2.2 \times 10^{-11} \left( \frac{ g_{R} }{ g_{L} } \right)^{2} 
  \left( \frac{1 \textrm{TeV} }{ M_{W_{R}} } \right)^{2} 
  \displaystyle\sum^{3}_{i=1} 
  \vt \left( {\bf U}^{R}_{\textrm{PMNS}} \right)_{\mu i} \vt^{2}, &  
  \quad 
  \textrm{if}\quad & \quad  M_{W_{R}}\gg m_{\nu_{R}} \\
  2.1 \times 10^{-11} \left( \frac{ g_{R} }{ g_{L} } \right)^{2} 
  \left(\frac{1 \textrm{TeV}}{M_{W_{R}}}\right)^{2} \displaystyle\sum^{3}_{i=1} 
  \vt \left({\bf U}^{R}_{\textrm{PMNS}}\right)_{\mu i} \vt^{2}, &  
  \quad \textrm{if}\quad & \quad  M_{W_{R}}\approx m_{\nu_{R}}.
 \end{array}
 \right.	
\end{equation}
The observable $\Delta a_{\mu}$ depends on the $W_R$ gauge-boson mass
and on the right-handed PMNS mixing matrix, with the sum running over
the three right-handed neutrino species, $i=1,2,3$.
In the present model, the right-handed leptonic mixing matrix is 
given by
\({\bf U}_{\rm PMNS}^{R} = {\bf O}_{e}^{\top}{\bf P}_{e}{\bf R}_{\pi/4}\).
Although ${\bf U}_{\rm PMNS}^{R}$ is in general nontrivial, it is a unitary
matrix. Therefore, the normalization of its muon row implies
\begin{equation}
 \sum_{i=1}^{3}
 \left| \left(U_{\rm PMNS}^{R}\right)_{\mu i} \right|^{2} =1.
\end{equation}
Furthermore, parity symmetry implies $g_R=g_L$. Consequently, within the two
limiting regimes considered in Eq.~(\ref{eq:115}), the contribution of the
right-handed neutrinos and the $W_R$ gauge boson to the muon anomalous
magnetic moment reduces to
\begin{equation}
 \Delta a_\mu(\nu_R,W_R) =
 \begin{cases}
  2.2\times10^{-11}
  \left( \dfrac{1~{\rm TeV}}{M_{W_R}} \right)^2, &
  M_{W_R}\gg m_{\nu_R}, \\[3mm]
  2.1\times10^{-11}
  \left(\dfrac{1~{\rm TeV}}{M_{W_R}}\right)^2, &
  M_{W_R}\simeq m_{\nu_R}.
 \end{cases}
\end{equation}
Thus, under the approximations entering in Eq.~(\ref{eq:115}), the dependence 
on the right-handed flavor-mixing parameters disappears after summing over the
three neutrino states. The resulting contribution is controlled only by the
right-handed gauge-boson mass and exhibits the characteristic scaling
$\Delta a_\mu\propto M_{W_R}^{-2}$. Figure~\ref{fig:amu_WR} shows this
behavior for $5~{\rm TeV}\leq M_{W_R}\leq15~{\rm TeV}$. The two limiting
cases follow nearly parallel curves and differ only slightly in their
normalization. In particular, for $M_{W_R}=5~{\rm TeV}$ one obtains
$\Delta a_\mu\simeq8.8\times10^{-13}$ and
$8.4\times10^{-13}$ for the $M_{W_R}\gg m_{\nu_R}$ and
$M_{W_R}\simeq m_{\nu_R}$ regimes, respectively, whereas for
$M_{W_R}=15~{\rm TeV}$ these contributions decrease to approximately
$9.8\times10^{-14}$ and $9.3\times10^{-14}$. Hence, the contribution
considered here becomes rapidly suppressed as the right-handed gauge boson
mass increases.

\begin{figure}
 \begin{tabular}{cc}
  \includegraphics[width=0.48\linewidth]{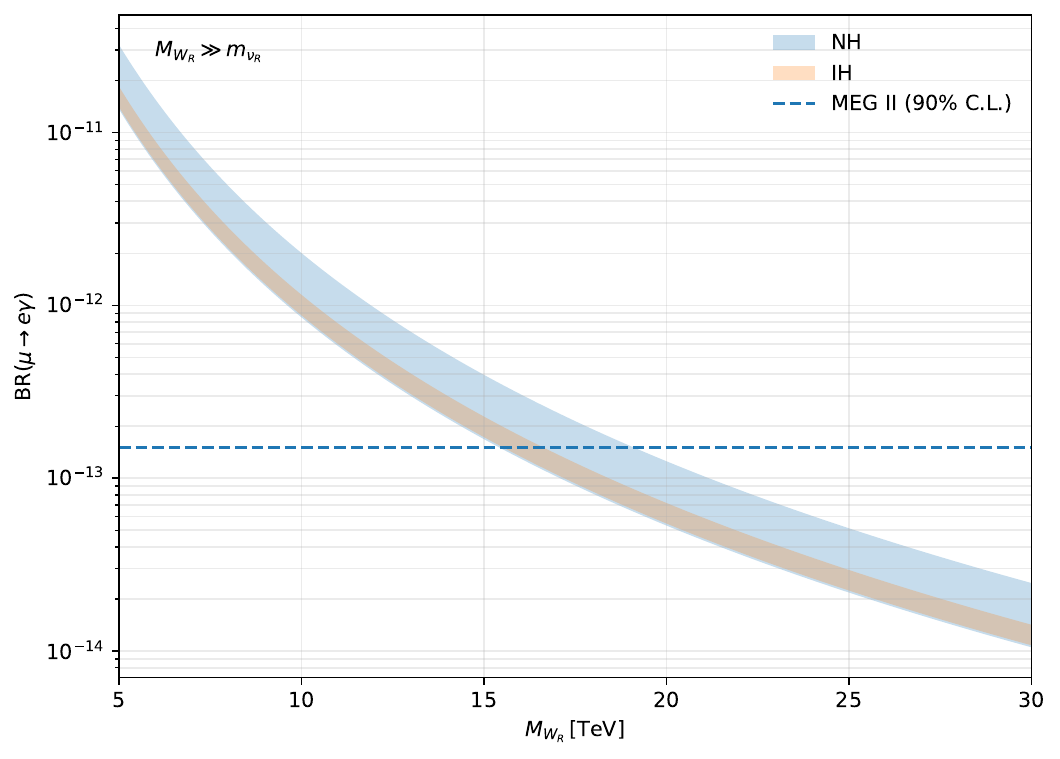} & 
  \includegraphics[width=0.48\linewidth]{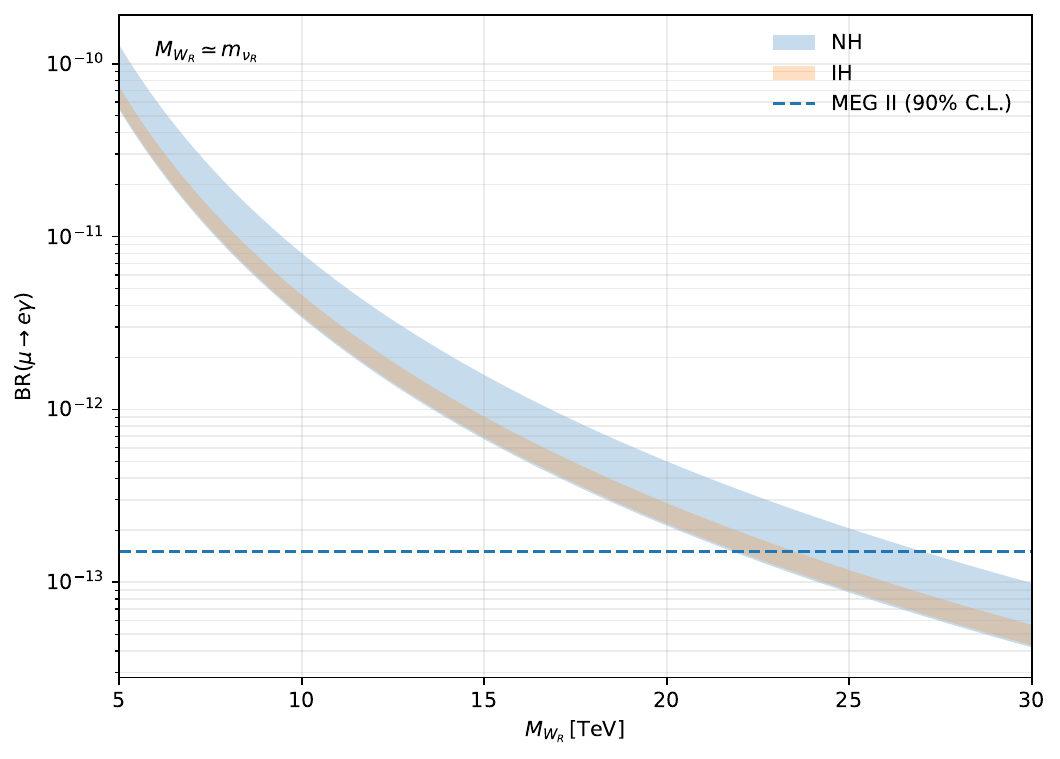}
 \end{tabular}
 \caption{Branching ratio for the lepton-flavor-violating decay
  $\mu\rightarrow e\gamma$ as a function of the right-handed charged
  gauge-boson mass $M_{W_R}$ for NH and IH. 
  The left panel corresponds to the regime
  $M_{W_R}\gg m_{\nu_R}$, whereas the right panel corresponds to
  $M_{W_R}\simeq m_{\nu_R}$. The shaded bands are obtained from the
  variation of the charged-lepton flavor factor $S_{e\mu}^{R}$ over the
  parameter configurations satisfying the numerical selection criterion
  $\chi^2<1$ in the lepton-sector analysis. Parity symmetry,
  $g_R=g_L$, has been assumed. The horizontal dashed line represents the
  current MEG~II upper bound,
  ${\rm BR}(\mu^+\rightarrow e^+\gamma)<1.5\times10^{-13}$ at
  $90\%$ C.L. Parameter configurations lying above this line are excluded
  by the present experimental limit. }
\label{fig:muegamma}
\end{figure} 

Lepton-flavor violation occurs in the neutrino sector through neutrino
oscillations, but has not been observed in charged-lepton processes.
The observation of charged-lepton flavor violation would therefore
provide a clear signal of physics beyond the SM. In particular, the
decay $\mu\rightarrow e\gamma$ has been extensively studied, and
stringent upper limits on its branching ratio have been established
(see Ref.~\cite{Lindner:2016bgg}).
Within the MLRSM, the branching ratio for this lepton-flavor-violating
process is given by
\begin{equation}\label{EQ:118-BR}
 \textrm{BR} \left( \mu\rightarrow e \gamma \right) = 
 \left\{ \begin{array}{lcc}
  5 \times 10^{-8} \left( \frac{g_{R}}{g_{L}} \right)^{4} 
  \left( \frac{1 \textrm{TeV}}{M_{W_{R}}} \right)^{4} 
  \displaystyle\sum^{3}_{i=1} 
   \vt \left({\bf U}^{\ast R}_{\textrm{PMNS}}\right)_{e i} 
   \left({\bf U}^{R}_{\textrm{PMNS}}\right)_{\mu i}\vt^{2}, &  
   \quad \textrm{if}\quad & 
   \quad  M_{W_{R}}\gg m_{\nu_{R}} \\
   2 \times 10^{-7} \left( \frac{g_{R}}{g_{L}} \right)^{4} 
   \left(\frac{1 \textrm{TeV}}{M_{W_{R}}}\right)^{4} 
   \displaystyle\sum^{3}_{i=1} 
   \vt \left({\bf U}^{\ast R}_{\textrm{PMNS}}\right)_{e i} 
   \left({\bf U}^{R}_{\textrm{PMNS}}\right)_{\mu i}\vt^{2}, &  
   \quad \textrm{if}\quad & 
   \quad  M_{W_{R}}\approx m_{\nu_{R}}.
  \end{array} \right.	
\end{equation}
For the lepton-flavor-violating decay $\mu\rightarrow e\gamma$, the
right-handed contribution depends on the combination
\begin{equation}
 {\cal S}_{e\mu}^{R}
 =
 \sum_{i=1}^{3}
 \left|
  \left({\bf U}_{\rm PMNS}^{R}\right)_{ei}^{*}
  \left({\bf U}_{\rm PMNS}^{R}\right)_{\mu i}
 \right|^2
 =
 \sum_{i=1}^{3}
 \left|
  \left({\bf U}_{\rm PMNS}^{R}\right)_{ei}
 \right|^2
 \left|
  \left({\bf U}_{\rm PMNS}^{R}\right)_{\mu i}
 \right|^2.
\end{equation}
In the MLRT scenario, using ${\bf U}_{\rm PMNS}^{R}={\bf O}_{e}^{\top}{\bf P}_{e}{\bf R}_{\pi/4}$
with ${\bf P}_{e}=\operatorname{diag}(1,1,e^{i\varphi_{ec}})$, this quantity
reduces to
\begin{equation}
\begin{aligned}
{\cal S}_{e\mu}^{R} ={}&
\frac14
(O_{11}+O_{21})^2
(O_{12}+O_{22})^2
+
\frac14
(O_{21}-O_{11})^2
(O_{22}-O_{12})^2
+
O_{31}^{2}O_{32}^{2}.
\end{aligned}
\end{equation}
Therefore, within the approximation adopted in Eq.~(\ref{EQ:118-BR}), the 
branching ratio can be written as
\begin{equation}
 {\rm BR}(\mu\rightarrow e\gamma)
 =
 \begin{cases}
  5\times10^{-8}
  \left(\dfrac{1~{\rm TeV}}{M_{W_R}}\right)^4
  {\cal S}_{e\mu}^{R}, &
  M_{W_R}\gg m_{\nu_R}, \\[3mm]
  2\times10^{-7}
  \left(\dfrac{1~{\rm TeV}}{M_{W_R}}\right)^4
  {\cal S}_{e\mu}^{R}, &
  M_{W_R}\simeq m_{\nu_R},
 \end{cases}
\end{equation}
where parity symmetry, $g_R=g_L$, has been used.
The predictions shown in Fig.~\ref{fig:muegamma} can be directly
confronted with the current MEG~II upper bound,
\(
 {\rm BR} (\mu^+\rightarrow e^+\gamma) < 1.5\times10^{-13}
 \quad (90\%~{\rm C.L.})
\)~\cite{MEGII:2025muegamma}.
Within the limiting approximation adopted in Eq.~(\ref{EQ:118-BR}), and using
parity symmetry, $g_R=g_L$, the branching ratio can be written as
\begin{equation}
 {\rm BR}(\mu\rightarrow e\gamma) =
 C
 \left( \frac{1~{\rm TeV}}{M_{W_R}} \right)^4 S_{e\mu}^{R},
\end{equation}
where $C=5\times10^{-8}$ for $M_{W_R}\gg m_{\nu_R}$ and
$C=2\times10^{-7}$ for $M_{W_R}\simeq m_{\nu_R}$. 
Therefore, the MEG~II limit translates into the lower bound
\begin{equation}
 M_{W_R} > 1~{\rm TeV}
 \left[ \frac{C\,S_{e\mu}^{R}}{1.5\times10^{-13}} \right]^{1/4}.
 \label{eq:MWR_MEGII}
\end{equation}
For the parameter configurations satisfying $\chi^2<1$, the charged-lepton
flavor factor lies in the intervals
\(0.1701\lesssim S_{e\mu}^{R}\lesssim0.4006 \quad{\rm (NH)} \),
and
\( 0.1748\lesssim S_{e\mu}^{R}\lesssim0.2298 \quad{\rm (IH)} \).
Consequently, in the regime $M_{W_R}\gg m_{\nu_R}$, the first allowed
parameter configurations appear at
$M_{W_R}\simeq15.43~{\rm TeV}$ for NH and
$M_{W_R}\simeq15.54~{\rm TeV}$ for IH. Requiring the entire flavor
band to satisfy the MEG~II bound gives the more conservative constraints
\( M_{W_R}\gtrsim19.12~{\rm TeV} \quad{\rm (NH)} \),
and
\( M_{W_R}\gtrsim16.64~{\rm TeV} \quad{\rm (IH)} \).
For the regime $M_{W_R}\simeq m_{\nu_R}$, the larger normalization of
the branching ratio strengthens these limits. The first allowed
configurations occur at approximately $21.82~{\rm TeV}$ for NH and
$21.97~{\rm TeV}$ for IH, whereas requiring the complete flavor bands
to lie below the MEG~II limit yields
\( M_{W_R}\gtrsim27.03~{\rm TeV} \quad{\rm (NH)} \),
and
\( M_{W_R}\gtrsim23.53~{\rm TeV} \quad{\rm (IH)} \).
Thus, within the approximation used in Eq.~(\ref{EQ:118-BR}), the present
$\mu\rightarrow e\gamma$ limit already provides a stringent constraint
on the right-handed scale of the model. In particular, the interval
$5~{\rm TeV}\leq M_{W_R}\leq15~{\rm TeV}$ previously considered lies
below the value at which even the first allowed flavor configurations
become compatible with MEG~II. The different widths of the NH and IH
bands originate from the different ranges of the charged-lepton mixing
matrix allowed by the corresponding lepton-sector fits.
%
\section{Conclusions}
%
We have investigated the origin of fermion mixing in a
non-minimal left-right symmetric framework supplemented by the
non-Abelian flavor symmetry $S_3$. The three fermion families are
organized as ${\bf 2}\oplus{\bf 1}_S$, while the scalar sector is
extended to three Higgs bidoublets together with one left-handed and
one right-handed triplet. After imposing parity symmetry and a
specific alignment of the bidoublet vacuum expectation values, the
quark and lepton mass matrices acquire a common structure that can be
treated analytically. An important consequence is that all fermion
mass matrices can be reduced to an effective two texture zeros form,
providing a unified description of the flavor sector in terms of
fermion masses, a small number of additional mass parameters, and
CP-violating phases.

In the pseudomanifest left-right theory (PLRT), the fermion mass 
matrices are complex symmetric, whereas in the manifest left-right 
theory (MLRT) they are Hermitian. 
Despite this difference, their diagonalization reveals a close 
connection between the two frameworks. In particular, the MLRT mixing 
structure emerges as a more restrictive realization of the PLRT one, 
with fewer independent phase factors. 
In the quark sector, the common $R_{\pi/4}$ rotation induced by 
the $S_3$ structure cancels from the CKM matrix, so that the observed 
quark mixing originates from fermion mass hierarchies and relative 
phases.
Our numerical analysis reproduces the observed CKM matrix, including
all nine magnitudes, in agreement with current global-fit values. In
the lepton sector, both NH and IH can reproduce the observed PMNS mixing 
pattern. Moreover, although
the PLRT and MLRT parametrizations have different underlying phase
structures, they contain the same number of effective phases entering
the left-handed PMNS matrix. Consequently, neutrino oscillation data
alone cannot distinguish between the two realizations.
At the level of the absolute neutrino-mass scale, however, the two
orderings behave quite differently. The NH best-fit point gives
$\Sigma_{\nu}\simeq0.063~{\rm eV}$ and is compatible with current
ordering-dependent cosmological constraints. In contrast, the IH
best-fit point yields $\Sigma_{\nu}\simeq0.204~{\rm eV}$, with the
local parameter region considered here corresponding approximately to
$\Sigma_{\nu}\simeq0.175$--$0.216~{\rm eV}$. This region is in strong
tension with current DESI+CMB limits within $\Lambda$CDM. Since these
bounds are cosmological-model dependent, we do not regard this as a
model-independent exclusion of the IH realization, but rather as an
important complementary constraint on this branch of the model.
The phenomenological analysis provides additional ways of probing the
model. For neutrinoless double-beta decay mediated by light Majorana
neutrinos, the parameter configurations satisfying $\chi^2<1$ predict
\[
 4.63~{\rm meV}\lesssim m_{\beta\beta}^{\rm NH}
 \lesssim7.74~{\rm meV},
 \qquad
 64.85~{\rm meV}\lesssim m_{\beta\beta}^{\rm IH}
 \lesssim77.52~{\rm meV},
\]
with best-fit values of approximately $5.10~{\rm meV}$ and
$71.00~{\rm meV}$, respectively. 
The strong separation between the
two orderings is one of the most distinctive predictions of the
model. In particular, the IH region lies within the
mass scale expected to be explored by next-generation
neutrinoless double-beta decay experiments, whereas testing the NH 
prediction would require sensitivity at the few-meV
level. 
The contribution from right-handed neutrinos is governed predominantly 
by the heavy-neutrino scale and the right-handed charged gauge-boson mass,
$|\eta_{\nu R}^{R}|\propto M_N^{-1}M_{W_R}^{-4}$. The numerical 
analysis further shows that the flavor-dependent correction remains 
close to unity over most of the allowed parameter space, explaining 
the nearly overlapping NH and IH predictions for this contribution. 

The same right-handed flavor structure also leaves characteristic
signatures in other low-energy observables. Within the limiting
expressions considered here, the $(\nu_R,W_R)$ contribution to the
muon anomalous magnetic moment scales as
$\Delta a_\mu\propto M_{W_R}^{-2}$ and becomes strongly suppressed as
the right-handed scale increases. In contrast,
$\mu\to e\gamma$ retains a direct sensitivity to the charged-lepton
flavor structure through
${\cal S}_{e\mu}^{R}$ and exhibits the stronger scaling
${\rm BR}(\mu\to e\gamma)\propto M_{W_R}^{-4}$.
The present MEG~II bound already imposes stringent constraints on the
right-handed scale. In the regime $M_{W_R}\gg m_{\nu_R}$, requiring the full 
flavor bands obtained from the lepton sector analysis
to satisfy the MEG~II limit leads to the conservative lower bounds
\( M_{W_R}\gtrsim 19.1~{\rm TeV} \quad {\rm (NH)} \),
\( M_{W_R}\gtrsim 16.6~{\rm TeV} \quad {\rm (IH)} \).
For the regime $M_{W_R}\simeq m_{\nu_R}$, the larger normalization of
the branching ratio strengthens these constraints to approximately
\( M_{W_R}\gtrsim 27.0~{\rm TeV} \quad {\rm (NH)} \),
\( M_{W_R}\gtrsim 23.5~{\rm TeV} \quad {\rm (IH)} \).
These bounds should be understood within the limiting approximation
adopted for the $\mu\to e\gamma$ rate and correspond to requiring the
complete flavor bands to lie below the present experimental upper
limit. Thus, the current MEG~II result already probes the right-handed
parameter space of the model beyond the $5$--$15~{\rm TeV}$ interval
initially explored. Future improvements in searches for
$0\nu\beta\beta$, $\mu\to e\gamma$, and right-handed charged currents
will provide complementary and increasingly stringent tests of the
flavored left-right symmetric framework developed here.
%
\section*{Acknowledgements}
%
This work was partially supported by SIP-IPN 20261005. FGC acknowledges the 
financial support from {\it SECIHTI} under Grant CBF-2025-G-1187.

%
\appendix
%
\section{The discrete \texorpdfstring{$\mathbf{S}_{3}$}{S3} flavor symmetry}
The non-Abelian group $S_{3}$ is the permutation group of three
objects~\cite{Ishimori:2010au} and has three irreducible representations: 
two 1-dimensional, ${\bf 1}_{S}$ and ${\bf 1}_{A}$, and one 2-dimensional 
representation, ${\bf 2}$. We list the multiplication rules among them
\begin{eqnarray}\label{rules}
 {\bf 1}_{S} & \otimes & {\bf 1}_{S}={\bf 1}_{S}, \qquad 
 {\bf 1}_{A} \otimes {\bf 1}_{S} = {\bf 1}_{A}, \qquad 
 {\bf 1}_{S} \otimes {\bf 1}_{A} = {\bf 1}_{A}, \nonumber \\
 {\bf 1}_{A} & \otimes & {\bf 1}_{A} = {\bf 1}_{S}, \qquad 
 {\bf 1}_{S} \otimes {\bf 2} = {\bf 2}, \qquad 
 {\bf 1}_{A} \otimes {\bf 2} = {\bf 2} \nonumber \\
 \begin{pmatrix}
  a_{1} \\ 
  a_{2}
 \end{pmatrix}_{{\bf 2}}
 & \otimes &
 \begin{pmatrix}
  b_{1} \\ 
  b_{2}
 \end{pmatrix}_{{\bf 2}}=
 \left( a_{1} b_{1} + a_{2} b_{2} \right)_{{\bf 1}_{S}} 
 \oplus  
 \left( a_{1} b_{2} - a_{2} b_{1} \right)_{{\bf 1}_{A}} 
 \oplus	
 \begin{pmatrix} 
  a_{1} b_{2} + a_{2}b_{1} \\ 
  a_{1} b_{1} - a_{2}b_{2}
 \end{pmatrix}_{{\bf 2}}. 
\end{eqnarray}

\bibliographystyle{bib_style_T1}
\bibliography{references.bib}

\end{document}